\documentclass[a4paper,UKenglish,cleveref, autoref, thm-restate]{lipics-v2021}
\nolinenumbers
\usepackage{amsmath,stmaryrd}
\usepackage{amsthm}
\usepackage{xcolor}
\usepackage{mathpartir}
\usepackage{listings}
\usepackage{mathtools}
\usepackage{cancel}
\usepackage{etoolbox}
\newcommand{\defeq}{\triangleq}
\newcommand{\dom}[1]{\mathrm{dom}(#1)}
\newcommand{\im}[1]{\mathrm{im}(#1)}

\newcommand{\emptymap}{\varepsilon}
\newcommand{\emptylist}{[]}

\newcommand{\supp}{\mathsf{supp}}
\newcommand{\disjoint}{\#}
\newcommand{\Names}{\mathrm{Names}}
\newcommand{\nn}{\mathtt{n}}

\newcommand{\freen}{\mathsf{supp}}%{\mathbf{fn}}

\newcommand{\op}{\mathsf{op}}
\newcommand{\opr}{\mathsf{red}}
\newcommand{\Terms}{\mathrm{Terms}}
\newcommand{\term}{\mathtt{M}}
\newcommand{\termb}{\mathtt{N}}
\newcommand{\opred}{\mapsto_{\opr}}
\newcommand{\Opred}{{|\mathrel{\mkern-8mu}\Rightarrow}_{\op}}

\newcommand{\var}{x}
\newcommand{\varb}{y}

\newcommand{\subst}[2]{\{#1 := #2 \}}
\newcommand{\substg}[1]{\{ #1 \}}
\newcommand{\Values}{\mathrm{Values}}

\newcommand{\val}{\mathtt{V}}
\newcommand{\valb}{\mathtt{W}}

\newcommand{\ectx}{\mathtt{E}}
\newcommand{\pectx}{\mathtt{D}} % evaluation contexts including par
\newcommand{\hole}{\bullet}

\newcommand{\ctx}{\mathtt{C}}

\newcommand{\Types}{\mathrm{Types}}
\newcommand{\type}{\sigma}
\newcommand{\typeb}{\tau}

\newcommand{\typectx}{\Gamma}
\newcommand{\emptyctx}{\varnothing}
\newcommand{\typectxb}{\Delta}
\newcommand{\covartype}[1]{\lnot#1}
\newcommand{\ectxtype}[2]{#1 \rightsquigarrow #2}

\newcommand{\Unit}{\mathtt{Unit}}
\newcommand{\unit}{()}
\newcommand{\Bool}{\mathtt{Bool}}
\newcommand{\bool}{\mathtt{b}}
\newcommand{\true}{\mathtt{true}}
\newcommand{\false}{\mathtt{false}}
\newcommand{\Int}{\mathtt{Int}}
\newcommand{\intv}{\mathtt{k}}
\newcommand{\baseval}{\mathtt{B}}
\newcommand{\gtype}{\iota}

\newcommand{\inop}{\mathtt{in}}
\newcommand{\ifop}{\mathtt{if}}
\newcommand{\thenop}{\mathtt{then}}
\newcommand{\elseop}{\mathtt{else}}
\newcommand{\ifte}[3]{\ifop \ #1 \ 
  \thenop \ #2 \ \elseop \ #3}

\newcommand*{\letrec}[3]{\mathtt{rec}\,#1(#2).#3}

\newcommand{\pairop}[1]{\langle #1 \rangle}

\newcommand{\letin}[3]{{\mathsf{let}}\, #1 \mathbin{=} #2\, {\mathsf{in}}\, #3}

\newcommand{\named}[2]{[#1]#2}

\newcommand*{\ctxle}{\preceq_{ctx}}
\newcommand{\ctxeq}{\simeq_{ctx}}

\newcommand{\ciule}{\preceq_{ciu}}
\newcommand{\ciueq}{\simeq_{ciu}}

\newcommand{\depst}{\mathbf{D}}
\newcommand{\depstsup}{|\depst|}
\newcommand{\depstjust}{\leq_{\depst}}
\newcommand{\depstdep}{\#_{\depst}}
\newcommand{\depsttr}[1][\tr]{\mathbf{D}_{#1}}
\newcommand{\depstsuptr}[1][\tr]{|\depsttr[#1]|}
\newcommand{\depstjusttr}[1][\tr]{\leq_{\depsttr[#1]}}
\newcommand{\depstdeptr}[1][\tr]{\#_{\depsttr[#1]}}
\newcommand{\depstTr}[1]{\mathbf{Tr}(#1)}
\newcommand{\depstCTr}[1]{\mathbf{CTr}(#1)}

\newcommand{\newref}[3]{\mathtt{let} \ #1 = \mathtt{ref}\ #2 \ \inop \ #3}
\newcommand{\refterm}[1]{\mathtt{ref}\ #1}
\newcommand{\reftype}[1]{\mathbf{ref}_{#1}}
\newcommand{\cas}[3]{\mathtt{CAS}\ #1\ #2\ #3}

\newcommand{\store}{\mathtt{S}}
\newcommand{\storeb}{\mathtt{T}}

\newcommand{\storectx}{\Sigma}

\newcommand{\loc}{\ell}
\newcommand{\storeP}{\store_{\Pl}}
\newcommand{\storeO}{\store_{\Op}}

\newcommand{\typeJudge}[4]{#1; #2 \vdash #3 : #4}
\newcommand*{\typeJudgeg}[3][\gamma]{#2 \vdash #1 : #3}
\newcommand*{\storeJudge}[2]{\vdash #1 : #2}

\newcommand{\aval}{\mathtt{A}}
\newcommand{\avalb}{\mathtt{B}}

\newcommand{\CNames}{\mathrm{CNames}}

\newcommand{\pn}{\mathsf{p}}
\newcommand{\qn}{\mathsf{q}}
\newcommand{\rn}{\mathsf{r}}
\newcommand{\fn}{\mathsf{f}}

\newcommand{\gn}{\mathsf{g}}
\newcommand{\hn}{\mathsf{h}}
\newcommand{\FNames}{\mathrm{FNames}}

\newcommand{\move}{\mathbf{m}}

\newcommand{\pmove}{\mathbf{p}}
\newcommand{\omove}{\mathbf{o}}

\newcommand{\imove}{\mathbf{i}}
\newcommand{\oplabel}{}
\newcommand{\Pl}{\mathsf{P}}
\newcommand{\Op}{\mathsf{O}}

\newcommand{\PQ}{\mathsf{PQ}}
\newcommand{\PA}{\mathsf{PA}}
\newcommand{\OQ}{\mathsf{OQ}}
\newcommand{\OA}{\mathsf{OA}}
\newcommand{\moveP}[2]{\overline{#1}(#2)}
\newcommand{\moveO}[2]{#1(#2)}
\newcommand{\movePI}[1]{?(#1)}
\newcommand{\moveOI}[1]{!(#1)}
\newcommand{\ansP}[2]{\overline{#1}(#2)}
\newcommand{\ansO}[2]{#1(#2)}
\newcommand{\questP}[2]{\overline{#1}(#2)}
\newcommand{\questO}[2]{#1(#2)}
\newcommand{\switchio}[1]{\overline{#1}}
\newcommand{\justif}[2]{\ensuremath{#1\curvearrowleft #2}}
\newcommand{\njustif}[2]{\ensuremath{#1\cancel{\curvearrowleft} #2}}

\newcommand{\justifstar}[2]{\ensuremath{#1\curvearrowleft^* #2}}
\newcommand{\lab}{\ensuremath{\mathbf{l}}}
\newcommand{\eqPl}{\equiv_\Pl}
\newcommand{\eqOp}{\equiv_\Op}  
\newcommand{\lts}{\mathcal{L}}

\newcommand{\act}{\mathbf{a}}
\newcommand{\ctuple}[1]{\langle #1 \rangle}

\newcommand{\Actions}{\mathrm{Actions}}

\newcommand{\LtsTr}[2]{\mathbf{Tr}_{#1}(#2)}
\newcommand{\LtsCTr}[2]{\mathbf{CTr}_{#1}(#2)}

\newcommand{\tr}{\mathbf{t}}

\newcommand{\trle}{\preceq_{\mathsf{tr}}}
\newcommand{\treq}{\simeq_{\mathsf{tr}}}
\newcommand{\lttr}{\preceq}
\newcommand*{\opLTS}{\lts_{{\mathtt{env}}}}
\newcommand{\Intera}{\mathsf{I}}
\newcommand{\ILTS}{\lts_{\Intera}}
\newcommand{\itrans}[1]{\xrightarrow{#1}_{\Intera}}
\newcommand{\iTrans}[1]{\xRightarrow{#1}_{\Intera}}

\newcommand{\iconf}{\mathbb{I}}
\newcommand{\iconfb}{\mathbb{J}}

\newcommand{\threadpool}{\mathtt{P}}
\newcommand{\threadpoolb}{\mathtt{Q}}
\newcommand{\threadpoolP}{\mathtt{P}_{\Pl}}

\newcommand{\threadpoolO}{\mathtt{P}_{\Op}}

\newcommand{\parcomp}{|}

\newcommand{\Ty}{\mathsf{Ty}}
\newcommand{\TyLTS}{\lts_{\Ty}}

\newcommand{\tconf}{\mathbb{S}}
\newcommand{\tconfb}{\mathbb{T}}
\newcommand{\tconfP}{\mathbb{S}_{\Pl}}
\newcommand{\tconfO}{\mathbb{S}_{\Op}}

\newcommand{\namectxP}{\Delta_{\Pl}}
\newcommand{\namectxO}{\Delta_{\Op}}

\newcommand{\namectx}{\Gamma}
\newcommand{\namectxb}{\Delta}
\newcommand{\typectxP}{\Delta_P}
\newcommand{\typectxO}{\Delta_O}

\newcommand{\ttrans}[1]{\xrightarrow{#1}_{\Ty}}
\newcommand{\tTrans}[1]{\xRightarrow{#1}_{\Ty}}
\newcommand{\dualTy}[2]{{#2}^{*_#1}}

\newcommand{\typeJudgePat}[4]{#2 \Vdash #3 : #4}
\newcommand{\lnamectx}{\Delta}

\newcommand{\Lvis}{\mathsf{vis}}
\newcommand{\VisLTS}{\lts_\mathsf{vis}}

\newcommand{\vtrans}[1]{\xrightarrow{#1}_{\Lvis}}
\newcommand{\vTrans}[1]{\xRightarrow{#1}_{\Lvis}}
\newcommand{\vconf}{\mathbb{V}}
\newcommand{\vconfb}{\mathbb{W}}

\newcommand{\view}{\mathcal{V}}
\newcommand{\viewF}{\mathcal{F}}

\newcommand{\viewFP}{\mathcal{F}_{\Pl}}

\newcommand{\viewFO}{\mathcal{F}_{\Op}}
\newcommand{\stackT}{\mathcal{T}}
\newcommand{\treeaux}{{\ensuremath{\mathtt{Tree\_aux}}}\,}
\newcommand{\tree}{{\ensuremath{\mathtt{Tree}}}\,}
\newcommand{\addleaf}[3]{\ensuremath{\mathtt{addLeaf}(#1,#2,#3)}}
\newcommand{\removeleaf}[2]{\ensuremath{\mathtt{removeLeaf}(#1,#2)}}
\newcommand{\some}[1]{\ensuremath{\mathtt{Some}}(#1)}
\newcommand{\none}{\ensuremath{\mathtt{None}}}
\newcommand{\viewG}{\mathcal{G}}
\newcommand{\stackTP}{\mathcal{T}_{\Pl}}
\newcommand{\viewGP}{\mathcal{G}_{\Pl}}
\newcommand{\stackTO}{\mathcal{T}_{\Op}}
\newcommand{\viewGO}{\mathcal{G}_{\Op}}

\newcommand{\OGS}{\mathsf{OGS}}
\newcommand{\ogs}{\mathsf{ogs}}
\newcommand{\OGSLTS}{\lts_{\OGS}}
\newcommand{\OGSConfs}{\mathrm{Confs}_{\ogs}}

\newcommand{\ogsconf}{\mathbb{G}}
\newcommand{\ogsconfb}{\mathbb{H}}
\newcommand{\ogsconfP}{\mathbb{G}_{\Pl}}
\newcommand{\ogsconfO}{\mathbb{G}_{\Op}}
\newcommand{\ogsconfPb}{\ogsconfb_{\Pl}}
\newcommand{\ogsconfOb}{\ogsconfb_{\Op}}
\newcommand{\ogsTrans}[1]%{\xRightarrow{#1}_{\ogs}}
{\mathrel{\stackrel{{\;\;#1\;\;}}{\Longrightarrow}_{\ogs}}} 
\newcommand{\ogstrans}[1]{\xrightarrow{#1}_{\ogs}}
\newcommand{\ogsTr}[1]{\LtsTr{\OGS}{#1}}
\newcommand{\ogsCTr}[1]{\LtsCTr{\OGS}{#1}}

\newcommand{\gammaP}{\gamma_{\Pl}}
\newcommand{\gammaO}{\gamma_{\Op}}

\newcommand{\soopred}{\mapsto^{\opr}_{\so}}
\newcommand{\sosynchred}{\mapsto^{\synchlabel}_{\so}}
\newcommand{\sored}{\mapsto_{\so}}
\newcommand{\parr}{\mathrel{||}}
\newcommand{\compose}[2]{\bigl(#1\parr#2\bigr)}
\newcommand{\Merge}{\mathrel{\wedge\mkern-8mu\wedge}}
\newcommand{\recons}[2]{\bigl(#1\,\Merge\,#2\bigr)}
\newcommand{\reconscont}[2]{[#1]_{#2}^{\mathtt{c}}}
\renewcommand{\so}{\mathsf{so}}
\newcommand{\synchlabel}{\mathsf{sync}}

\newcommand{\wbarb}[2]{#1\!\Downarrow#2}

\newcommand{\acc}[2]{\mathtt{Acc}(#1,#2)} %set of accessible names
\newcommand{\accmeas}[1]{m_{#1}} %measure of accessible names

\usepackage{macros-ken}

\newtoggle{showcomments}

\togglefalse{showcomments} %uncomment to hide comments

\newcommand{\gj}[1]{\iftoggle{showcomments}{{\color{magenta}GJ:\footnotesize#1}}{}}
\newcommand*{\sk}[1]{\iftoggle{showcomments}{\textcolor{teal}{\footnotesize Ken: #1}}{}}
\newcommand*{\iq}[1]{\iftoggle{showcomments}{\textcolor{violet}{\footnotesize Iwan: #1}}{}}
\newcommand*{\ds}[1]{\iftoggle{showcomments}{\textcolor{blue}{\footnotesize DS: #1}}{}}

\newcommand*{\iqr}[1]{#1}

\newtoggle{fullappendix}
\togglefalse{fullappendix}

\title{Concurrent Visibility: higher-order concurrency with first-order store} %TODO Please add

\titlerunning{Concurrent Visibility} %TODO optional, please use if title is longer than one line

\author{Iwan Quémerais}{ENS de Lyon \& University of Bologna}{}{}{}
\author{Guilhème Jaber}{Nantes Université}{}{https://orcid.org/0009-0006-7149-5073}{}
\author{Ken Sakayori}{The University of Tokyo}{}{https://orcid.org/0000-0003-3238-9279}{JSPS KAKENHI Grant Number JP24K20731}
\author{Davide Sangiorgi}{University of Bologna \& INRIA}{}{}{}%TODO mandatory, please use full name; only 1 author per \author macro; first two parameters are mandatory, other parameters can be empty. Please provide at least the name of the affiliation and the country. The full address is optional

\authorrunning{I. Quémerais,  G. Jaber, K. Sakayori, D. Sangiorgi} %TODO mandatory. First: Use abbreviated first/middle names. Second (only in severe cases): Use first author plus 'et al.'

\Copyright{Iwan Quémerais, Guilhème Jaber, Ken Sakayori and Davide Sangiorgi} %TODO mandatory, please use full first names. LIPIcs license is "CC-BY";  http://creativecommons.org/licenses/by/3.0/


 %TODO mandatory; please add comma-separated list of keywords

\EventEditors{Ana Sokolova and Patrick Totzke}
\EventNoEds{2}
\EventLongTitle{37th International Conference on Concurrency Theory (CONCUR 2026)}
\EventShortTitle{CONCUR 2026}
\EventAcronym{CONCUR}
\EventYear{2026}
\EventDate{September 1--4, 2026}
\EventLocation{Liverpool, UK}
\EventLogo{}
\SeriesVolume{391}
\ArticleNo{15}
\begin{document}

\maketitle

\begin{abstract}
  We propose an Operational Game Semantics for a (call-by-value)
  concurrent higher-order language with first-order store (references
  may contain other references or first-order values such as integers
  or booleans; however they may not store higher-order values such as   functions).  We
  adapt the game-semantic notion of \emph{visibility},
  which semantically captures the absence of higher-order references and  developed for sequential
  higher-order languages,
  to a concurrent setting.  We thus define a
  \emph{complete-trace} preorder, 
% (a complete trace intuitively 
%   representing a `completed' sequence of interactions between a term and
%   its environment).  
and  prove it sound for 
 the contextual preorder, by introducing a
  synchronization-based composition of semantic configurations and
  establishing an observational adequacy result.  We also prove
  completeness for the subset of the language in which functions
  return first-order values.

In contrast to the  case of  sequential visibility, 
in the labeled
transition semantics we have to account for the presence of multiple
active threads with possibly different visibilities, and of a
tree-like structure for managing the dependencies among the threads so
created. Moreover, % in the completeness proof 
we have to reason on families
of traces, rather than single traces, as in concurrent setting the
order among certain actions %of a trace 
cannot be enforced.

\end{abstract}

\section{Introduction}

A central question in programming language semantics is how the presence or absence of a
given program construct, such as mutable store, control operators, parallelism, or
nondeterministic choice, is reflected in the semantics. From the point of view of
\emph{contextual equivalence}, this question can be approached by asking whether adding a
construct gives contexts more power to distinguish programs. For instance, in sequential
higher-order languages, local store
allows a context to distinguish $\lambda f.\,f()$ from $\lambda
f.\,f();f()$, by counting 
how many times $f$ is called.  

In interactive models of computation, programs are described through their possible
interactions with the environment, such as the calls to $f$ in the example above. The
absence of a program construct can then often be characterized semantically as a
restriction on the allowed interactions. \emph{Game
  semantics}~\cite{HylandOng00,AbramskyJM00}, which emerged and grew in
the realm of sequential higher-order languages,  has provided several
striking examples: % of this phenomenon: 
the absence of non-local control operators
corresponds to \emph{well-bracketing}~\cite{Laird97}, the absence of higher-order store to \emph{visibility}~\cite{AbramskyMcCusker96}, and
stronger restrictions on state access give rise to conditions such as \emph{innocence}~\cite{HylandOng00}. In
operational presentations of game semantics~\cite{JaberMurawski21,Laird07}, based on labelled transition systems, such
conditions are typically enforced by enriching the states with additional components
recording suitable information about  past interactions. 

When contexts have access to parallelism, they can distinguish more programs than in the sequential setting. Consider the following example:
\begin{example}
  \label{exa:ex1}
\newcommand*{\termfalse}{\term_{\false}}
\begin{align*}
  \term_1 &\defeq \newref \var \true {\app f {(\lambda \_. \writeop \var \false); \false}} \\
  \term_2 &\defeq \newref \var \true {\app f {(\lambda \_. \writeop \var \false); \readop \var \mathrel{\&\&} \mathtt{not}\,\readop \var }}\\
  % \iqr{\ectx}&\iqr{\defeq \parop{(\letin \var \hole {\ifte \var \unit \Omega})}{\readop \loc \unit}}\\
  % \iqr{f}&\iqr{\defeq \lambda \var. \writeop \loc \var}
\end{align*}
\end{example}
Both terms first create a boolean reference stored in $\var$, before calling a higher-order function $f$ with a setter function as an argument.
In a sequential setting, the two terms are contextually equivalent, since the expression
$\readop \var \mathrel{\&\&} \mathtt{not}\,\readop \var$ always evaluates to $\false$: 
the value stored in  the reference  $\var$ may not be changed between the two  reads of   $\var$. %  In
% contrast, in a language combining parallelism with higher-order references, a context could
% distinguish $  \term_1$ and $  \term_2$, 
% by first
% storing the setter function $\lambda \_. \writeop \var \false$ passed to $f$ in a
% higher-order reference, and then invoking it in a separate thread between the two reads
% performed by the program.
\iqr{In contrast, in a language combining parallelism with higher-order references, the following context $\ectx$ and instance of $f$
distinguishes $\term_1$ and $  \term_2$, by 
 storing the setter function $\lambda \_. \writeop \var \false$ passed to $f$ in the
higher-order reference $\loc$, and invoking it in a separate thread; this invocation can be performed
between the two reads performed by the program.
\[\ectx \defeq \parop{(\letin \var \hole {\ifte \var \unit \Omega})}{\readop \loc \unit} \hspace{3em} f \defeq \lambda \var. \writeop \loc \var\]
Thus
$\ectx[\term_1]$ always diverges while $\ectx[\term_2]$ can terminate.}
Higher-order storage is essential here:
once the call to $f$ has returned, the setter is no longer in scope and would not
otherwise be available. This suggests that,
even in the presence of concurrency, the terms $  \term_1$ and $  \term_2$
 should remain equivalent provided that higher-order
references are unavailable. The question is then how to capture this semantically. 

\ds{The goal of this paper is precisely how 
to answer such a  question~---
how to 
to capture the constraint to  first-order store
in a   concurrent higher-order language~---
 in the setting of \emph{operational game
semantics} (OGS).}
As a language, we adopt a call-by-value higher-order
language with parallel composition and full ground references
 (full ground meaning that references
  may contain other references or first-order values such as integers
  or booleans; however they may not contain higher-order values such as   functions).
We then
  adapt the game-semantic notion of \emph{visibility}
  to a concurrent setting. 
For this we first define an \emph{interactive 
 labelled transition system} (LTS), expressing possible interaction between a term and its
 environment. Following the tradition of operational game semantics as well as of
 higher-order concurrency, the action of this LTS may not contain higher-order values; 
 higher-order values   are replaced by abstract values, technically,
 names, that provide
 access to the actual values.  
In this setting, the \emph{visibility} of a thread is intuitively the set of names of
functions  that the
thread is allowed to call (i.e., the functions that are in its scope).  
We then couple this interactive LTS  with two 
further components: 
 a \emph{typing LTS},   controlling the introduction of
fresh names together with their types; a \emph{visibility LTS}, tracking the 
 possibly different visibilities  of the current active threads.  
The use of the  visibility LTS allows us to 
take into account  the absence
of higher-order references in the model. Such LTS  also records the call-return well-bracketed
discipline of interaction among terms of the language, 
reflecting the fact that the language has no non-local control
operators. 
In the sequential case,  well-bracketing is managed using a stack; 
in our concurrent case,  we have to adopt 
 a
tree-like structure, recording  dependencies among the threads
created.

On top of the composite LTS,
we define a
  \emph{complete-trace} preorder (a complete trace intuitively 
  representing a `completed' sequence of interactions between a term and
  its environment).  Our first main result is  soundness of the 
complete-trace preorder
 with respect
  to the contextual preorder. 
This is obtained  
 by introducing a
  synchronization-based composition of semantic configurations and
  establishing an observational adequacy result. 
Our second main result 
is   completeness for the subset of the language in which functions
  return first-order values.
The usual strategy of  completeness proofs is to  consider two terms in the context
and a complete trace
for one of them. Then one 
constructs  a context that produces the complement trace,  followed by a success signal.
From the contextual relation between the two terms one then concludes that the
second term also has that trace.
The difficulty in our concurrent setting  is that 
the
order among certain actions of a trace (produced by a context or a term) cannot be enforced.
Indeed, certain consecutive actions of traces may always be commuted. 
Thus we have to reason about families of traces, rather than single traces. 
This complicates the proof significantly, and brings in the requirement, discussed in the paper, that
  functions   return first-order values.

In this paper, we do not consider the possibility for the program and its environment to share a store.
While game semantics provides a way to represent such shared store by enriching the actions of the LTS~\cite{Laird07,MurawskiTzevelekos11},
this complexifies the models in a direction that is mostly orthogonal to the treatment of visibility.

% We the  define a labelled transition semantics for this language
%  a call-by-value higher-order
% language with parallel composition and full ground references
%  (full ground meaning that references
%   may contain other references or first-order values such as integers
%   or booleans; however they may not contain higher-order values such as   functions).
% The semantics combines
% three components: an interactive component recording the exchange of abstract values
% between a program and its environment, a typing component controlling the introduction of
% fresh names together with their types, and a visibility component tracking the names that
% are dynamically available across threads. This last component is what enforces the absence
% of higher-order references in the model. It also records the call-return well-bracketed
% discipline of interaction, reflecting the fact that our language has no non-local control
% operators.

\section{Target Language and Operational Semantics}

%\subsection{Types and Syntax}
\label{sec:type_syntax}
We work with a call-by-value \( \lambda \)-calculus extended with
 references and parallel compositions.
A reference may contain references to other
references~---allowing programmers to write cyclic data
structures---~but may not store functions; such references are called
\emph{full ground}~\cite{MurawskiTzevelekos12}.
Types,  values, terms and (sequential) evaluation contexts are defined by the following grammars:
\begin{align*}
  \Gtypes \quad \gtype &\Coloneqq \Unit \mid \Bool \mid \Int \\
  \Rtypes \quad \rtype &\Coloneqq \reftype \gtype \mid \reftype \rtype \\
  \Types \quad \type, \typeb &\Coloneqq \gtype \mid \rtype \mid \type \times \typeb \mid \type \to \typeb
% \end{align*}
% % The set of values, terms and (sequential) evaluation contexts are
% % given as follows: 
% \begin{align*}
\\
  \Values \quad  \val, \valb &\Coloneqq x \mid \loc \mid \unit \mid \bool \mid \intv \mid \pairop {\val, \valb} \mid \lambda x. \term \\
  \Terms \quad \term, \termb &\Coloneqq \val \mid \app \val \term \mid \projop \val \mid \ifte \val \term \termb \mid \letrec f x \term \\
  &\quad \mid \refterm\val \mid \readop \val \mid \writeop \val \valb \mid \parop \term \termb \mid \cas \val \val \val \\
  \text{Sequential contexts} \quad \ectx &\Coloneqq \hole \mid \app \val \ectx \\
  \text{Evaluation contexts} \quad \pectx &\Coloneqq \hole \mid \app \val \pectx \mid  \parop \pectx \term \mid  \parop \term \pectx \\
 % \text{Evaluation contexts} \quad \pectx &\Coloneqq \ectx \mid \par \val \ectx \\
\end{align*}
We let  \( x \) range over  variables, \( \loc \) over
locations,   \( \bool \in \{ \true, \false \} \), and  \( \intv \) over
 integers. 
The term \( \parop \term \termb \) is a `parallel pair' that
executes the first and the second element in parallel.
Such parallel pair construct can be found in~\cite{BirkedalST12},
which was a source of inspiration for our language.
A compare-and-swap operation \( \cas \loc \val_1 \val_2 \)
 \emph{atomically}  compares the value stored at the location \( \loc \) with \(
\val_1 \),  and swaps it with \( \val_2 \) if they are the same. 
% Note that the compare-and-swap operation is \emph{atomic} in that
% the comparison and the swap cannot be interrupted by other terms
% running in parallel. 
Other constructs, from 
 \( \lambda \)-calculus with store,
 are standard.
By restricting the place where we can have terms rather than values in the syntax,
we force the order of evaluation to be fully explicit, as is done with administrative normal forms~\cite{DBLP:conf/lfp/SabryF92}.
Doing so, sequential evaluation contexts $\ectx$ and parallel ones $\pectx$, used in defining the operational semantics and the OGS, have a particularly simple definition.
As a syntax sugar, we sometimes write:  \( \letin x \term \termb \)
to mean
 \( \app {(\lambda x . \termb)} \term \); and
 \( \app \term \termb \) (resp.\  
 \( \pairop{\term, \termb} \))  to mean \( \letin x \term
{\app x \termb} \) (resp.\  \( \letin x \term {\letin y \termb {\pairop{x,
      y}}} \)). 
% We write \( \letin x \term \termb \) as a syntax sugar of \( \app
% {(\lambda x . \termb)} \term \). 
% By abuse of notation, we may write expressions such as \( \app \term
% \termb \) or \( \pairop{\term, \termb} \) to mean \( \letin x \term
% {\app x \termb} \) and \( \letin x \term {\letin y \termb {\pairop{x,
%       y}}} \). 
%Evaluation contexts will be used in defining the operational semantics and the OGS; defining both general and sequential evaluation contexts is for technical convenience.
%\sk{I defined  recursion using the ``let-rec'' operation because it was employed in~\cite{JaberMurawski21}.}
%\ds{for the conference paper, maybe omit seq evl contexts in the main text?}
%\sk{Omitted concurrent evaluation contexts instead.}
%\iq{For the compare-and-swap : I think we need it in the proof of completeness in order to control the number of times a function may be called}
% I agree (Ken)

The type judgment for terms is of the form \( \typeJudge \storectx \typectx \term \type \), where \( \typectx \) (resp.\ \( \storectx \)) is a \emph{typing context} (resp.~\emph{store typing context}), which is a finite map from variables (resp.\ locations) to types.
Similarly, the type judgment for evaluation contexts is of the form \( \typeJudge \storectx \typectx \pectx {\ectxtype{\typeb}{\type}} \), where \( \typeb \) represents the type of the hole and \( \type \) represents the type of the result.
%The typing rules are shown in Figure~\ref{fig:typing}.
The typing rules are standard, see Appendix~\ref{app:typing}

%\subsection{Operational Semantics and Contextual Equivalance}
\label{sec:opsem}
We now define the main operational semantics and behavioural
equivalence, namely  reduction semantics and contextual equivalence. 
%
%
% This section defines two operational semantics: (1) a standard
% reduction semantics to define the internal dynamics of a program and
% (2) a (na\"ive) labeled transition semantics (LTS) to express how a
% program interacts with the environment. 
% The labeled transition semantics is a ``warm-up'' that serves as a motive for the operational game semantics, which is considerably more intricate.
% We also formally define the contextual equivalence that we capture
% in this paper. 
% %\sk{I think it makes sense to bring the definition of CIU in this
% section (after) defining the reduction.} 
%
%\paragraph{Reduction semantics}
As  standard for reduction relations of  languages with
references, we use  \emph{stores} (or heaps), as maps from
locations to values. 
We write \( [\loc \mapsto \val ] \) for the singleton store mapping \( \loc \) to \( \val \); \( \store \cdot \storeb \)  for the disjoint concatenation of two stores; and \( \store [\loc \mapsto \val] \)  for the store that maps \( \loc \) to \( \val \) and \( \loc' \neq \loc \) to \( S(\loc') \).
The judgment \( \storeJudge \store \storectx \) denotes that, for each \( \loc  \in \dom \store \), we have \( \storectx \vdash \store(\loc) : \storectx(\loc) \).
A (reduction) \emph{configuration} is a pair of a term and a store written \( (\term ; \store) \).\footnote{
Several notions of configuration will be introduced in this paper, but we will usually just call them configuration when the meaning is clear from the context.}
We write \( \typeJudge \storectx \typectx {(\term; \store)} \type \) if \( \storeJudge \storectx \store \) and \( \typeJudge \storectx \typectx \term \type \).
The reduction rules are shown in  Figure~\ref{fig:reduction}.
 The left column shows standard operational rules for a sequential programming language with stores.
 Rules on the right column are related with concurrency. 
% These include
%  the rules for compare-and-swap and  those for parallel pairs.
% Among the latter rules,  the
% first shows 
A  parallel pair is treated as  an  ordinary pair
when both  components are  values;
otherwise, the two  components may reduce in any order.
We write \( (\term, \store) \Downarrow \val \) if \( (\term, \store) \opred^* (\val, \store') \) for some \( \store' \).
%
% Most of the rules are standard for a call-by-value sequential
% programming language with stores. 
% The less-standard rules for the reduction relation are given in Figure~\ref{fig:reduction}. We leave the other rules to Appendix~\ref{app:reduction}.
% The rule for compare-and-swap formalizes the intended behavior we
% explained above. 
% The rule for parallel pairs on the left says that when each
% component of the parallel pair reduces to a value we treat the
% parallel pair as a normal pair. 
% Otherwise, we may reduce the component of the pair in any order we want.
%\sk{I'm following the convention of~\cite{BirkedalST12}. This means that we do not allow \( \projop[1] (\parop \val \Omega) \opred \val \). }
%\ds{we should mention ~\cite{BirkedalST12} somewhere, as our language
%  was inspired by this}
%\gj{I have added a sentence about it}

\begin{figure}[t]
\begin{minipage}{0.4\linewidth}
\begin{gather*}
  (\app {(\lambda x. M)} \val; \store) \opred (M \subst x \val; \store) \\
  (\ifte \true \term \termb; \store) \opred (\term; \store) \\
  (\ifte \false \term \termb; \store) \opred (\termb; \store) \\
  (\refterm\val; \store) \opred (\loc; \store \cdot [\loc \mapsto \val]) \tag{\text{if \( \loc \notin \dom{\store}\)}} \\
  (\readop \loc; \store) \opred (\store(\loc); \store) \quad(\text{if \( \loc \in \dom{\store}\)}) \\
  (\writeop \loc \val; \store) \opred (\unit; \store \cdot [\loc \mapsto \val]) \\ 
\end{gather*}
\end{minipage}
\quad
\begin{minipage}{.5\linewidth}
\begin{gather*}
  (\projop \pairop{\val_1, \val_2}; \store) \opred (\val_i; \store) \\
  (\cas \loc {\val_1} {\val_2}; \store \cdot [\loc \mapsto \val_1]) \opred (\true; \store \cdot [\loc \mapsto \val_2])\\
  (\cas \loc {\val_1} {\val_2}; \store \cdot [\loc \mapsto \val_3]) \opred (\false; \store \cdot [\loc \mapsto \val_3]) \tag{\text{if $\val_1\neq \val_3$}}\\
  (\parop \val \valb; \store) \opred (\pairop{\val, \valb}; \store) \\
    \infer{(\term; \store) \opred (\term'; \storeb)}{(\pectx[\term]; \store) \opred (\pectx[\term']; \storeb)} \\
  %\infer{(\term; \store) \opred (\term'; \storeb)}{(\parop \term \termb; \store) \opred (\parop {\term'} \termb; \storeb)} \\[1em]
  %\infer{(\termb; \store) \opred (\termb'; \storeb)}{(\parop \term \termb; \store) \opred (\parop \term {\termb'}; \storeb)}
\end{gather*}
\end{minipage}
\caption{Reduction semantics.}
\label{fig:reduction}
\end{figure}

%\sk{Some notes for the design choice of the rules. I could avoid using the evaluation context and write every rule as a derivation rule. Having sequential evaluation context, however, is handy when we define the \( \PQ \) rule.
%  On the other hand I can consider a evaluation context \( \ectx \Coloneqq \cdots \mid \parop \ectx \term \mid \parop \term \ectx \) and avoid using a derivation rule. But using this definition is inconvenient when we define the \( \PQ \) rule because we have to extract the ``sequential'' part from the evaluation context. Hence, I'm using a ``hybrid approach''.}

As usual, contextual equivalence relates terms that have the same termination
properties in all closing contexts.
%In the formal definition, we have to make sure that typing of all components involved is respected.
A \((\typectx/\typeb)\)-\emph{testing-context} is a term with a hole that (1) accepts a term \( \typeJudge \emptyctx \typectx \term \typeb \), and (2) becomes a closed term of type \( \Unit \) once the hole is filled.

\begin{definition}[Contextual equivalence]
\label{d:ce}
Suppose that \( \typeJudge \storectx \typectx {\term_1,\term_2} \typeb \). The \emph{contextual preorder} \( \typeJudge \storectx \typectx {\term_1 \ctxle \term_2} \typeb \) holds if, for all \( (\typectx/\typeb) \)-testing-context \( \ctx \)
and $\store:\storectx$, \( (\ctx[\term_1];\store) \Downarrow () \) implies \( (\ctx[\term_2];\store) \Downarrow () \).
We write \( \typeJudge \storectx \typectx {\term_1 \ctxeq \term_2} \typeb \) if \( \term_1 \) is below \( \term_2\) and vice versa.
\end{definition}

%\ds{i'd suggest to have in the appendix a table summarising all
%  notations (symbols, typing judgments, configurations, etc}

A standard simplification of  contextual equivalence
is 
\emph{closed instances of uses} (CIU)
equivalence~\cite{MasonTalcott91}, which  
 makes use   
of \emph{evaluation contexts}, rather than general contexts.
%
%
%
% %\paragraph{CIU equivalence}
% We now formally define the program equivalence that we aim to characterize using OGS.
% In short, it is may-testing equivalence.
% Instead of employing the standard Morris-style definition for
% contextual-equivalence we adopt an equivalent definition known as the
% \emph{closed instances of uses} (CIU)
% equivalence~\cite{MasonTalcott91}. 
% CIU equivalence compares terms under arbitrary \emph{evaluation
% contexts} (rather than general contexts) after closing the free
% variables of the terms using a value assignment. 
For an assignment \( \gamma \) from variables to values, we write \( \typeJudgeg \storectx \typectx \) if for any \( x \in \dom \typectx \), %the term \( \gamma(x) \) is a \emph{value} and 
\( \typeJudge \storectx \emptyset {\gamma(x)} {\typectx(x)}\).
We write \( \term \substg \gamma \) for the term obtained by substituting  in parallel  in \( \term \) all the \( x \in \dom{\gamma} \) by $\gamma(x)$.

%\sk{I haven't checked the CIU theorem, but I believe it holds even in a concurrent setting.}

\begin{definition}
  \label{def:ciu}
  For two terms \( \typeJudge \storectx \typectx {\term_1,\term_2} \typeb \) we say that \( \term_1 \) is a \emph{CIU approximation} of \( \term_2 \), written \( \typeJudge \storectx \typectx {\term_1\ciule \term_2} \typeb \) (or simply \( \term_1 \ciule \term_2 \)), if for all \( \storectx' \), \( \store \), \( \gamma \) and \( \pectx \), \ such that
  \begin{itemize}
  \item \(\storectx \subseteq \storectx' \) and \( \storeJudge {\store} {\storectx'} \);
  \item \( \typeJudgeg {\storectx'} \typectx \)
 and %  \item 
\( \typeJudge {\storectx'} \emptyset \pectx {\ectxtype{\typeb}{\Unit}} \); %\sk{In the previous version only sequential evaluation contexts were used, but I think we need to quantify over concurrent evaluation contexts, which for the time being is ranged over by \( \pectx \).}
  %\iq{I don't think it would change much, the other branches wouldn't have access to any name of player, the only effects would be on the store of the context which could be done by adding these branches after the first question from player}
  %\sk{Probably you are right. I'll try to prove it later (without any reference to the game semantics). Anyway the fact that sequential evaluation contexts suffice may be a bit surprising, and this should be explained in the text.}
  \end{itemize}
  $(\pectx[\term_1\{\gamma\}];\store)\Downarrow\unit$ implies $(\pectx[\term_2\{\gamma\}];\store)\Downarrow\unit$.
  If both \( \term_1 \ciule \term_2 \) and \( \term_2 \ciule \term_1 \) hold, then we say that \( \term_1 \) and \( \term_2 \) are \emph{CIU equivalent}, written \( \typeJudge \storectx \typectx {\term_1 \ciueq \term_2} \typeb \) (or simply \( \term_1 \ciueq \term_2\)).
\end{definition}
%Contexts  in CIU equivalence  forbids  holes 
%not only underneath  lambda abstractions, but also in parallel pairs.
%Nonetheless, 
CIU preorder coincides with contextual preorder. The proof is done by showing that the CIU preorder is a congruence, following a standard technique established in~\cite{honsell1995variable}.

\begin{proposition}
%  Let \( \typeJudge \emptyctx \typectx {\term_1, \term_2} \typeb \) be ref-free terms.
  We have \( \typeJudge \emptyctx \typectx {\term_1\ctxle \term_2} \typeb \) if and only if \( \typeJudge \emptyctx \typectx {\term_1\ciule \term_2} \typeb \).
\end{proposition}
\sk{I'll prove this. Here we should not need the ref-free assumption. But anyway we are interested in ctx-equivalence that do not share locations.}

In the following, we work only on interactions between
a program and a context that do not share any locations.
To preserve this property during the reduction, we enforce that the typing
at their interface do not contain any $\reftype{}$ type constructor.
% In contextual equivalence we consider in this paper only considers
% the case where a program and a context do not share any locations. 
So we say that a term \( \typeJudge \storectx \typectx \term \typeb \) is \emph{ref-free} if no reference type occurs (even as a subexpression) in \( \typectx \) and \( \typeb \).

\section{Operational Game Semantics}\label{s:ogs}

\subsection{LTSs}
%\label{ss:}
             
Contextual equivalence, as well as  CIU equivalence, makes use of a heavy
universal quantification on contexts. 
Our goal is  to  follow the ideas of Operational Game Semantics so to 
derive a Labeled Transition System (LTS) and,  on top of
this, 
 a trace semantics, which
can be used to reason about contextual equivalence.  
We will obtain this through a few steps. 
\begin{enumerate}
\item
First we define an 
LTS
 \( \opLTS \) that describes how terms interact with the environment.
The LTS does not account for reductions internal to terms, for which we still refer to the
basic reduction semantics (it is common in operational game semantics to  
 separate internal
reductions  from interactions with the environment).

\item 
In the LTS \( \opLTS \), actions 
may  carry  higher-order values. It is well-known that such kinds of
actions   complicate definitions of satisfying behavioural
relations (see e.g.~\cite{San923,SangiorgiKS11}). In  OGS such higher-order
values are transformed into abstract values, intuitively pointers to
the actual values.  
This is accomplished in the second step, namely 
 LTS $\ILTS$ in Figure~\ref{fig:def-ilts-new}.
In 
$\ILTS$ we also 
 structure the states of the  LTS as a pool
of threads; each thread, say 
$\named \qn \term $, is a running term $\term$ that will compute
a value to be returned to the environment at $\qn$.

\item 
The two aspects that still remain to be taken into account are typing
and visibility; that is, we have to make sure that only well-typed
actions are performed, and that actions respect the
constraints produced by visibility 
(the impossibility for the environment  of storing higher-order
  values). 
To account for these two aspects 
we  introduce, respectively, the LTS $\TyLTS$
  (Figure~\ref{fig:def-tylts}), ensuring well-typedness, and 
the LTS  $\VisLTS$ (Figure~\ref{fig:def-vislts}), that keeps track of
the evolution of the visibility view of the threads in the
environment. 
\sk{I think we should say that the visible LTS also takes care of well-bracketing. Game semanticists might expect a well-bracketing LTS as well}

\item The final LTS is the combination of the above three LTS: 
$\ILTS$, 
$\TyLTS$, $\VisLTS$; that is, the legal actions are those produced by 
 $\ILTS$ and allowed by the typing LTS $\TyLTS$ and the visibility LTS
 $\VisLTS$.

 \end{enumerate}

 \paragraph*{Partially suspended terms and the LTS $\opLTS$}
 \label{sec:suspend}
The LTS \( \opLTS \) describes how terms interact with the environment.
% \sk{\( \op \) may look like ``opponent'', so it might be better to
% change the name} 
The states of the LTS are extensions of terms
 called \emph{partially suspended terms}, as they may
include terms that, having interrogated the environment,  are now
waiting for   an answer:
\begin{align*}
  \mathrm{Partially\ suspended\ terms} \quad \psterm, \pstermb &\Coloneqq \term \mid \suspend \ectx \pn \mid \ectx[\parop \psterm \pstermb]
  %\mathrm{Named\ contexts} \quad \nctx &\Coloneqq \ectx \mid \named \pn \ectx
\end{align*}
We call names such as  
 \( \pn \)  \emph{continuation names}.
Term \( \suspend \ectx \pn \) is suspended 
waiting for the environment to provide a value at \( \pn \).
%  can
% be thought of as a channel that is used to pass the value. 
We write \( \psterm[(\termb, \pn)]\) for the (partially suspended)
term obtained by replacing
the subterm of  \( \psterm \) suspended on \( \pn \), say,
 \( \suspend \ectx \pn \), 
with  \( \ectx[\termb] \).
% \( \suspend \ectx \pn \) in \( \psterm \)  with \( \ectx[\termb] \).
% \gj{We should introduce the notation
% $\suspend{\term}[(\termb,\pn)]$
% for the filling by $\termb$ of the hole named $\pn$ in $\term$.}
% \sk{Is the above description/notation fine? I know that there's a
% more formal definition in Section~\ref{ss:soundness}, but that IMO
% writing it here is too much.} 
The transition rules are shown  in Figure~\ref{fig:def-oplts}.
A transition is of the form \( \psterm \pstrans{\mu} \pstermb \) where
the action \( \mu \) is either of the form \( \fn(\val, \pn) \) or \(
\pn(\val) \). 
We omit the   straightforward
extension of the  reduction relation $ \opred$ to partially suspended terms,
which will be used in the following LTSs.

\begin{figure}[t]
\begin{mathpar}
\infer{\text{\( \pn \) is fresh}}{\ectx[\app f \val] \pstrans{f(\val, \pn)} \suspend \ectx \pn}\and
\infer{ }{\suspend \ectx \pn \pstrans{\pn(\val)} \ectx[\val]} 
\and 
\infer{\psterm \pstrans{\mu} \psterm'}{\parop \psterm \pstermb
  \pstrans{\mu} \parop {\psterm'} \pstermb} \\ 
%\and
\infer{\pstermb \pstrans{\mu} \pstermb'}{\parop \psterm \pstermb \pstrans{\mu} \parop \psterm {\pstermb'}} \and
\infer{\parop {\psterm_1} {\psterm_2} \pstrans{\mu} \pstermb}{\ectx[\parop {\psterm_1} {\psterm_2}] \pstrans{\mu} \ \ectx[\pstermb]}
\end{mathpar}
\caption{Definition of \( \opLTS \).}
\label{fig:def-oplts}
\end{figure}

\paragraph*{Abstract values and the LTS $\ILTS$}
\label{sec:abstract_values}
In OGS,  values exchanged between a term (\emph{Proponent}) and the an environment (\emph{Opponent}) are
abstracted away, by transforming them into
 \emph{abstract values} (aka \emph{ultimate patterns}~\cite{LassenLevy08})
$\aval$
%$\aval \in \AValues$ 
defined thus:
\[\aval \Coloneqq \baseval \mid \fn % \mid \pn
\mid \pairop{\aval,\aval}\]
where $\fn \in \FNames$ is  a function name,%  $\pn \in \CNames$
% is a continuation name,
and $\baseval$ is a base value.
Each syntactic value \( \val \) can be decomposed into an abstract
value \( \aval \) (its positive skeleton) together with a
corresponding name assignment map \( \gamma \)  (its negative filling)
such that \( \gamma(\aval) = \val \). 
Typing rules of abstract values are reported in Figure~\ref{fig:nup-type}.
\sk{(TODO): continuation type needs to be introduced before here. Maybe we can introduce them wheb we define \( \opLTS\). The continuation name already appears there.}%
\gj{Should we merge $\supp$ and $\freen$ macros ?}%
\sk{I agree, althouh I do not have a preference}%
\begin{figure}
\begin{mathpar}
  \inferrule*
    { }
	{\typeJudgePat
	  {\namectx}
	  {\fn:\type \rightarrow \typeb}
	  {\fn}
	  {\type \rightarrow \typeb}
	}
	\and
  %       \inferrule*
  %   { }
  %       {\typeJudgePat
  %         {\namectx}
  %         {\pn:\covartype{\type}}
  %         {\pn}
  %         {\covartype{\type}}
  %       }
  % \and
  \inferrule*
	{\baseval \text{ a base value of ground type } \gtype }
	{\typeJudgePat{\namectx}{\emptyctx}{\baseval}{\gtype}}
  \and
  \inferrule*
	{\typeJudgePat{\namectx}{\lnamectx_1}{\aval_1}{\type_1}
	    \and
	    \typeJudgePat{\namectx}{\lnamectx_2}{\aval_2}{\type_2} 
	}
	{\typeJudgePat{\namectx}{\lnamectx_1\cdot\lnamectx_2}
		{\pairop{\aval_1;\aval_2}}{\type_1 \times \type_2}
	}
\end{mathpar}
\caption{Definition of the type judgment for abstract values}
\label{fig:nup-type}
\end{figure}
Abstract values are exchanged between Proponent and Opponent through \emph{moves}
$\move$ of the following shape:
\[\begin{array}{|l|l|l|l|}\hline
    %\text{Move kind} & 
    \text{P-question} & \text{P-answer}
      &\text{O-question} & \text{O-answer}
    \\
	\hline
    %\text{Move form}\raisebox{4mm}{} & 
    \questP{\fn}{\aval,\pn}_\qn & \ansP{\pn}{\aval}
      & \questO{\fn}{\aval,\pn}_\qn & \ansO{\pn}{\aval}
	\\
	\hline
\end{array}\]
%\sk{I guess now you chose to use the term moves for moves with labels? That's what I see in Figure~\ref{fig:def-ilts-new}. We need some explanation for that}
Such moves %, plus $\tau$
% \ds{why not $\tau$ now? also: is an extension the reduction semantics
%   here implicit?}
 are the labels of the 
%\subsection{Interactive LTS}
\emph{Interactive LTS} $\ILTS$,
in
Figure~\ref{fig:def-ilts-new}.
Question moves are indexed with continuation names, which is not standard in OGS. The indices resemble thread identifiers sometimes used in labelled transition semantics~\cite{JeffreyRathke04,JeffreyRathke05}, allowing us to identify the thread from which an action originates.
They are introduced for technical convenience to define well-bracketing, but are treated as unobservable and removed when considering traces.

% is then defined as the triple
% $(\IConfs,\Actions,\itrans{})$ with $\itrans{}$ defined in
% Figure~\ref{fig:def-ilts-new}.
The states of the LTS are \emph{interactive configurations} \(
\iconf, \iconfb  \)  %\in \IConfs \)  
which are of the shape \(
\ctuple{\threadpool;\store;\gamma} \), with $\threadpool$ a \emph{pool
  of threads}, $\store$ a store, and $\gamma$ an assignment from names in $\FNames$ to either variables or $\lambda$-abstractions.
%  (The
% domain of \( \gamma \) can be restricted to \( \FNames \).) 
A pool is a parallel composition of threads,  of the form \( \named {\pn_1} {\psterm_1} \parcomp \cdots \parcomp \named {\pn_n} {\psterm_n} \).
Thus there are two levels of parallelism: the pool runs multiple
threads in parallel, and each thread may contain parallelism
caused by the parallel-pair construct. 
A new thread is created when a \( \OQ \)-transition is performed.
This intuitively corresponds to invoking a closure in parallel with the other threads.

\gj{We should introduce the notation $\threadpool(\pn)$ to retrieve
the thread named by $\pn$.}

\begin{figure}
\begin{mathpar}
	\inferrule*[Left=$\oplabel$]
	{(\psterm;\store) \opred
		(\pstermb;\storeb)}
	{\ctuple{\threadpool \parcomp \named{\pn}{\psterm} \parcomp \threadpoolb;
		  \store;\gamma}
		\itrans{\oplabel}
		\ctuple{\threadpool \parcomp \named{\pn}{\pstermb} \parcomp \threadpoolb;
		  \storeb;\gamma}
	}
	\\
	\inferrule*[Left=$\PQ$]
	{\val = \aval\substg{\gamma'} \and \psterm \pstrans{\fn(\val, \pn)} \pstermb}
	{\ctuple{\threadpool \parcomp \named \qn \psterm \parcomp \threadpoolb;
	      \store;\gamma}
		\itrans{\questP{\fn}{\aval,\pn}_\qn}
		\ctuple{\threadpool \parcomp  \named \qn \pstermb \parcomp \threadpoolb;
		  \store;\gamma\cdot\gamma'}}
	\and
	\inferrule*[Right=$\PA$]
	{\val = \aval\substg{\gamma'}}
	{\ctuple{\threadpool \parcomp \named{\pn}{\val} \parcomp \threadpoolb;
	       \store;\gamma}
		\itrans{\ansP{\pn}{\aval}}
		\ctuple{\threadpool \parcomp \threadpoolb;
		  \store;\gamma\cdot\gamma'}
	}
	\\
	\inferrule*[Left=$\OQ$]
	{  }
	{\ctuple{\threadpool;\store;\gamma}
		\itrans{\questO{\fn}{\aval,\pn}_\qn}
		\ctuple{\threadpool \parcomp \named{\pn}{(\app {\gamma(\fn)} \aval)};
            \store;\gamma}}
	\and
	\inferrule*[Right=$\OA$]
	{ \psterm \pstrans{\pn(\aval)} \pstermb}
	{\ctuple{\threadpool \parcomp \named \qn \psterm;\store;\gamma}
		\itrans{\ansO{\pn}{\aval}}
		\ctuple{\threadpool \parcomp \named \qn \pstermb;\store;\gamma}}
\end{mathpar}
\caption{Definition of $\ILTS$}
\label{fig:def-ilts-new}
\end{figure}

\paragraph*{The typing LTS $\TyLTS$ and
the visible  LTS  $\VisLTS$}
\label{sec:tylts_vislts}
\ds{how about placing $\TyLTS$ in an appendix?}
\sk{+1, if we don't have space. It's the least important one.}

The typing LTS $\TyLTS$ tracks the type of new names introduced and
ensures that everything remains well-typed. 
 \emph{Type-context configurations}, ranged over by 
$\tconf,\tconfb$,
%$\tconf,\tconfb \in \TConfs$  
are 
of the shape
$\ctuple{\namectxO \mid \namectxP}$,
where $\namectxO,\namectxP$ are 
disjoint typing contexts 
 mapping function names to function types 
 $\type\rightarrow\typeb$,
and continuation names to context types $\lnot \type$.
The typing LTS $\TyLTS$, whose states are type-context configurations, 
is in Figure~\ref{fig:def-tylts}.
Names in $\namectxO$, called Opponent names, are the names introduced by Opponent.
Symmetrically, names in $\namectxP$ are the one introduced by Proponent.

\begin{figure}[t]
\begin{mathpar}
	\inferrule*[Left=PQ]
		{\namectxO(\fn) = \type \rightarrow\typeb \quad 
		\typeJudgePat{\namectxO}{\lnamectx}{\aval}
			{\type}}
		{\ctuple{\namectxO \mid \namectxP} 
			\ttrans{\questP{\fn}{\aval,\pn}_\qn} 
			\ctuple{\namectxO \mid \namectxP,\lnamectx,\pn:\lnot\typeb}}
	\and
	\inferrule*[Right=PA]
		{\namectxO(\pn) = \lnot\type \and \typeJudgePat{\namectxO}{\lnamectx}{\aval}
			{\type}}
		{\ctuple{\namectxO \mid \namectxP} 
			\ttrans{\ansP{\pn}{\aval}} 
			\ctuple{\namectxO \mid \namectxP,\lnamectx}}
	\and
	\inferrule*[Left=OQ]
		{\namectxP(\fn) = \type \rightarrow\typeb \quad 
		\typeJudgePat{\namectxP}{\lnamectx}{\aval}
			{\type}}
		{\ctuple{\namectxO \mid \namectxP} 
			\ttrans{\questO{\fn}{\aval,\pn}_\qn} 
			\ctuple{\namectxO,\lnamectx,\pn:\lnot\typeb \mid \namectxP}}
	\and
	\inferrule*[Right=OA]
		{\namectxP(\pn) = \lnot\type \and \typeJudgePat{\namectxO}{\lnamectx}{\aval}
			{\type}}
		{\ctuple{\namectxO \mid \namectxP}
			\ttrans{\ansO{\pn}{\aval}} 
			\ctuple{\namectxO,\lnamectx \mid \namectxP}}
\end{mathpar}
\caption{Definition of $\TyLTS$}
\label{fig:def-tylts}
\end{figure}

%\subsection{Visible LTS}

%\newcommand{\mA}{\mathcal{A}}
%\newcommand{\mF}{\mathcal{F}}
%\newcommand{\mG}{\mathcal {G}}
%\newcommand{\view}{\mathcal{V}}
%\newcommand{\mT}{\mathcal{T}}
\newcommand{\pf}{\mathcal{P}_{\mathrm{fin}}}

\ds{not sure below `Opponent' and `Proponent names' have been introduced }

The  \emph{Visible}  LTS  $\VisLTS$,
in Figure~\ref{fig:def-vislts},
 allows us to manage the different
visibilities of the Opponent names. 
% In the \emph{Visible} %History
%  LTS $\VisLTS$
%$\VisLTS = (\VisConfs,\Actions,\vtrans{})$
States are 
%with $\VisConfs$ formed by 
configurations  $\ctuple{\viewF, \viewG, \stackT}$, 
ranged over by $\vconf,\vconfb$,  where:
% with $\VisConfs$ formed by 
% configurations $\ctuple{\viewF, \viewG, \stackT}$, where:
\begin{itemize}
\item $\viewF \colon \Names \to \pf(\Names)$
is a function from Opponent names 
to a set of Proponent names which tells us the visibility 
of each  Opponent name (intuitively, which Proponent
 names were known by the Opponent name when it was introduced; 
or, in other words, which 
functions 
introduced by   Proponent
may be used when the Opponent name is called);
\item $\viewG \colon \CNames \to \pf(\Names)$
is a function from Proponent continuation names,
seen as thread identifiers, 
to a set of Proponent names, and tells us the visibility of each thread.
The premise \( \fn \in \viewG(\qn) \) of the rule OQ is the key condition that forces the environment to respect visibility.
\item $\stackT \colon \tree $
is a tree that represents the `multi-threaded stack'
of continuation names, and accounts for the `well-bracketed' structure
of return-call interactions  among terms. 
A  well-bracketing    discipline is 
common  
in 
higher-order  languages
 without  control operators. 
In sequential languages, the discipline is 
formalised 
by means of a stack of continuation names, and operations for
handling them. 
Since our language is concurrent, we need 
tree-shaped stacks, 
with operations for dealing with
such trees. 
The grammar for the trees is: 
\[
\tree\  \defeq \treeaux \mid \emptymap
\hspace{2em}
 \treeaux \defeq \CNames \times \left(\mathtt{list}\ \treeaux
 \right) \] 
% \gj{I don't think we need to provide $\CNames$ as
% an argument to $\treeaux$ and $\tree$.}
We use two functions to manipulate trees:
\begin{itemize}
\item $\addleaf{\stackT}{\pn}{\qn}$ that adds $\pn$  to the tree
  $\stackT$ as a leaf with parent $\qn$; 
\item $\removeleaf{\stackT}{\pn}$ that returns an option type;
  intuitively, used to remove the leaf $p$ from $\stackT$ and obtain
$p$'s parent. Formally: 
  \begin{itemize}
  \item if $\pn$ is a leaf in $\stackT$, the function
 returns $\some{\stackT',\qn^?}$ where $\stackT'$ is 
tree with $\pn$  removed; 
 and $\qn^?= \none$ if $\pn$ is the root of $\stackT$, 
otherwise
 $\qn^? = \some \qn$ where $\qn$ is $\pn$'s parent;
  \item
if $\pn$ is not  a leaf
 the function returns $\none$.
  \end{itemize}
\end{itemize}
\end{itemize}
% The Visible LTS  is  defined in Figure~\ref{fig:def-vislts}.

\begin{figure}
\begin{mathpar}
  \inferrule*[Left=PQ]
	{\,}
	{\ctuple{\viewF; \viewG; \stackT} 
	\vtrans{\questP{\fn}{\aval,\pn}_\qn} 
	\ctuple{
	  \viewF; 
	  \viewG \cdot [\pn \mapsto \viewF(\fn) \cup \supp(\aval)];
	  \addleaf{\stackT}{\pn}{\qn}}
	}
  \and
  \inferrule*[Left=PA]
	{\removeleaf{\stackT}{\pn} = \some{\stackT',\qn^?}}
	{\ctuple{\viewF; \viewG; \stackT} 
	\vtrans{\ansP{\pn}{\aval}} 
	\ctuple{
		\viewF; 
		\viewG [\qn^? \mapsto \viewG(\qn) \cup \supp(\aval)];
		\stackT'
	}}
  \and
  \inferrule*[Left=OQ]
	{\fn \in \viewG(\qn)}
	{\ctuple{\viewF; \viewG; \stackT} 
	\vtrans{\questO{\fn}{\aval,\pn}_\qn} 
	\ctuple{
	  \viewF \cdot \left(\prod_{\gn \in \supp(\aval)}[\gn \mapsto \viewG(\qn)]\right)
	    % \cdot [\pn \mapsto \viewG(\qn)]
            ; 
	  \viewG;
	  \addleaf{\stackT}{\pn}{\qn}}
	}
  \and
  \inferrule*[Left=OA]
	{\removeleaf{\stackT}{\pn}=\some{\stackT',\qn^?}}
	{\ctuple{\viewF; \viewG \cdot [\pn \mapsto \view]; \stackT} 
	\vtrans{\ansO{\pn}{\aval}} 
	\ctuple{
		\viewF \cdot \prod_{\fn \in \supp(\aval)}[\fn \mapsto \view]; 
		\viewG ;
		\stackT'
	}}
\end{mathpar}
\caption{Definition of $\VisLTS$}
\label{fig:def-vislts}
\end{figure}

\paragraph*{The OGS LTS $\lts_{\OGS}$}
\label{sec:ogs}
The \emph{Operational Game Semantics} (OGS) LTS 
$\lts_{\OGS} \defeq (\OGSConfs,\Actions,\ogstrans{\act})$
is defined over configurations 
$\ogsconf,\ogsconfb \in \OGSConfs$ of the 
shape $(\iconf,\tconf,\vconf)$,
and its transition relation, the conjunction of the transitions of the
three previous LTSs, is defined by the rules:
	\[
	\inferrule*{
	  \iconf \itrans{\act} \iconfb 
	  \and
	  \tconf \ttrans{\act} \tconfb
	  \and
	  \vconf \vtrans{\act} \vconfb
	}{
	  (\iconf,\tconf,\vconf) 
	  \ogstrans{\act} (\iconfb,\tconfb,\vconfb)}
\hskip 1cm 
	\inferrule*{
	  \iconf \itrans{ } \iconfb 
	}{
	  (\iconf,\tconf,\vconf) 
	  \ogstrans{ } (\iconfb,\tconf,\vconf)}
	\]

 An $\OGS$ configuration  is \emph{valid} when all its components are
 consistent with each other, including types and structure of the
 tree-stack of continuation names with respect to the pool of threads;
see Appendix~\ref{app:def-ogs}, where we also show 
that validity is preserved by transition.

\begin{example}[Example of transitions]
\label{exa:trans}
From $\val \defeq \lambda \fn. \gn(\lambda\_.\fn\unit)$,
 we have the following execution, omitting  the
store component of interactive configurations
and the typing configurations:
\[\begin{array}{@{}l@{}ll}
	\begin{array}{l}
	\ctuple{\named{\pn_0}{\val};\emptymap};
	\\
	\ctuple{
		\viewF_0;
          \viewG_0;
          (\qn_0,[(\pn_0,[])])
	}
	\end{array}
	& \ogstrans{\ansP{\pn_0}{\hn}}
	&
	\begin{array}{l}
	\ctuple{\emptymap;\overbrace{[\hn \mapsto \val]}^{\gamma_1}};
	\\
	\ctuple{
		\viewF_0;
		\underbrace{
			[\qn_0 \mapsto \{\hn\}]
		}_{\viewG_1};
		(\qn_0,[])
	}
	\end{array}
	\\
    & \ogstrans{\questO{\hn}{\fn;\pn_1}_{\qn_0}}
	& \begin{array}{l}
		\ctuple{\named{\pn_1}\val\ \fn;\gamma_1}
		\\
		\ctuple{
			\underbrace{
			  \viewF_0\cdot[\fn \mapsto \{\hn\}]}
			_{\viewF_1};
			\viewG_1;
			(\qn_0,[(\pn_1,[])])
		}
		\text{ since } \hn \in \viewG_1(\qn_0)
	\end{array} 
	\\
	& \ogstrans{\oplabel}
	& \begin{array}{l}
       \ctuple{\named{\pn_1}\gn\  (\lambda\_.\fn\unit);\gamma_1}
	   \\
	   \ctuple{
		\viewF_1;
          \viewG_1;
          (\qn_0,[(\pn_1,[])])
	   }
	\end{array}   
	\\
	& \ogstrans{\questP{\gn}{v;\qn_1}_{\pn_1}}
	& \begin{array}{l}
		\ctuple{\named{\pn_1}{\suspend \bullet {\qn_1}};\overbrace{\gamma_1 \cdot [v \mapsto \lambda\_.\fn\unit]
		}^{\gamma_2}}
		\\
		\ctuple{
			\viewF_1;
			\viewG_1
          \cdot [\qn_1 \mapsto \{v\}];
          (\qn_0,[(\pn_1,[(\qn_1,[])])])
		}
	\end{array}
\end{array}\]

with $\viewF_0 \defeq [\gn,\pn_0 \mapsto \emptyset]$
and $\viewG_0 \defeq [\qn_0 \mapsto \emptyset]$.
\end{example}

 \subsection{Complete-trace preorder}
\label{sec:complete_trace}

%\ds{in abstract and intro we talk about preorder, here equivalence}
 In order to define OGS-traces of terms, we have also to
 add initial moves, so to transform  typed terms into appropriate 
configurations of the $\lts_{\OGS}$ LTS. This is done in the rules of
 Figure~\ref{fig:init-move}. In rule \textsc{PI}, $\aval_i$'s are the abstract
 values that replace the free variables of the initial term; 
$\pn$ is the name along which the final return value will be
delivered; the tree is initialised with just $\pn$ as a root; the
visibility components are initially empty. Rule \textsc{OI} is similar, for
environment used in the ciu preorder, and will be useful in the soundness proofs.
To transform a parallel evaluation context $\pectx$ into a suspended term,
we add to the LTS $\opLTS$ the rule $\ectx \pstrans{?(\pn)} \suspend \ectx \pn$,
so that we can suspend the maximal sequential evaluation context $\ectx$
appearing around the hole of a parallel context. This rule also uses a distinguished
continuation name $\pn_{f}$ used by Opponent to perform the final answer.
 \begin{figure}
   \centering
   \begin{mathpar}
     \inferrule*[Left=PI]
     {\namectx= \overrightarrow{(\var_i:\typeb_i)}
	  \and \typectxb_i \Vdash \aval_i:\typeb_i}
     {\langle \namectx \vdash (\term,\store):\typeb\rangle \ogstrans{\movePI{\overrightarrow{\aval_i},\pn}}
       \left(
         \begin{array}[c]{l}
           \ctuple{\named{\pn}{\term\{ \overrightarrow{\aval_i / \var_i}\}};\store;\emptymap};\cr
           \ctuple{\overrightarrow{\typectxb_i},\pn:\neg \typeb \parcomp \emptymap};\cr
           \ctuple{\prod_{\fn \in \supp(\overrightarrow{\aval_i})}[\fn \mapsto \emptyset] ;\emptymap; (\pn,\emptylist)}
         \end{array}
     \right)
   }
     \and
     \inferrule*[Left=OI]
     {\namectx = \overrightarrow{(\var_i:\typeb_i)}
	  \and \typectxb_i\Vdash \aval_i:\typeb_i
	  \and \aval_i\{\gamma_i\} = \delta(\var_i)
	  \and \pectx \pstrans{?(\pn)}  \psterm} 
     {\langle \vdash (\pectx,\delta,\store):(\ectxtype{\typeb}{\Unit};\namectx)\rangle
       \ogstrans{\moveOI{\overrightarrow{\aval_i},\pn}}
       \left(
       \begin{array}[c]{l}
         \ctuple{\named{\pn_f}{\psterm};\store;\overrightarrow{\gamma_i}};\cr
         \ctuple{\pn_f:\neg \Unit \parcomp \overrightarrow{\typectxb_i},p:\neg\typeb};\cr
         \ctuple{\emptymap; [\pn\mapsto \supp(\overrightarrow{\aval_i})]; (\pn_f, [(\pn,\emptylist)])}
       \end{array}
       \right)
       }
   \end{mathpar}
   \caption{Initial moves for $\OGSLTS$}
   \label{fig:init-move}
 \end{figure}
 
While defining traces, we erase the thread labels as they are not
supposed to be observed; for instance $\questO{\fn}{\aval,\pn}_\qn$ becomes $\questO{\fn}{\aval,\pn}$ and $\questP{\fn}{\aval,\pn}_\qn$ becomes $\questP{\fn}{\aval,\pn}$.
Weak transitions are defined in the expected manner. 
Thus
$
\ogsTrans{}$ is the reflexive and transitive closure of 
$
\ogstrans{}$; and  $
{\ogsTrans{\act}}\defeq 
{\ogsTrans{} \ogstrans{\act} \ogsTrans{}}$; finally
for a trace  $\tr = \act_1 \ldots \act_n$, we write 
$\ogsTrans{\tr}$  for  $\ogsTrans{\act_1} \cdots \ogsTrans{\act_n}$.
A  \emph{complete trace} is a trace in which `all questions have been
answered'; formally, at the end of the trace, the tree-stack should be
empty.

 \begin{definition}
   A complete trace $\tr$ of $\langle \namectx \vdash (\term,\store):\typeb\rangle$ is a trace of $(\term,\store)$ such that there exists $\iconf,\tconf,\viewF,\viewG$ such that $\langle \namectx \vdash (\term,\store):\typeb\rangle \ogsTrans{\tr} (\iconf;\tconf;\ctuple{\viewF,\viewG,\emptymap})$.\\
   We write $\ogsCTr{\langle \namectx \vdash (\term,\store):\typeb\rangle}$ for the set of complete traces of $\langle \namectx \vdash (\term,\store):\typeb\rangle$.
 \end{definition}

 \begin{definition}
Given $\storectx,\namectx \vdash \term_1,\term_2 : \typeb$, 
term $\term_1$ is a \emph{trace approximation} of $\term_2$, written
$\storectx,\namectx \vdash \term_1\trle \term_2 : \typeb$, if for all
$\store$ with
 $\vdash \store:\storectx$, we have $\ogsCTr{\langle \namectx \vdash (\term_1,\store):\typeb\rangle}\subseteq \ogsCTr{\langle \namectx \vdash (\term_2,\store):\typeb\rangle}$.
 \end{definition}

 \paragraph*{Examples}

% Note that an action of \( \opLTS \) can carry a higher-order value.
% As it is well-known~\cite{HOpi?Environmentalbisim?}, comparison of
% higher-order transitions is tricky since there is no canonical way
% to compare higher-order values, which in turn prevent us from
% (naively) defining the notion of trace equivalence or bisimilarity
% out of the LTS. 
% Operational game semantics, which we introduce in the subsequent
% section, can be thought of a machinery to express \( \opLTS \)
% without actually passing any higher-order values. 
% Another problem with the \( \opLTS \) is that it does not adequately
% restrict the actions the environment can perform: 

Example~\ref{exa:gc} shows that the visibility constraints can be used
to perform some garbage collection, exploiting the property that 
the store is not higher-order, hence certain function names, while
free, cannot be used by the environment.

 \begin{example}
\label{exa:gc}
   Consider the interactive configuration $\iconf \defeq
   \ctuple{\named{\pn}{\letin \var {\fn\,(\lambda \varb. \term)}
       {\termb}};\store;\gamma}$:  for any typing configuration
   $\tconf$ and visible configuration $\vconf$, if
   $(\iconf,\tconf,\vconf) \ogstrans{\questP{\fn}{\gn,\qn}_\pn}
   \ogstrans{\ansO{\qn}{n}}
   (\ctuple{\threadpool;\store';\gamma'},\tconf',\vconf')$, then 
   $\ogsTr{\ctuple{\threadpool;\store';\gamma'},\tconf',\vconf'} = 
   \ogsTr{\ctuple{\threadpool;\store';\gamma'\setminus\{\gn\}},\tconf'\setminus\{\gn\},\vconf'}$. This holds because visibility ensures us
 that $\gn$ is forgotten after the answer $\ansP{\qn}{n}$. 
   
   More generally considering a configuration $(\ctuple{\threadpool;\store;\gamma},\tconf,\ctuple{\viewF;\viewG;\stackT})$, and a proponent name $\gn\in \dom{\gamma}$, if $\gn \notin \freen(\viewF,\viewG)$ then $\ogsTr{\ctuple{\threadpool;\store;\gamma},\tconf,\ctuple{\viewF;\viewG;\stackT}}=\ogsTr{\ctuple{\threadpool;\store;\gamma\setminus\{\gn\}},\tconf\setminus\{\gn\},\ctuple{\viewF;\viewG;\stackT}}$
%    This example and this result can be used to perform some garbage
%    collection, for instance if one wants to mechanize a test for
%    trace equivalence. 
 \end{example}

%\ds{example below: to be polished and refined, and probably positioned
%  somewhere else } 
\begin{example}
  Back to the terms in Example~\ref{exa:ex1} of the introduction, with
  $\readop \var \mathrel{\&\&} \mathtt{not}\,\readop \var$ defined as
  $\ifte {\readop \var} {(\ifte {\readop \var} \false \true)}
  {\false}$: in our concurrent 
language,  $\term_1$ and $\term_2$ are trace equivalent. 
If we had higher-order store, then 
a  distinguishing
  trace could begin thus: $ \movePI{(),\rn}.\questP{\fn}{\gn,\pn}.\ansO{\pn}{}.\questO{\gn}{\qn}.\ansP{\qn}{}$.
  However, as seen in Example~\ref{exa:gc}, after the answer $\ansO{\pn}{}$ we can garbage collect $\gn$ meaning that the question $\questO{\gn}{\qn}$
  is not allowed by the visibility LTS. More generally, after the
  answer to $\questP{\fn}{\gn,\pn}$, name $\gn$ cannot be used anymore and thus 
  Opponent cannot access $\var$; hence $\readop \var
  \mathrel{\&\&} \mathtt{not}\,\readop \var$ always returns $\false$. 

% As a running example, we consider the following terms:
% \newcommand*{\termfalse}{\term_{\false}}
% \begin{align*}
%   \term_1 &\defeq \newref \loc \true {\app f {(\lambda \_. \writeop \loc \false); \false}} \\
%   \term_2 &\defeq \newref \loc \true {\app f {(\lambda \_. \writeop \loc \false); \termfalse}} \\
%   \termfalse &\defeq \ifte {\readop \loc} {(\ifte {\readop \loc} \false \true)} {\false}
% \end{align*}
% In a sequential setting, the term \( \termfalse \) is equivalent to $\false$, as it can be seen as a macro for \( \readop \loc \mathrel{\&\&} \readop \mathtt{not}\, \loc \).
% \sk{(TODO) write the trace that distinguishes these two terms (assuming that we do not care about visibility).}
% $\tr = \questP{\fn}{\gn,\pn}.\ansO{\pn}{}.\questO{\gn}{\qn}.\ansP{\qn}{}$
% We have both $\term_2 \Trans{\tr} \true$ and $\term_2 \Trans{\tr} \false$ but only $\term_1 \Trans{\tr} \false$. 
\end{example}
% \sk{Think for a good example that uses the parallel pair.}

% \iq{Proposition of example, putting two instances of the previous example in parallel using the same reference.}
% \begin{example}
%   \newcommand*{\termfalse}{\term_{\false}}
%   \newcommand*{\termtrue}{\term_{\true}}
%   \begin{align*}
%     \term_1 &\defeq \newref \loc \true {\parop {\left(\app f {(\lambda \_. \writeop \loc \true)};\true\right)} {\left( \app f {(\lambda \_. \writeop \loc \false)}; \false \right)}}\\
%     \term_2 &\defeq \newref \loc \true {\parop {\left(\app f {(\lambda \_. \writeop \loc \true)};\termtrue\right)} {\left( \app f {(\lambda \_. \writeop \loc \false)}; \termfalse\right)}}\\
%     \termtrue &\defeq \ifte {\readop \loc} {\true} {(\ifte {\readop \loc} \false \true)}
%   \end{align*}
% \end{example}

\section{Soundness of OGS}

The goal of this section is to sketch the proof of the following theorem relating the trace-semantics preorder with the ciu-preorder. 
\begin{theorem}[Soundness]
  \label{thm:soundness}
  % For all $\storectx,\namectx,\typeb,\term_1,\term_2$ if $\storectx,\namectx \vdash \term_1\trle \term_2 : \typeb$ then $\storectx,\namectx \vdash \term_1\ciule \term_2 : \typeb$
  %(Ken) I made the quantification implicit
  If $\emptyctx,\namectx \vdash \term_1\trle \term_2 : \typeb$ with $\namectx,\typeb$ ref-free, then $\emptyctx,\namectx \vdash \term_1\ciule \term_2 : \typeb$.
\end{theorem}

\paragraph*{Composition via synchronization}

The first step to prove this theorem is to introduce a way to compose the embedding of a term $\term$ as an active OGS configuration, with the embedding of an evaluation context $\pectx$
as a passive configuration. This composition is defined by a form of synchronization between transitions of $\OGSLTS$ triggered respectively by a passive and an active configuration.
As  standard in process algebra, this synchronization is defined via a form of parallel composition.
Since our goal is to compose two OGS configurations that are `closing' each other (as in the composition of $\pectx[\term\substg{\gamma}]$ that provides a closed term), we do not allow
the composed configuration to interact with the outside world. Thus the parallel composition is defined as a reduction relation.

We reflect this idea that the two OGS configurations are closing each other
by introducing the notion of pairs of \emph{composable} configurations $(\ogsconfP,\ogsconfO)$.
It enforces, among other things, that each suspended contexts present in one configuration corresponds to a thread in the other,
and that their respective typing configurations can be written as
$\ctuple{\typectxO\mid\typectxP}$
and $\ctuple{\typectxP,\pn_f:\covartype{\Unit}\mid \typectxO}$.
The extra continuation name $\pn_f$ in Opponent typing configuration
is used to perform the final Opponent answer.
\iftoggle{fullappendix}{Its full definition can be found in Appendix~\ref{ap:subsec-composability}}{}

\begin{definition}[Parallel composition]
  \label{def:parcomp}
  We define the reduction relation
  $\sored$ 
  on pairs of $\OGS$
  composable configurations 
  $\compose{\ogsconf_1}{\ogsconf_2}$,
  defined as $\soopred \cup \sosynchred$,
  where:
{\small   \begin{mathpar}
  \inferrule*{
    \ogsconfP \ogstrans{\oplabel} \ogsconfPb 
    \and 
    \supp(\ogsconfPb) \cap \supp(\ogsconfO)
    \subseteq \supp(\ogsconfP) \cap \supp(\ogsconfO)
    }{
      \compose{\ogsconfP}{\ogsconfO} \soopred
      \compose{\ogsconfPb}{\ogsconfO}
    }
  \and
  \inferrule*{
    \ogsconfO \ogstrans{\oplabel} \ogsconfOb 
    \and 
    \supp(\ogsconfOb) \cap \supp(\ogsconfP)
    \subseteq \supp(\ogsconfO) \cap \supp(\ogsconfP)
    }{
      \compose{\ogsconfP}{\ogsconfO} \soopred
      \compose{\ogsconfP}{\ogsconfOb}
    }
  \and
  \inferrule*{
    \ogsconfP \ogstrans{\move} \ogsconfPb
    \and
    \ogsconfO \ogstrans{\switchio{\move}} \ogsconfOb
  }{
    \compose{\ogsconfP}{\ogsconfO} 
    \sosynchred
    \compose{\ogsconfPb}{\ogsconfOb}}
  \end{mathpar}
}\end{definition}
Following~\cite{Laird20},
the nominal side-condition for $\soopred$
ensures us that freshly-allocated locations
in one configuration
do not appear in the other configuration.
Thus composability is preserved by $\sosynchred$.
We use parallel composition to observe the synchronization process between two OGS configurations.
So we write 
$\wbarb{\compose{\ogsconfP}{\ogsconfO}}{\pn}$,
when there exist two OGS configurations $\ogsconfPb,\ogsconfOb$
such that
$\compose{\ogsconfP}{\ogsconfO} 
 \sored^{*} 
 \compose{\ogsconfPb}{\ogsconfOb}$,
 with the interactive configuration of $\ogsconfPb$ of the shape $\ctuple{\storeP;\gammaP}$,
 and  the interactive configuration of $\ogsconfOb$ of the shape 
$\ctuple{\named{\pn}{\unit};\storeO;\gammaO}$.

\paragraph*
%\subsection
{Observational adequacy}
\label{sec:obs_adeq}

%We now state that parallel composition of OGS configurations satisfies a form of observational adequacy.
%
%For this, we define another way to compose composable configurations : $\recons{\ogsconf}{\ogsconfb}$ \\
%such that if $\compose{\langle \storectx \vdash (\term,\store):\typeb\rangle}{\langle \vdash (\pectx,\gamma,\store'):(\ectxtype{\typeb}{\Unit};\namectx)\rangle} \sosynchred \compose{\ogsconf}{\ogsconfb}$ 
%then $\recons{\ogsconf}{\ogsconfb}= (\pectx[\term\{\gamma\}],\store\cdot\store')$.

As the parallel composition $\compose{\ogsconfP}{\ogsconfO}$ is formed
by two composable configurations that are closing each other,
we introduce a way to merge them  into an operational configuration, written $\recons{\ogsconfP}{\ogsconfO}$, 
which is of the shape
$(\term\{\gamma\},\store)$ with:
\begin{itemize}
\item $\term$ is defined by shuffling the thread pools of $\ogsconfP$ and $\ogsconfO$ into one, 
feeding their respective holes. It uses the trees $\stackTP$ and $\stackTO$ of the visible configuration
to substitute the threads in their corresponding suspended contexts.
\item $\gamma$ is defined by shuffling the value assignments of $\ogsconfP$ and $\ogsconfO$ into one,
following the order in which the function names in their domain have been introduced.
One thus needs to enforce that such an order does exist, which is done in the
definition of composability, and preserved by the reduction semantics of parallel composition.
\item $\store$ is simply the concatenation $\storeP\cdot\storeO$ of the respective stores of 
$\ogsconfP$ and $\ogsconfO$, as they are disjoint since we do not allow shared store,
and hold only first-order values so that we do not need to use $\gamma$ on them.
\end{itemize}

While the shuffling of $\gammaP$ and $\gammaO$ is the same construction as the one used in sequential OGS (e.g.~\cite{JaberMurawski21,DBLP:conf/esop/BorthelleHJZ25}), the one for the shuffling of thread pools is new.
Indeed, in a sequential well-bracketed setting, rather than two tree structures, we simply have
two stacks of same length. % that can be directly shuffled. 
\iftoggle{fullappendix}{The  definition  can be found in Appendix~\ref{apx:obs_adeq}.}{}

Our goal is then to relate the observation over $\compose{\ogsconfP}{\ogsconfO}$ defined above via
the reduction semantics of the parallel composition,
with the termination observation on $\recons{\ogsconfP}{\ogsconfO}$ defined via the operational reduction.

\begin{theorem}
  \label{thm:obs-adequacy}
Taking $\ogsconfP,\ogsconfO$ two composable OGS configurations, then $\compose{\ogsconfP}{\ogsconfO} \Downarrow \pn_f$
if and only if $\recons{\ogsconfP}{\ogsconfO} \Downarrow \unit$.
\end{theorem}

The left-to-right direction follows the standard proof of OGS, namely that if 
$\compose{\ogsconfP}{\ogsconfO}\sosynchred \compose{\ogsconfPb}{\ogsconfOb}$ then $\recons{\ogsconfP}{\ogsconfO}=\recons{\ogsconfPb}{\ogsconfOb}$
\iftoggle{fullappendix}{(Lemma~\ref{lm:sync_eq} in the Appendix)}{}.
The right-to-left direction is more challenging in presence of parallelism in the programming language. Indeed,
when an operational reduction is available for $\recons{\ogsconfP}{\ogsconfO}$, some synchronisation steps may be required for $\compose{\ogsconfP}{\ogsconfO}$ to perform the same reduction.
We then prove that we can always perform the synchronization steps eagerly, without losing the possibility of making a reduction step when available \iftoggle{fullappendix}{(Lemma~\ref{lm:synchro-eager})}{}.
Finally, we show that there are always only a finite number of synchronization steps between two reduction steps \iftoggle{fullappendix}{(Lemma~\ref{lm:saturate_sync})}{}.

From $\compose{\langle \storectx \vdash
  (\term,\store):\typeb\rangle}{\langle \vdash
  (\pectx,\gamma,\store'):(\ectxtype{\typeb}{\Unit};\namectx)\rangle}
\sosynchred \compose{\ogsconfP}{\ogsconfO}$, we derive
$\recons{\ogsconfP}{\ogsconfO}=
(\pectx[\term\{\gamma\}],\store\cdot\store')$; we then deduce  
the following corollary of Theorem~\ref{thm:obs-adequacy} on initial configurations.

\begin{corollary}
  \label{cor:ciu_par}
  $(\pectx[\term\{\gamma\}],\store\cdot\store')\Downarrow\unit$ if and only if $\compose{\langle \storectx \vdash (\term,\store):\typeb\rangle}{\langle \vdash (\pectx,\gamma,\store'):(\ectxtype{\typeb}{\Unit};\namectx)\rangle}\Downarrow \pn_f$.
\end{corollary}

\paragraph*
%\subsubsection
{Proof of Soundness}

The final ingredient to prove the soundness statement 
is to relate the observation defined from the reduction relation of parallel composition with the existence of a complete trace.

\begin{theorem}
  \label{thm:wbarb-to-ctrace}
Taking $\ogsconfP,\ogsconfO$ two composable OGS configurations, then $\compose{\ogsconfP}{\ogsconfO} \Downarrow \pn_f$
if and only if there exists a complete trace $\tr$ such that $\tr \in \ogsCTr{\ogsconfP}$ and $\switchio{\tr \cdot \ansO{\pn_f}{}} \in \ogsCTr{\ogsconfO}$.
\end{theorem}

\begin{corollary}
  \label{cor:par_tr}
  $\compose{\langle \storectx \vdash (\term,\store):\typeb\rangle}{\langle \vdash (\pectx,\gamma,\store'):(\ectxtype{\typeb}{\Unit};\namectx)\rangle}\Downarrow \pn_f$ if and only if there exist a trace $\tr$ such that $\tr \in \ogsCTr{\langle \storectx \vdash (\term,\store):\typeb\rangle}$ and $\overline{\tr\cdot\pn_f\unit}\in \ogsCTr{\langle \vdash (\pectx,\gamma,\store'):(\ectxtype{\typeb}{\Unit};\namectx)\rangle}$
\end{corollary}

We now have all is needed to prove Theorem~\ref{thm:soundness}.

\begin{proof}
Take $\storectx,\namectx,\typeb,\term_1,\term_2$ such that $\storectx,\namectx \vdash \term_1\trle \term_2 : \typeb$. Let $\storectx',\gamma,\pectx,\store,\store'$
  as in Definition~\ref{def:ciu} of the ciu-preorder, with
%  such that
%  \begin{itemize}
%  \item $\dom{\storectx'}\cap\dom{\storectx}=\varnothing$ and $\storectx';\emptyset\vdash \gamma:\namectx$;
%  \item $\storectx';\emptyset \vdash \pectx:\ectxtype{\typeb}{\Unit}$; 
%  \item $\vdash \store:\storectx$; $\vdash \store':\storectx'$;
%  \item 
$(\pectx[\term_1\{\gamma\}];\store\cdot\store')\Downarrow\unit$.
%  \end{itemize}
  We show that $(\pectx[\term_2\{\gamma\}];\store\cdot\store')\Downarrow\unit$.
  By Corollary~\ref{cor:ciu_par}, we have $\compose{\langle \storectx \vdash (\term_1,\store):\typeb\rangle}{\langle \vdash (\pectx,\gamma,\store'):(\ectxtype{\typeb}{\Unit};\namectx)\rangle}\Downarrow \pn_f$.
From Corollary~\ref{cor:par_tr}, we infer that there is a trace $\tr$ such that $\tr \in \ogsCTr{\langle \storectx \vdash (\term_1,\store):\typeb\rangle}$ and $\overline{\tr\cdot\pn_f\unit}\in \ogsCTr{\langle \vdash (\pectx,\gamma,\store'):(\ectxtype{\typeb}{\Unit};\namectx)\rangle}$.

  We have  $\storectx,\namectx \vdash \term_1\trle \term_2 : \typeb$, therefore $\tr \in \ogsCTr{\langle \storectx \vdash (\term_2,\store):\typeb\rangle}$.
  By Corollary~\ref{cor:par_tr},  $\compose{\langle \storectx \vdash (\term_2,\store):\typeb\rangle}{\langle \vdash (\pectx,\gamma,\store'):(\ectxtype{\typeb}{\Unit};\namectx)\rangle}\Downarrow \pn_f$;
 finally, by Corollary~\ref{cor:ciu_par},  $(\pectx[\term_2\{\gamma\}];\store\cdot\store')\Downarrow\unit$.
\end{proof}

\section{Completeness}
\label{sec:completeness}
We prove completeness for a subset of our language.
Before delving into the result, 
we discuss the problem for completeness on the whole language.
In the usual strategy for completeness in OGS,
one takes two
contextually-equivalent term, and a trace $\tr$ for one of them.
Then one builds a
context that generates the complement trace; thus from the 
 contextual equivalence between 
the two terms combined with Corollaries~\ref{cor:ciu_par} and \ref{cor:par_tr}, 
one infers that also the other term can produce the
trace $\tr$.
Now, suppose 
$\tr$ is as follows,
using $"\val"$ to designate an abstract value such that questions on names in $"\val"$ would trigger questions on names in $\val$
% using $``\val"$ to indicate
% the abstract value corresponding to value $\val$ 
(and 
$\_$  or $\dots$ for non-relevant subterms):
$$\questO{\fn}{\_,\pn}\cdot \questO{\fn}{\_,\qn}\cdot \ansP{\pn}{\gn}\cdot \questO{\fn}{``\gn",\rn}\cdot\ansP{\qn}{\mathsf{h}}\cdot\questO{\fn}{``\langle\gn,\mathsf{h}\rangle",\mathsf{s}}\cdot \ansP{\rn}{}\cdot \ansP{\mathsf{s}}{}$$
\gj{A minor issue is that due to typing, $\fn$ cannot use both a pair $`\langle\gn,\mathsf{h}\rangle"$
and a single function name $\gn$ as arguments.}
Trace $\tr$  can indeed be generated by a configuration. We explain
why its dual trace $\overline\tr$
cannot be generated by a context.
The two first actions of  $\overline\tr$ should 
 necessarily emanate from parallel components, 
 because, in $\tr$,
 the second question
 is asked before the first one is answered. That is, 
the  context should be  of the form 
$\parop{\letin{\gn}{\fn\, \_}{\dots}}{\letin{\mathsf{h}}{\fn\,
    \_}{\dots}}.$
\newline
Moreover, in $\tr$,  question $\questO{\fn}{``\gn",\rn}$
precedes the answer at $\qn$, and uses a name introduced in the answer at
$\pn$;   this forces  
 the context to  have the form  $\parop{\letin{\gn}{\fn\, \_}{\fn\; \gn}}{\letin{\mathsf{h}}{\fn\, \_}{\dots}}$.
Correspondingly, $\overline\tr$
should exhibit the 
action
$\questP{\fn}{``\langle\gn,\mathsf{h}\rangle",\mathsf{s}}$;
this is impossible, as 
 no subterm of the context knows 
 both   $\gn$ and $\mathsf{h}$.
%We conclude that  
%no  context can generate the dual of the trace above.
%We leave the question of completeness on the whole language for future work. 

%\ds{explain the use of divergence}

In the remainder of the section we focus
on the restricted language, obtained by  imposing that
functions \iqr{at the boundary }only return first-order values, which we call the \emph{fo-returning fragment}.
Notwithstanding the restriction, 
 concurrency  introduces 
 significant challenges. 
First of all, 
the order in which two consecutive  actions occur in a trace cannot always be enforced. 
That is, a term, or a context, 
 that produces a given trace will also produce other traces,
obtained by commuting certain actions. 
% As a consequence, in the completeness proof 
%  we have to reason on \emph{families} of traces.
A commutation
is possible, intuitively, when the two actions are
 independent; that is, they are  
syntactically unrelated and there is no internal causality between
them (i.e., via use of references).
Concretely this may occur  when the two actions:
 are   both Proponent moves;  
are both Opponent moves; the first  is a Proponent move $\pmove$ and the second 
 an Opponent move $\omove$ that  
is not \emph{justified} by  the Proponent move, written $\njustif{\pmove}{\omove}$,
in that
the Opponent move  
does not use names introduced in the Proponent move\footnote{This is
 referred to as 
`courtesy' in concurrent game semantics \cite{CastellanCW19}}.
We write $\tr' \lttr \tr $ if $\tr'$ is a legal permutation of $\tr$,
where  
 $\lttr$  is the least precongruence on traces induced by the following 
relation between consecutive actions: 
{\small \begin{mathpar}
  \inferrule*{\,}{ \omove\cdot\omove' \lttr  \omove'\cdot \omove}
  \and
  \inferrule*{\,}{ \pmove\cdot\pmove'\lttr \pmove'\cdot \pmove}
  \and
  \inferrule*[Left=Crt]{\njustif{\pmove}{\omove}}{ \omove\cdot \pmove  \lttr  \pmove\cdot \omove}
\end{mathpar}
}
% The usual idea of such proofs is to consider a trace $\tr$ of $\term_1$ and to build a
% context that generates exactly $\overline{\tr\cdot\pn_f\unit}$, and thus we can deduce
% that $\tr$ is also a trace of $\term_2$. However, in a concurrent setting, some traces are
% indistinguishable as the order in which two independant actions occur cannot be
% enforced. This means that for a trace $\tr$ we also need to consider some permutations of
% $\tr$. For instance, two consecutive player moves are necessarily independant, if one
% depended on the other there should be at least an opponent answer between them. Similarly,
% two consecutive opponent moves are necessarily independant. And last, if an opponent move
% $\omove$ follows a player move $\pmove$, unless $\omove$ uses names introduced by
% $\pmove$, player cannot assume any dependency between these moves, therefore $\omove$ can
% be done before $\pmove$, this notion is known as courtesy in concurrent game semantics.\\
% More formally, we define the notion of justification for a move using names introduced by
 % another move :

\begin{remark}
\label{r:courtesy} 
The dual of the courtesy rule {\tt Crt} does not  
appear, meaning that  an Opponent  move followed by a Proponent move cannot be 
commuted. The reason is that, in these cases, 
 an internal causality may be created
using references~\cite{GhicaMurawski08}.  To see an example, consider
a pool of threads $\named{\pn}{\suspend \ectx \rn} \parcomp
\named{\qn} \fn \, \val$: the answer on $\rn$ and the question on
$\fn$ may look independent; however
 we can enforce an order between them  thus:
{\small\[\named{\pn}{\suspend {(\lambda \var.\ \writeop \loc 1;\var) \ectx} \rn} \parcomp \named{\qn} \ifte {\readop \loc = 1}{\fn \, \val}{\Omega}\]}
If the $\ifte{}{}{}$ is evaluated before the answer on $\rn$, then 
 the process diverges; this is avoided
if the question on $\fn$  happens after the answer on $\rn$.
\end{remark}

\begin{remark}[Divergence]
  As in the previous remark, it is a standard practice to use divergence to throw away undesired behaviours of some terms.
  This takes advantage of the use of complete traces instead of general traces, when any part of a term has a divergence,
  the term as a whole cannot terminate and the execution cannot be completed into a complete trace and this execution does not
  need to be considered.
\end{remark}

We also have to take into account, however, that our language does not
allow higher-order store, and hence only the \emph{visible traces}, that
is, intuitively, traces produced  by the visible LTS on some given
visible configuration, are relevant \iftoggle{fullappendix}{(see Appendix~\ref{sec:app_completeness} for details)}{}.
%  A trace
% can be visible \emph{from some visible configuration $\vconf$}, or
% begin with an initial move in which case the visible configuration
% considered is the one obtain after that initial move.
%
% \begin{definition}[Visible traces]
%  $ $  
% \begin{itemize}
% \item
% A  trace $\tr$  is \emph{visible from (a visible configuration)
%   $\vconf$}  if  $\vconf \vTrans{\tr}$
%
% \item 
% A  trace $\tr = \imove \cdot\tr'  $ (where $\imove$ is an initial
% move)
% is 
% \emph{visible} if $\tr'$ is visible from the (unique) visible configuration
% resulting from $\imove$.
% \end{itemize}
% %   Let $\vconf$ a visible configuration and $\tr$ a trace, we say
% %   that $\tr$ is visible from $\vconf$ if $\vconf \vTrans{\tr}$\\ 
% %   % Let $\vconf$ a visible configuration, $\tr$ a trace and $\lab$ a
% %   labelling, we say that $\tr_\lab$ is visible from $\vconf$ if
% %   $\vconf \vTrans{\tr_\lab}$\\ 
% %   Let $\tr$ a trace starting with an initial move, if $\tr_{\geq 1}$
% %   is visible from the visible configuration after that initial move,
% %   we say that $\tr$ is visible. 
% \end{definition}
%
Lemma~\ref{lm:CTr_lttr} shows that, on  visible traces,
the  permutations defined above are  indeed possible.
\begin{lemma}
  \label{lm:CTr_lttr}
  If $\tr \in \ogsCTr{\ogsconf}$ then for all $\tr'\lttr\tr$ such that $\tr'$ is visible from $\vconf_\ogsconf$, $\tr'\in \ogsCTr{\ogsconf}$.
\end{lemma}

In the proof of the lemma, 
we  reason
 by   induction  on the proof of 
 $\tr'\lttr \tr$.
The difficult case  is the courtesy rule {\tt Crt} where, sometimes, 
a permutation  may imply a modification 
on the   Opponent actions involved, such as changing the thread label
$\qn$ in  $\questO{\fn}{\aval,\pn}_\qn$.
The final and crucial step is to show 
that,  for any given complete visible trace,  
 a context exists  that
generates its complement trace
% $\overline{\tr\cdot \pn_f}$
and all its family of legal permutations.

\begin{lemma}
  \label{lm:ex_context}
  For all $\term,\store,\namectx,\typeb$ in the fo-returning fragment with $\namectx,\typeb$ ref-free, for all complete trace $\tr \in \ogsCTr{\langle \namectx \vdash (\term,\store):\typeb\rangle}$, there exists a context $\ectx$, a value assignment $\gamma$ and a store $\store'$ disjoint from $\store$ such that $\ogsCTr{\langle \vdash (\ectx,\gamma,\store'):(\typeb\to \Unit;\namectx)\rangle} = \{ \tr' | \tr'\lttr \overline{\tr\cdot\pn_f}\ \text{ with }\tr' \text{ visible }\}$.
\end{lemma}
% Interestingly, the context built in this definability result is always sequential.  %%% previous version
\iqr{The context built in this definability result is a sequential context, which does not use any parallel composition before the first interaction with the term.}
We sketch the reasoning,  
 taking the point of view of the environment $\ectx,\gamma$ we  build.
% thus
% a Proponent move $\questP{\fn}{\aval,\pn}$ means that 
%$\fn\ \val$ is a subterm of $\ectx$, for some $\val$. 
%\gj{Shouldn't $\gamma$ be also taken into account ?}
%
Some care is needed for the following issues:

\begin{itemize}
\item As shown in Remark~\ref{r:courtesy}, 
 an Opponent move  cannot be commuted with a following
 Proponent move. This can be    enforced
 along the lines of 
what suggested   in that remark.
%   they should not be commuted and we need to this order. This will
%   be done using  
% %   Some order of moves have to be strict even if the moves are
% independant, otherwise our context will generate too many traces :
% If an opponent move $\omove$ appears before a player move $\pmove$
% then they should not be commuted. 
% % \iq{maybe rewrite a bit above: refer now to Remark~\ref{r:courtesy} }
\item For each Opponent function name $\fn$, if there are n questions with subject $\fn$, then $\fn$ should be called exactly n times.
\end{itemize}

We handle these issues using
 locks,  modeled by means of references. A lock may be tested to know
 its state (open/closed); and its state may be atomically changed,
 exploiting the  $\mathtt{CAS}$ construct.  
When  something unexpected is encountered in a branch, we force a
divergence, so to rule out the possibility of taking that branch.

\begin{itemize}
\item A lock is modeled by means of a reference,  set to $0$ 
to mean that the lock is closed, or $1$ if open. For each Opponent
move $\omove$ we create a lock $\loc_\omove$, initially
closed. After the move $\omove$ is performed, we open the lock. For
each Proponent move $\pmove$, we consider the set of Opponent moves that 
  should occur before $\pmove$ and the set $L_\pmove$ of their
  locks. Before performing any Proponent
  move $\pmove$, we check that the references in $L_\pmove$ are all
  set to $1$; if they are not, we diverge:
$\mathtt{check\_lock}_\pmove \defeq \Parn{\loc \in L_\pmove}{\ifte{!\loc=1}{()}{\Omega}}$
\item Using the same locks, we ensure that each Opponent question
  occurs exactly once. After an Opponent question $\omove$, and
  before any Proponent action triggered by $\omove$, we check
  that the lock $\loc_\omove$ is set to $0$: if it is, we set it
  to $1$ and continue, otherwise we diverge. 
To ensure
  that no interleaving can bypass the test, we use the atomic
  construct $\cas {\loc_\omove} 0 1$. 
\end{itemize}

A further technicality  to handle is about multiple Opponent questions for the same function
 name. For this  we introduce, in the definition of these functions, a separate branch for each of them; 
 to choose the branch we start by inspecting the structure of the abstract value received; when  some abstract values are
 indistinguishable, we set a non-deterministic branching using a race between the different choices.
\iftoggle{fullappendix}{For more details, we refer to Appendix~\ref{sec:app_completeness}.}{}
 
Exploiting the above ideas, 
 we build 
an
evaluation context
 $\ectx_\tr$, a value assignment
$
\delta_\tr$, and 
a store $\store_\tr$ that can produce  the expected trace.
%  $\tr$, where $\ectx_\tr$ is an
% evaluation context, $\delta_\tr$ is a value assignment and $\store_\tr$ is
We then need to show three results:
 the context  has the expected type; the context 
 generates the trace;  the context does not generate unexpected complete
 traces (i.e., traces  $\tr'$ with $\neg (\tr'%\cancel{
\lttr%}
 \tr)$). 
For the last result, intuitively, 
the
 use of locks makes sure that the context  diverges
as soon 
 an unwanted prefix trace is generated.
\iftoggle{fullappendix}{Again, we refer to 
 Appendix~\ref{sec:app_completeness} for more details.}{}
% The typing of the context is left to
% Appendix~\ref{sec:app_completeness}. We need to check that 
% the context
% generates the intended trace. 
% \begin{lemma}
%   \label{lm:mk_trace}
%   For all $\term,\store,\typeb$, given a complete trace $\tr \in
%   \ogsCTr{\langle \namectx \vdash (\term,\store):\typeb\rangle}$,
%   we have $\overline{\tr\cdot \pn_f}\in \ogsCTr{\langle \vdash
%   (\ectx_{\overline{\tr\cdot \pn_f}},\delta_{\overline{\tr\cdot
%   \pn_f}},\store_{\overline{\tr\cdot
%   \pn_f}}):(\typeb\to\Unit;\namectx)\rangle}$ 
% \end{lemma}
% \begin{proof}
%   The proof is split into three parts,  for each component of the
%   configuration: 
%   \begin{itemize}
%   \item For the interactive configuration, we build interactive
%   configurations for each step of the trace and show that they can
%   make the next step. 
%   \item For the typing configuration, we show by induction on traces
%   that if $\ctuple{\namectxO\mid \namectxP}\tTrans{\tr}$ then
%   $\ctuple{\namectxP\mid \namectxO} \tTrans{\overline{\tr}}$. 
%   \item For the visible configuration we need to show that
%   $\overline{\tr\cdot \pn_f}$ is visible. To prove this we build a
%   sequence of visible configurations that will validate each step of
%   the reduction $\ogsCTr{\langle \namectx \vdash
%   (\term,\store):\typeb\rangle} \tTrans{\tr}$. 
%   \end{itemize}
% \end{proof}
%
% Having carried out the above construction,
%  we can finally prove completeness

\begin{theorem}[Completeness wrt fo-returning fragment]
  \label{thm:completeness}
  If $\emptyctx,\namectx \vdash \term_1\ciule \term_2 : \typeb$ with $\term_1$, $\term_2$ being terms of the fo-returning fragment and $\namectx,\typeb$ ref-free, then $\emptyctx,\namectx \vdash \term_1\trle \term_2 : \typeb$.
%  For all $\storectx,\namectx,\typeb,\term_1,\term_2$ if $\storectx,\namectx \vdash \term_1\ciule \term_2 : \typeb$ then $\storectx,\namectx \vdash \term_1\trle \term_2 : \typeb$.
\end{theorem}

\begin{proof}
We take a trace $\tr$ generated by $\term_1$. Using the existence of a
context $(\ectx,\gamma,\store')$ generating $\{ \tr' | \tr'\lttr \overline{\tr\cdot\pn_f}\
\text{ with }\tr' \text{ visible }\}$,  ciu approximation, 
 Corollaries~\ref{cor:ciu_par} and \ref{cor:par_tr} in both directions, 
we have $\compose{\langle \namectx \vdash
(\term_2,\store):\typeb\rangle}{\langle \vdash
(\ectx,\gamma,\store'):(\ectxtype{\typeb}{\Unit};\namectx)\rangle}\Downarrow \pn_f$.
Hence there is  some visible $\tr'$  generated by $\term_2$ such
  that $\overline{\tr'\cdot \pn_f}$ is generated by the context. This
  means that $\overline{\tr'\cdot \pn_f}\lttr\overline{\tr\cdot
    \pn_f}$; we can conclude that $\tr \lttr \tr'$ thus $\tr$ is a complete trace generated by $\term_2$.
\end{proof}

\section{ Conclusions, Related and Future Work}
\label{sec:related-work}
%\sk{Very rough draft...}
%\paragraph*{Conclusions}ma
\vspace{-.5em}
We have investigated how  visibility, a  central notion  in operational game
semantics (OGS),  can be accommodated within a concurrent setting. 
\ds{Thus we have constructed a complete trace semantics for a 
 (call-by-value)
  concurrent higher-order language with
full ground references.
The constraint on 
full ground references, and hence the absence of higher-order store and the need of taking
visibility into account, is  precisely the crux of the study in the paper. 
We have  shown the semantics  to be sound with respect to
 the contextual preorder, and  have also established completeness under the additional assumptional  that
 functions   return first-order values.}

We currently do not know whether  completeness holds on
the full language.
%\paragraph*{Related work}
Our OGS can be thought of as a concurrent extension of~\cite{JaberMurawski21}, which gives a systematic development of visibility and well-bracketing for OGS in the sequential case.
To our knowledge, the first OGS that took (O-)visibility into account is~\cite{Jaber15}.
%\sk{@Guilhem is my claim correct?}\gj{it is :)}
 A difference between~\cite{Jaber15} and our work (as well
 as~\cite{JaberMurawski21}) is that we do not allow locations to be
 exchanged between the program and the environment. 
 To handle location exchange, and the resulting shared stores, one
 needs to generalise moves to
 moves-with-store~\cite{Laird07,MurawskiTzevelekos11}, where the
 attached store contains the values for the disclosed locations. 
\sk{I should also mention~\cite{Laird01}. The idea of concurrency pointer seems to be related to how we identify different threads.}
In our concurrent setting, we  leave such issues for future work. 
Although not an instance of OGS, the trace semantics for concurrent objects~\cite{JeffreyRathke05} also employs concepts used in game semantics such as well-bracketing and courtesy.
Their definition of well-bracketing makes essential use of thread identifiers and requires that each subtrace with the same identifier is well-bracketed in the standard sense.
%that are given to every parallel component of a program.

One of the closest denotational game models related to our OGS is that by Ghica and Murawski~\cite{GhicaMurawski08}.
They developed a game semantics for Idealised Parallel Algol~---
essentially the call-by-name variant of our target language~--- that is fully abstract with respect to may-testing equivalence.
While the relationship between operational and denotational game semantics has been made formal in some cases (for example, \cite{Jaber15} relates \cite{Laird07} to  \cite{MurawskiTzevelekos11}) there is no general method to relate  operational and denotational approaches.
It would be an interesting future work to formally relate our OGS with the denotational model of Ghica and Murawski.
The Ghica-Murawski model has also been extended to a causal model based on event structures~\cite{CastellanCW19,CastellanClairambault24}.
Both in the sequential and the causal case, no notion of visibility is explicitly given; we note that the visibility condition of causal models~\cite{CastellanClairambault24} is a concept closer to innocence rather than the absence of higher-order references.
On the other hand, Laird has introduced a notion of visibility in a concurrent setting by introducing \emph{concurrency pointers}~\cite{Laird01}, an internal thread information attached to plays.
Investigating the relationship between concurrent pointers and our labeling by continuation names is left for future work.

The concept of visibility has been incorporated into the
 \( \pi \)-calculus as a type system~\cite{HirschkoffQS25}. 
 However, 
% although the \( \pi \)-calculus is used, 
% only sequential computation
%  is considered in this work, and how to capture
%  concurrent visibility as a type system remains open. 
 computation is purely
 sequential, and how to capture
  concurrent visibility as a type system remains open. 
 This makes it hard to link our OGS with process encodings, unlike the
 case for pure \( \lambda \)-calculus against which a formal
 connection is given~\cite{JaberSangiorgi22}.
%, or possibly to build
% models of our language within the $\pi$-calculus itself  \cite{}.

Other operational techniques to reason about program equivalence of
higher-order languages with references or asynchronous channels
%\footnote{Here, we also view
%  asynchronous channels as a form of (shared) references. \sk{This
%    footnote is to include~\cite{JeffreyRathke04}, which doesn't have
%    an explicit reference cell. Not sure if this footnote is needed.}}
and/or concurrency includes Kripke logical
relation~\cite{BirkedalST12,DreyerNB12} and various form of
bisimulation~--- such as contextual and 
normal~\cite{San923,JeffreyRathke04,LengletSS11}, normal
form~\cite{BiernackiLP19} and
environmental~\cite{SangiorgiKS11,KoutavasWand06}  bisimulations~--- or
their hybrids~\cite{HurDNV12,JaberTabareu15,DBLP:conf/lics/JaberM21}. 
In these works, references, if considered, are usually
 full higher-order references rather than ground references.
An exception is~\cite{DreyerNB12} where the absence of higher-order references in a higher-order sequential language is captured as the ability to `backtrack' the Kripke frame. This technique was imported to Kripke normal-form
bisimulations in~\cite{DBLP:conf/lics/JaberM21}, via a formal connection with
the notion of visibility. Extending this `backtrack' proof technique to a concurrent setting and
relating it to our notion of concurrent visibility is an exciting research direction.
%\gj{?}

%to the best of our knowledge, this idea has not been combined with concurrency.

\sk{TODO: Consider what to list. Here some are typed and some are not. There are other works that has references, but whose main focus is not reference (e.g. the seminal work by  Ahmed, Dreyer \& Rossberg). It's a endless list. Also~\cite{JaberTabareu15} can be categorized as OGS.}
%The restriction to first-order references corresponds to the possibility to backtrack in the world.

% While our language includes references, these 
% cannot  not be
% exchanged 
% between a term and  its
% environment  (as abstract values may not be references). 
% In operational game semantics for sequential higher-order 
% languages, the exchange of references requires additional and  orthogonal
% techniques ~\cite{Jaber15,Laird07,MurawskiTzevelekos11}, such as 
%  generalising moves to moves-with-store, where actions of traces
%  include  the values for disclosed locations.
% In our concurrent setting, we  leave such issues for future work. 
% Finally, the techniques  developed in this paper could be useful also in a 
% pure name-passing language such as the $\pi$-calculus
% by separating first-order name and higher-order names. 
% Some initial work in this direction exists~\cite{HirschkoffQS25};
% however
% only purely  sequential computations are considered, in  
% a calculus  extended with reference names to model
%  store. 

% This makes it hard to link our OGS with process encodings unlike the
% case for pure \( \lambda \)-calculus against which a formal
% connection is given~\cite{JaberSangiorgi22}. 

% %We leave a  detailed investigation for future work.

%%
%% Bibliography
%%

%% Please use bibtex, 

\bibliography{main}
\clearpage
\appendix

\section{Summary of main notations}
\label{app:nota}

\begin{tabular}[h]{|l|l|l|}
  \hline
 {\bf Notation} &  {\bf Meaning} & {\bf Definition}\\
  \hline
  $\term,\termb$ & Terms &  Section~\ref{sec:type_syntax} \\
  \hline
  $\type,\typeb$ & Types &  Section~\ref{sec:type_syntax} \\
  \hline
  $\typectx$, $\storectx$ & Typing context and store typing context &  Section~\ref{sec:type_syntax} \\
  \hline
  $\store, \store [\loc \mapsto \val], \store\cdot \storeb$ & Store, store update and disjoint concatenation of stores &  Section~\ref{sec:opsem}\\
  \hline
  $\opred$& Reduction semantic of terms with store&  Figure~\ref{fig:reduction}\\
  \hline
  $\ciule,\ciueq$ & CIU approximation and CIU equivalence &  Definition~\ref{def:ciu}\\
  \hline
  $\suspend \ectx \pn, \psterm$ & Suspended context and partially suspended term &  Section~\ref{sec:suspend}\\
  \hline
  $\pstrans \mu, \opLTS$ & Reduction and LTS for interactions with the environment &  Figure~\ref{fig:def-oplts}\\
  \hline
  $\aval$ & Abstract values &  Section~\ref{sec:abstract_values}\\
  \hline
  $\move,\questP{\fn}{\aval,\pn},\ansP{\pn}{\aval},\questO{\fn}{\aval,\pn},\ansO{\pn}{\aval}$ & Moves &  Section~\ref{sec:abstract_values}\\
  \hline
  $\gamma$ & Substitutions & Section~\ref{sec:abstract_values}\\
  \hline
  $\threadpool,\threadpoolb,\named \pn \psterm$ & Pool of threads & Section~\ref{sec:abstract_values}\\
  \hline
  $\iconf,\iconfb,\ctuple{\threadpool;\store;\gamma}$& Interactive configurations & Section~\ref{sec:abstract_values}\\
  \hline
  $\itrans \move, \ILTS$ & Interactive reduction and LTS &  Figure~\ref{fig:def-ilts-new}\\
  \hline
  $\tconf,\tconfb,\ctuple{\namectxO \mid \namectxP}$ & Typing configurations &  Section~\ref{sec:tylts_vislts}\\
  \hline
  $\ttrans,\TyLTS$& Typing reduction and LTS&  Figure~\ref{fig:def-tylts}\\
  \hline
  $\vconf,\vconfb,\ctuple{\viewF;\viewG;\stackT}$& Visible configurations&  Section~\ref{sec:tylts_vislts}\\
  \hline
  $\vtrans,\VisLTS$& Visible reduction and LTS&  Figure~\ref{fig:def-vislts}\\
  \hline
  $\ogsconf,\ogsconfb,(\iconf,\tconf,\vconf)$ & OGS configurations &  Section~\ref{sec:ogs}\\
  \hline
  $\ogstrans{},\OGSLTS$& OGS reduction and LTS &  Section~\ref{sec:ogs}\\
  \hline
  $\movePI{\overrightarrow{\aval_i},\pn},\moveOI{\overrightarrow{\aval_i},\pn}$& Initial OGS moves & Figure~\ref{fig:init-move}\\
  \hline
  \( \ctxle, \ctxeq \) & Contextual preorder and contextual equivalence &  Section~\ref{sec:opsem} \\
  \hline
  $\trle,\treq$& Trace approximation and trace equivalence &  Section~\ref{sec:complete_trace}\\
  \hline
  $\parr,\soopred,\sosynchred$ & Parallel composition, operational reduction and synchronisations &  Definition~\ref{def:parcomp}\\
  \hline
  $\Merge$ & Composition by merging of configurations &  Section~\ref{sec:obs_adeq}\\
  \hline
  $\lttr$ & Preorder for commutation of traces &  Section~\ref{sec:completeness}\\
  \hline
\end{tabular}

\section{Target Language and Operational Semantics}

\subsection{Typing}
\label{app:typing}

The typing rules are shown in Figure~\ref{fig:typing}
\begin{figure}[t]
\begin{flushleft}
  \fbox{\( \typeJudge \storectx \typectx \term \type \)}
\end{flushleft}
  \begin{mathpar}
	\inferrule[Var]
		{ \typectx(x) = \type }
		{\typeJudge \storectx \typectx x \type}
    \and
  \inferrule[Loc]
		{ \storectx(\loc) = \type }
		{\typeJudge \storectx \typectx \loc \reftype\type}
    \and
  \inferrule[Unit]
		{ }
		{\typeJudge \storectx \typectx \unit \Unit}
    \and
  \inferrule[Bool]
		{ }
		{\typeJudge \storectx \typectx \bool \Bool}
    \and
  \inferrule[Int]
		{ }
		{\typeJudge \storectx \typectx \intv \Int}
    \and
    \inferrule[Pair]
		{\typeJudge \storectx \typectx \val \type \\ \typeJudge \storectx \typectx \valb \typeb}
		{\typeJudge \storectx \typectx {\pairop{\val, \valb}} {\type \times \typeb}}
    \and
    \inferrule[Proj]
		{\typeJudge \storectx \typectx \val {\type_1 \times \type_2} \\ i \in \{ 1, 2 \}}
		{\typeJudge \storectx \typectx {\projop \val} {\type_i }}
\\
    \inferrule[Abs]
		{\typeJudge \storectx {\typectx, x : \type} \term \typeb}
		{\typeJudge \storectx \typectx {\lambda x. \term} {\type \to \typeb}}
    \and
    \inferrule[App]
		{\typeJudge \storectx \typectx \val {\type \to \typeb} \\ \typeJudge \storectx \typectx \term \type}
		{\typeJudge \storectx \typectx {\app \val \term} \typeb}
    \and
    \inferrule[Rec]
		{\typeJudge \storectx {\typectx, f : \type \to \typeb, x : \type} \term \typeb}
		{\typeJudge \storectx \typectx {\letrec f x \term} \typeb}
    \and
    \inferrule[Ite]
		{\typeJudge \storectx \typectx \val \Bool \\ \typeJudge \storectx \typectx \term \type \\ \typeJudge \storectx \typectx \termb \type}
		{\typeJudge \storectx \typectx {\ifte \val \term \termb} \type}
    \\
   \inferrule[New]
		{\typeJudge \storectx \typectx \val \type}
		{\typeJudge \storectx \typectx {\refterm{\val}} \reftype \type}
    \and
    \inferrule[Deref]
		{\typeJudge \storectx \typectx \val {\reftype \type}}
		{\typeJudge \storectx \typectx {\readop \val} \type}
    \and
    \inferrule[Assign]
		{\typeJudge \storectx \typectx \val {\reftype \type} \\ \typeJudge \storectx \typectx \valb \type}
		{\typeJudge \storectx \typectx {\writeop \val \valb} \Unit}
    \and
    \inferrule[Par]
		{\typeJudge \storectx \typectx \term \type \\ \typeJudge \storectx \typectx \termb \typeb}
		{\typeJudge \storectx \typectx {\parop \term \termb} {\type \times \typeb}}
    \and
    \inferrule[Cas]
		{\typeJudge \storectx \typectx \valb {\reftype \type} \\ \typeJudge \storectx \typectx {\val_1} {\type} \\ \typeJudge \storectx \typectx {\val_2} \type}
		{\typeJudge \storectx \typectx {\cas \valb {\val_1} {\val_2}} \Bool}
\end{mathpar}
\begin{flushleft}
\fbox{\( \typeJudge \storectx \typectx \pectx {\ectxtype{\typeb}{\type}}  \)}
\end{flushleft}
\begin{mathpar}
  \inferrule[Hole]{ }{\typeJudge \storectx \typectx \hole {\ectxtype \type \type}}
  \and
  \inferrule{\typeJudge \storectx \typectx \pectx {\ectxtype \typeb \type} \\ \typeJudge \storectx \typectx \val {\type \to \type'}}{\typeJudge \storectx \typectx {\app \val \pectx} {\ectxtype \typeb {\type'}}}
  \and
  \inferrule{\typeJudge \storectx \typectx \pectx {\ectxtype \typeb \type_1} \\ \typeJudge \storectx \typectx \term {\type_2}}{\typeJudge \storectx \typectx {\parop \pectx \term} {\ectxtype \typeb {\type_1 \times \type_2}}}
    \and
  \inferrule{ \typeJudge \storectx \typectx \term {\type_1}\\ \typeJudge \storectx \typectx \pectx {\ectxtype \typeb \type_2}}{\typeJudge \storectx \typectx {\parop \term \pectx} {\ectxtype \typeb {\type_1 \times \type_2}}}
\end{mathpar}
\caption{Typing rules.}
\label{fig:typing}
\end{figure}

\iftoggle{fullappendix}{\section{Operational semantics}

\begin{lemma}
  \label{lm:opred-preserve-threadsubst}
  Let $\psterm,\pstermb,\store,\storeb$ such that $(\psterm,\store)\opred (\pstermb,\storeb)$,
  then for any $\overrightarrow{(\term_i,\pn_i)}$, $(\psterm[\overrightarrow{(\term_i,\pn_i)}],\store)\opred (\pstermb[\overrightarrow{(\term_i,\pn_i)}],\storeb)$.
\end{lemma}

\begin{proof}
By induction on $\psterm$.
\end{proof}

\begin{lemma}
  \label{lm:opred-preserve-varsubst}
  Let $\psterm,\pstermb,\store,\storeb$ such that $(\psterm,\store)\opred (\pstermb,\storeb)$,
  then for any $\gamma$ \gj{Need to add typing constraints}, 
  $(\psterm[\gamma],\store)\opred (\pstermb[\gamma],\storeb)$.
\end{lemma}

\begin{lemma}
  \label{lm:opred-reflect-varsubst}
Let $\gamma$ and $\psterm$ such that $\psterm \cancel{\pstrans{\mu}}$ for Proponent question $\mu$. 
Let $\store$ a store, if $(\psterm[\gamma],\store) \opred (\pstermb,\storeb)$ then there exists $\psterm'$ such that $(\psterm,\store)\opred (\psterm',\storeb)$ and $\pstermb = \psterm'[\gamma]$.
\end{lemma}

\begin{proof}
  We start by proving it for the case $\psterm = \term$ by induction on $\term$, we only consider 3 cases : $\val\,\termb, \refterm\val\, \parop {\termb_1} {\termb_2}$ the other ones are similar trivial or similar :
  \begin{itemize}
    \item $\term = \val\,\termb$, if $\termb$ is a value and $\val = \fn\in \dom{\gamma}$ then $\psterm \pstrans{\fn(\val,\pn)} \suspend \hole \pn$ which contradicts our hypothesis. \\
      Otherwise, if $\termb$ is a value and $\val = \lambda \var.\termb'$ then the only reduction possible of $\term[\gamma]$ is $(((\lambda x.\termb')\ \termb)[\gamma];\store) = ((\lambda x.\termb'[\gamma])\ \termb[\gamma];\store) \opred (\termb'[\gamma]\subst{\var}{\termb[\gamma]};\store) = ((\termb'\subst \var \termb)[\gamma];\store)$.\\
      Otherwise, if $\termb$ is a value then $\term$ does not reduce.\\
      Otherwise, $\termb$ is not a value, then the only possible reductions are $(\term[\gamma],\store) = (\val[\gamma]\, \termb[\gamma],\store) \opred (\val[\gamma]\, \termb',\storeb)$ where $(\termb[\gamma],\store) \opred (\termb',\storeb)$. By induction hypothesis we have that there exists $\term'$ such that $(\termb,\store)\opred(\term',\storeb)$ and $\termb'=\term'[\gamma]$. Hence $(\term,\store)\opred(\val\,\term',\storeb)$ and $\val[\gamma]\,\termb' = (\val\,\term')[\gamma]$
    \item $\term = \refterm\val$, then $\val$ is a ground value so that $\term[\gamma]=\term$
    \item $\term = \parop{\termb_1}{\termb_2}$ If $(\term[\gamma],\store)\opred (\term',\storeb)$ then without loss of generality assume $(\termb_1[\gamma],\store)\opred(\term_1',\storeb)$ and $\term'=\parop{\term_1'}{\termb_2[\gamma]}$, by induction there exists $\termb_1'$ such that $(\termb_1,\store)\opred (\termb_1',\storeb)$ and $\term_1'=\termb_1'[\gamma]$ therefore $(\term,\store)\opred(\parop{\termb_1'}{\termb_2},\storeb)$ and $\term'= (\parop{\termb_1'}{\termb_2})[\gamma]$
  \end{itemize}
  The general case is straightforward by induction on $\psterm$.
\end{proof}

\begin{lemma}
  \label{lm:opred:reflect:threadsubst}
   Let $\psterm,\store,\term_1,\dots,\term_n,\pn_1,\dots,\pn_n$ such that none of the $\term_i$ are values, 
   therefore if $(\psterm[\overrightarrow{(\term_i,\pn_i)}],\store)\opred (\pstermb,\storeb)$ 
   then either there exists $\pstermb'$ such that $(\psterm,\store)\opred(\pstermb',\storeb)$ and $\pstermb'[\overrightarrow{(\term_i,\pn_i)}]=\pstermb$ or there exists $\termb_i$ such that $(\term_i,\store)\opred (\termb_i,\storeb)$ and $\psterm[\overrightarrow{(\term_j,\pn_j)}_{j\neq i}, (\termb_i,\pn_i)] = \pstermb$.
\end{lemma}

\begin{proof}
    This result is straightforward by induction on $\psterm[\overrightarrow{(\term_i,\pn_i)}]$.
\end{proof}
}{}

\section{Operational Game Semantics}
\label{app:def-ogs}

\begin{definition}
  An $\OGS$ configuration
  $(\iconf,\tconf,\vconf)$,
  with:
  \begin{itemize}
	\item $\iconf = \ctuple{\named{\pn_1}{\psterm_1}
	\parcomp \cdots \parcomp \named{\pn_k}{\psterm_k};
	\store;\gamma}$;
	\item $\tconf = \ctuple{\namectxO \mid \namectxP}$;
	\item $\vconf = \ctuple{\viewF; \viewG; \stackT}$;
  \end{itemize}
  is said to be \emph{valid} when:
  \begin{itemize}
	\item there exists a store typing context
	$\storectx$ such that:
	\begin{itemize}
	  \item $\storectx;\namectxO \vdash \gamma : \namectxP$;
	  \item $\vdash \store:\storectx$;
	  \item for all $i \in \{1,\ldots,k\}$,
		$\storectx;\namectxO \vdash \psterm_i:\type_i$
		with $\namectxO(\pn_i) = \covartype{\type_i}$.
	\end{itemize}
	\item $\dom{\viewF} = \dom{\namectxO}$
	  and $\im{\viewF} \subseteq \pf(\dom{\namectxP})$;
	\item $\dom{\viewG} \subseteq \dom{\namectxP}$
	  and $\im{\viewG} \subseteq \pf(\dom{\namectxP})$;
	\item The tree $\stackT$ is alternating,
	  that is for all $\pn \in \stackT$:
          \begin{itemize}
          \item either $\pn$ is the root and $\pn \in \dom{\namectxO}$
          \item or $\pn \in \dom{\namectxO}$ and
		$\qn \in \dom{\namectxP}$ where $\qn$ is the parent of $\pn$;
	  \item or $\pn \in \dom{\namectxP}$ and
		$\qn \in \dom{\namectxO}$ where $\qn$ is the parent of $\pn$;
	\end{itemize}
	\item The set of nodes of $\stackT$ that are	
	  in $\namectxO$ is equal to 
	  $\{\pn_1,\ldots,\pn_k\}$;
        \item $\suspend \ectx {\pn_j}$ appears in $\psterm_i$ if and only if $\pn_j$ is a child of $\pn_i$.
	\end{itemize}
\end{definition}

\begin{lemma}
	Taking $\ogsconf$ a valid $\OGS$
	configuration such that
	$\ogsconf \ogstrans{\act} \ogsconfb$
	then $\ogsconf$ is valid.
\end{lemma}

\iftoggle{fullappendix}{\section{Soundness of the $\OGS$ LTS}

The goal of this section is give the proof of Theorem~\ref{thm:soundness}.

\subsection{Composability}
\label{ap:subsec-composability}

\begin{definition}
  Taking a $\Ty$ configuration $\tconf$
  and $\pn$ a continuation name,
  we define its $\pn$-dual configuration,
  written $\dualTy{\pn}{\tconf}$, as:
  \[\dualTy{\pn}{\ctuple{\typectxO\mid\typectxP}}
    \defeq 
    \ctuple{\typectxP,\pn:\covartype{\Unit}\mid \typectxO}
  \]
\end{definition}

\begin{definition}
  An environment $\gamma$
  is said to be \emph{validated} by
  a view function $\viewF$,
  when
  for all $\nn \in \dom{\gamma}, 
    \freen(\gamma(\nn)) \subseteq \viewF(\nn)$.
  Similarly, a thread pool $\threadpool$
  is validated by a view function $\viewG$
  when for all $\pn \in \dom{\threadpool},
  \freen(\threadpool(\pn)) \subseteq \viewG(\pn)$.
\end{definition}
  
\begin{definition}
  Two
  $\OGS$ configurations 
  $(\ctuple{\threadpoolP;\storeP;\gammaP};
    \tconfP;
    \ctuple{\viewFP; \viewGP; \stackTP})$
  and \\
  $(\ctuple{\threadpoolO;\storeO;\gammaO};
    \tconfO;
    \ctuple{\viewFO; \viewGO; \stackTO})$
  are said to be $\pn$-\emph{composable}
  when:
  \begin{itemize}
    \item $\tconfO = \dualTy{\pn}{(\tconfP)}$;
    \item $\gammaP$ is validated by $\viewFO$,
    and $\gammaO$ by $\viewFP$;
    \item $\threadpoolP$ is validated by $\viewGO$,
    and $\threadpoolO$ by $\viewGP$;
    \item $\stackTO = (\pn,[\stackTP])$;
    \item There exists an order $<_{\gammaP\cdot\gammaO}$ such that $\fn\in \freen(\gammaP\cdot \gammaO(\gn)) \Rightarrow \fn <_{\gammaP\cdot \gammaO} \gn$
      \iq{we need a strict order on $\dom{\gammaP}\cup\dom{\gammaO}$ such that $f\in \gamma(g) \Rightarrow f<g$. We can either require the existence here, or (for a less ad hoc way) have a condition on $\ctuple{\viewFP; \viewGP; \stackTP}$ and $\ctuple{\viewFO; \viewGO; \stackTO}$, but for now the condition I have is a bit complicated (linking each function name to a continuation name and using the tree as an order)}
  \end{itemize}
\end{definition}

\begin{lemma}
  Let $\pn$ a continuation, and $\ogsconfP,\ogsconfO$ two $\pn$-composable $\OGS$ configurations :
  \begin{itemize}
  \item If $\ogsconfP \ogstrans{\oplabel} \ogsconfPb$ and  $\supp(\ogsconfPb) \cap \supp(\ogsconfO) \subseteq \supp(\ogsconfP) \cap \supp(\ogsconfO)$ then $\ogsconfPb$ and $\ogsconfO$ are $\pn$-composable
  \item If $\ogsconfO \ogstrans{\oplabel} \ogsconfOb$ and  $\supp(\ogsconfOb) \cap \supp(\ogsconfP) \subseteq \supp(\ogsconfO) \cap \supp(\ogsconfP)$ then $\ogsconfP$ and $\ogsconfOb$ are $\pn$-composable
  \item If $\ogsconfP \ogstrans{\move} \ogsconfPb$ and $\ogsconfO \ogstrans{\switchio{\move}} \ogsconfOb$ then $\ogsconfPb$ and $\ogsconfOb$ are $\pn$-composable
  \end{itemize}
\end{lemma}

 \begin{lemma}
   Let $\namectx, \term,\store,\storeb,\typeb,\overrightarrow{\aval_i},\pn,\pectx,\delta$, if $\langle \namectx \vdash (\term,\store):\typeb\rangle \ogstrans{\movePI{\overrightarrow{\aval_i},\pn}} \ogsconfP$ and $\langle \vdash (\pectx,\delta,\storeb):(\ectxtype{\typeb}{\Unit};\namectx)\rangle \ogstrans{\moveOI{\overrightarrow{\aval_i},\pn}} \ogsconfO$, then $\ogsconfP$ and $\ogsconfO$ are composable.
 \end{lemma}

\subsection{Observational adequacy}
\label{apx:obs_adeq}

To merge the thread pools, we use the tree $\stackT$ of the visible configuration  Then we use the order on the function names to apply the substitutions of $\gammaP$ and $\gammaO$ :
\begin{itemize}
\item We start by matching the continuations :
  \begin{align*}
    \reconscont{\threadpool}{\stackT} \defeq& \mathtt{match}\ \stackT\ \mathtt{with}\\
                                            &  | (\pn,[\stackT_1\dots \stackT_n]) \Rightarrow (\threadpool(\pn)[ \overrightarrow{\reconscont{\threadpool}{\stackT_i}}],\pn)\\
    % \psterm[\overrightarrow{(\termb_i,\pn_i)}] \defeq& \mathtt{match}\ \psterm\ \mathtt{with}\\
    %                                         & | \term \Rightarrow \term\\
    %                                         & | \suspend \ectx {\pn_i} \Rightarrow \ectx[\termb_i]\\
    %                                         & | \suspend \ectx \qn \Rightarrow \suspend \ectx \qn\\
    %                                          & | \parop {\psterm_1} {\psterm_2}\Rightarrow
    %                                            \parop {\psterm_1[\overrightarrow{(\termb_i,\pn_i)}]}{\psterm_2[\overrightarrow{(\termb_i,\pn_i)}]}\\
    %                                         & | \ectx[\psterm]\Rightarrow
    %                                           \ectx[\psterm[\overrightarrow{(\termb_i,\pn_i)}]]\\
    \term[\overrightarrow{(\termb_i,\pn_i)}] \defeq & \term \\
    \suspend \ectx {\pn_i}[\overrightarrow{(\termb_i,\pn_i)}] \defeq & \ectx[ \termb_i]\\
    \suspend \ectx \qn [\overrightarrow{(\termb_i,\pn_i)}] \defeq & \suspend \ectx \qn \text{ with } \qn \notin \overrightarrow{\pn_i}\\
    (\parop {\psterm_1} {\psterm_2}) [\overrightarrow{(\termb_i,\pn_i)}] \defeq & \parop {\psterm_1[\overrightarrow{(\termb_i,\pn_i)}]}{\psterm_2[\overrightarrow{(\termb_i,\pn_i)}]}\\
    (\ectx[\psterm]) [\overrightarrow{(\termb_i,\pn_i)}] \defeq & \ectx[\psterm[\overrightarrow{(\termb_i,\pn_i)}]]\\
  \end{align*}
  For this to be well-defined we need $\reconscont{\threadpool}{\stackT}$ to be a couple of a term and a continuation (no suspended term), this is ensured by the last condition of validity of a configuration : $\suspend \ectx {\pn_j}$ appears in $\psterm_i$ if and only if $\pn_j$ is a child of $\pn_i$.\\
  When the continuation is unnecessary, to simplify notations, we will write $\reconscont{\threadpool}{\stackT}$ for $\mathtt{fst}\ \reconscont{\threadpool}{\stackT}$
\item We then define an ordered value assignment, taking a value assignment $\gamma$ and a strict total order $<_\gamma$ such that $\fn\in \freen(\gamma(\gn)) \Rightarrow \fn <_\gamma \gn$ :\\
  \begin{align*}
    \threadpool[\gamma,<_\gamma] \defeq & \mathtt{if}\ \gamma = \emptyset \ \mathtt{then} (\term,\store)\\
                                           & \mathtt{else}\ \mathtt{let}\ \fn = \mathtt{max}(\dom{\gamma})\ \mathtt{in}\\
                                           &\hspace{2em} (\threadpool[\gamma(\fn)/\fn])[\gamma\setminus\{\fn\},<_\gamma\setminus\{(\gn,\fn)| \gn<_\gamma \fn\}]
  \end{align*}
\item We can now combine the two, for two compatible configurations $\ogsconfP,\ogsconfO$
  \[\recons{\ogsconfP}{\ogsconfO} \defeq \left(\reconscont{(\threadpoolP|\threadpoolO)[\gammaP\cdot\gammaO,<_{\gammaP\cdot\gammaO}]}{\stackTO},\storeP\cdot\storeO\right)  \]
\end{itemize}

For the ordered value assignment to be well-behaved we want the value assignment to not depend on the choice of order.
This is proven in the following lemma.

\begin{lemma}
  \label{lm:indep-order-subst}
  Let $\threadpool,\gamma,<_{1,\gamma},<_{2,\gamma}$ 
  that satifies that forall $\fn,\gn$ with $\fn\in \freen(\gamma(\gn))$,
  we have that $\fn <_{1,\gamma} \gn$ and $\fn <_{2,\gamma} \gn$.
  Then $\threadpool[\gamma,<_{1,\gamma}] = \threadpool[\gamma,<_{2,\gamma}]$.
\end{lemma}
\begin{proof}
  We start by showing that for $\fn,\gn,\term,\termb$ if $\fn\notin \freen(\termb)$ and $\gn\notin \freen(\term)$ then for all pool of thread $\threadpool$, $\threadpool[\term/\fn][\termb/\gn] = \threadpool[\termb/\gn][\term/\fn]$. This is direct by induction on $\threadpool$.\\
  For the rest of the proof, we look at the total order as a decreasing (for the order) list of all the elements in $\dom{\gamma}$. We define an equivalence relation $\equiv_\gamma$ between orders such that
  \begin{mathpar}
    \inferrule*{\fn \notin \freen(\gamma(\gn))\and \gn \notin \freen(\gamma(\fn))}{\fn :: \gn :: l \equiv_\gamma \gn :: \fn :: l}
    \and
    \inferrule*{l\equiv_\gamma l'\and \fn \notin l}{\fn ::l \equiv_{\gamma[\fn\mapsto \val]} \fn ::l'}
    \and
    \inferrule*{\,}{l\equiv_\gamma l}
    \and
    \inferrule*{l\equiv_\gamma l' \and l'\equiv_\gamma l''}{l\equiv_\gamma l''}
  \end{mathpar}
  By induction on $\equiv_\gamma$, we have that if $<_{1,\gamma} \equiv_\gamma <_{2,\gamma}$ then $\threadpool[\gamma,<_{1,\gamma}] = \threadpool[\gamma,<_{2,\gamma}]$\\
  We now need to show that for any $<_{1,\gamma},<_{2,\gamma}$ such that $\fn\in \freen(\gamma(\gn)) \Rightarrow \fn <_{1,\gamma} \gn \wedge \fn <_{2,\gamma} \gn$ we have $<_{1,\gamma}\equiv_\gamma <_{2,\gamma}$, we do so by induction on the size of $\dom{\gamma}$.\\
  If $\dom{\gamma} = \emptyset$, we have $<_{1,\gamma}=<_{2,\gamma} = []$\\
  Otherwise there exists $\fn,l$ such that $<_{1,\gamma} = \fn::l$, we have that for any $\gn\in l$, $\fn\notin \freen(\gamma(\gn))$. $<_{2,\gamma}$ being a total order on $\dom{\gamma}$, there exists $l_1,l_2$ such that $<_{2,\gamma} = l_1::\fn::l_2$, we have that for any $\gn\in l_1,\gn\notin \freen(\gamma(\fn))$ therefore we can use the commutation rule (inductively on $l_1$) to have that $<_{2,\gamma} \equiv_\gamma \fn :: l_1++l_2$ and $l,l_1++l_2$ are strict total orders on $\gamma\setminus\{\fn\}$ such that $\gn\in \freen(\gamma(\gn')) \Rightarrow \gn (l) \gn' \wedge \gn (l_1++l_2) \gn'$ hence by induction hypothesis we have $l\equiv_{\gamma\setminus\{\fn\}}l_1++l_2$ and therefore $<_{1,\gamma}\equiv_\gamma <_{2,\gamma}$
  
\end{proof}

We now write $\threadpool[\gamma]$ for the ordered value assignment $\threadpool[\gamma,<_\gamma]$ with $<_\gamma$ any strict total order on $\dom{\gamma}$ such that $\fn\in \freen(\gamma(\gn))\Rightarrow \fn <_\gamma \gn$

\begin{lemma}
  Let $\gamma_1,\gamma_2$ such that $\freen(\im{\gamma_1})\cap \dom{\gamma_2} = \emptyset$, for any pool of thread $\threadpool$, $\threadpool[\gamma_1\cdot\gamma_2]= \threadpool[\gamma_2][\gamma_1]$.
\end{lemma}
\begin{proof}
  This proof is done by exhibiting an order $<_{\gamma_1\cdot\gamma_2}$ such that if $\fn\in \dom{\gamma_1}$ and $\gn\in \dom{\gamma_2}$ then $\fn <_{\gamma_1\cdot\gamma_2} \gn$. We then proceed by induction on $\threadpool[\gamma_1\cdot\gamma_2,<_{\gamma_1\cdot\gamma_2}]$.
\end{proof}

\gj{The previous lemma does not seem to be used afterward.}

\begin{lemma}
  \label{lm:sync_eq}
  For any composable configurations $\ogsconf,\ogsconfb$, if $\compose{\ogsconf}{\ogsconfb}\sosynchred \compose{\ogsconf'}{\ogsconfb'}$ then $\recons{\ogsconf}{\ogsconfb}=\recons{\ogsconf'}{\ogsconfb'}$.
\end{lemma}

\begin{proof}
  There exists $\move$ such that $\ogsconf \ogstrans{\move} \ogsconf'$ and $\ogsconfb \ogstrans{\switchio{\move}} \ogsconfb'$, we proceed by cases on $\move$, there are 4 cases :$\questO{\fn}{\aval,\pn}_\qn$,$\questP{\fn}{\aval,\pn}_\qn$,$\ansO{\pn}{\aval}$ or $\ansP{\pn}{\aval}$. We only consider the case $\questP{\fn}{\aval,\pn}_\qn$ as the other ones will be similar.\\
  If $\move = \questP{\fn}{\aval,\pn}_\qn$, then there exists $\threadpool,\psterm,\pstermb,\val,\gamma'$ such that $\threadpool_\ogsconfb = \threadpool \parcomp \named \qn \psterm$, $\psterm \pstrans{\questP{\fn}{\val,\pn}} \pstermb$, $\val=\aval\substg{\gamma'}$, $\threadpool_{\ogsconfb'}=\threadpool \parcomp \named \qn \pstermb$ and $\threadpool_{\ogsconf'} = \threadpool_\ogsconf \parcomp \named \pn (\gamma_\ogsconf (\fn)\ \aval)$\\
    The store being unchanged by synchronisations all we have to show is \[\reconscont{(\threadpool_\ogsconf \parcomp \threadpool \parcomp \named \qn \psterm)[\gamma_\ogsconf\cdot \gamma_\ogsconfb]}{\stackT_\ogsconfb} = \reconscont{(\threadpool_\ogsconf \parcomp \named \pn (\gamma_\ogsconf (\fn)\ \aval) \parcomp \threadpool \parcomp \named \qn \pstermb)[\gamma_\ogsconf\cdot \gamma_\ogsconfb \cdot \gamma']}{\addleaf{\stackT_{\ogsconfb}}{\pn}{\qn}}\]
    As the names in $\dom{\gamma'}$ only appear in $\aval$, we have that
    \begin{align*}
      (\threadpool_\ogsconf \parcomp \named \pn (\gamma_\ogsconf (\fn)\ \aval) \parcomp \threadpool \parcomp \named \qn \pstermb)[\gamma_\ogsconf\cdot \gamma_\ogsconfb \cdot \gamma']
      &= \threadpool_\ogsconf[\gamma_\ogsconf\cdot \gamma_\ogsconfb] \parcomp (\named \pn (\gamma_\ogsconf (\fn)\ \aval))[\gamma_\ogsconf\cdot \gamma_\ogsconfb\cdot \gamma'] \parcomp (\threadpool \parcomp \named \qn \pstermb)[\gamma_\ogsconf\cdot \gamma_\ogsconfb]\\
      &= \threadpool_\ogsconf[\gamma_\ogsconf\cdot \gamma_\ogsconfb] \parcomp (\named \pn (\gamma_\ogsconf (\fn)\ \aval\substg{\gamma'}))[\gamma_\ogsconf\cdot \gamma_\ogsconfb] \parcomp (\threadpool \parcomp \named \qn \pstermb)[\gamma_\ogsconf\cdot \gamma_\ogsconfb]\\
      &= \threadpool_\ogsconf[\gamma_\ogsconf\cdot \gamma_\ogsconfb] \parcomp (\named \pn (\fn\ \val))[\gamma_\ogsconf\cdot \gamma_\ogsconfb] \parcomp (\threadpool \parcomp \named \qn \pstermb)[\gamma_\ogsconf\cdot \gamma_\ogsconfb]\\
      &=(\threadpool_\ogsconf \parcomp \named \pn (\fn\ \val)\parcomp \threadpool \parcomp \named \qn \pstermb)[\gamma_\ogsconf\cdot \gamma_\ogsconfb]\\
    \end{align*}
    We have by easy induction on $\pstrans{\questP{\fn}{\val,\pn}}$ that $\psterm = \pstermb[(\fn\ \val,\pn)]$.\\
    We show by induction on $\stackT_\ogsconfb$ that : \[\reconscont{(\threadpool_\ogsconf \parcomp \threadpool \parcomp \named \qn \psterm)[\gamma_\ogsconf\cdot \gamma_\ogsconfb]}{\stackT_\ogsconfb} = \reconscont{(\threadpool_\ogsconf \parcomp \named \pn (\fn\ \val) \parcomp \threadpool \parcomp \named \qn \pstermb)[\gamma_\ogsconf\cdot \gamma_\ogsconfb]}{\addleaf{\stackT_{\ogsconfb}}{\pn}{\qn}}\]
    $\stackT_\ogsconfb = (\rn,[\stackT_1,\dots,\stackT_n])$, there are two cases depending on $\rn$ :
    \begin{itemize}
    \item If $\rn=\qn$, then $\addleaf{\stackT_\ogsconfb}{\pn}{\qn} = (\qn,[\stackT_1,\dots,\stackT_2,(\pn,[])])$ therefore    
      \begin{align*}
        \reconscont{(\threadpool_\ogsconf \parcomp \named \pn (\fn\ \val) \parcomp \threadpool \parcomp \named \qn \pstermb)[\gamma_\ogsconf\cdot \gamma_\ogsconfb]}{\addleaf{\stackT_{\ogsconfb}}{\pn}{\qn}}
        &= ((\threadpool_\ogsconf \parcomp \named \pn (\fn\ \val) \parcomp \threadpool \parcomp \named \qn \pstermb)[\gamma_\ogsconf\cdot \gamma_\ogsconfb] (\qn) [\overrightarrow{\reconscont{\threadpool'}{\stackT_i}},\reconscont{\threadpool'}{(\pn,[])}],\qn)\\
        &= \pstermb [\overrightarrow{\reconscont{\threadpool'}{\stackT_i}},((f\ \val),\pn)]\\
        &= \psterm [\overrightarrow{\reconscont{\threadpool'}{\stackT_i}}]\\
        &= (\threadpool_\ogsconf \parcomp \threadpool \parcomp \named \qn \psterm)[\gamma_\ogsconf\cdot \gamma_\ogsconfb](\qn)[\overrightarrow{\reconscont{\threadpool'}{\stackT_i}}]
      \end{align*}
      where $\threadpool' = (\threadpool_\ogsconf \parcomp \named \pn (\fn\ \val) \parcomp \threadpool \parcomp \named \qn \pstermb)[\gamma_\ogsconf\cdot \gamma_\ogsconfb]$, by induction hypothesis we have
      $\reconscont{\threadpool'}{\stackT_i} = \reconscont{(\threadpool_\ogsconf \parcomp \threadpool \parcomp \named \qn \psterm)[\gamma_\ogsconf\cdot \gamma_\ogsconfb]}{\stackT_i}$ therefore we have the equality we wanted
    \item If $\rn\neq\qn$, we have that 
      \[(\threadpool_\ogsconf \parcomp \named \pn (\fn\ \val) \parcomp \threadpool \parcomp \named \qn \pstermb)[\gamma_\ogsconf\cdot \gamma_\ogsconfb](\rn) = (\threadpool_\ogsconf \parcomp \named \pn (\fn\ \val) \parcomp \threadpool \parcomp \named \qn \pstermb)[\gamma_\ogsconf\cdot \gamma_\ogsconfb](\rn) \]
      And we can conclude using the induction hypotheses      
    \end{itemize}
\end{proof}

\begin{lemma}
  \label{lm:par_recons}
  For any pair of composable configurations $(\ogsconf,\ogsconfb)$ and $(\ogsconf',\ogsconfb')$, if $\compose{\ogsconf}{\ogsconfb} \soopred \compose{\ogsconf'}{\ogsconfb'}$ then $\recons{\ogsconf}{\ogsconfb} \opred \recons{\ogsconf'}{\ogsconfb'}$
\end{lemma}
\begin{proof}
   without loss of generality let us assume that $\ogsconf \ogstrans{\oplabel} \ogsconf'$, $\ogsconfb=\ogsconfb'$ and $\supp(\ogsconf') \cap \supp(\ogsconfb) \subseteq \supp(\ogsconf) \cap \supp(\ogsconfb)$. Therefore, there exists $\threadpool,\pn,\psterm,\pstermb$ such that $\threadpool_\ogsconf = \threadpool \parcomp \named \pn \psterm$, $\threadpool_{\ogsconf'} = \threadpool \parcomp \named \pn \pstermb$ and $(\psterm,\store_\ogsconf)\opred (\pstermb,\store_{\ogsconf'})$.

  From $\supp(\ogsconf') \cap \supp(\ogsconfb) \subseteq \supp(\ogsconf) \cap \supp(\ogsconfb)$, we get that 
  $\dom{\store_{\ogsconf'}} \cap \dom{\store_{\ogsconfb}}
   \subseteq \dom{\store_{\ogsconf}}$, so that
  $(\psterm,\store_\ogsconf \cdot \store_{\ogsconfb})
   \opred 
   (\pstermb,\store_{\ogsconf'}\cdot\store_{\ogsconfb})$. 

  From Lemma~\ref{lm:opred-preserve-varsubst},
  $((\threadpool\parcomp \named \pn \psterm \parcomp \threadpool_\ogsconfb)[\gamma_\ogsconf\cdot\gamma_\ogsconfb],\store_\ogsconf\cdot\store_\ogsconfb)
    \opred 
    ((\threadpool\parcomp \named \pn \pstermb \parcomp \threadpool_\ogsconfb)[\gamma_{\ogsconf}\cdot\gamma_\ogsconfb],\store_{\ogsconf'}\cdot\store_\ogsconfb)$,
  which is equal to $((\threadpool\parcomp \named \pn \pstermb \parcomp \threadpool_\ogsconfb)[\gamma_{\ogsconf'}\cdot\gamma_\ogsconfb],\store_{\ogsconf'}\cdot\store_\ogsconfb)$ 
  since $\gamma_{\ogsconf'}=\gamma_\ogsconf$.
  We can conclude by induction on $\stackT_\ogsconfb$ using Lemma~\ref{lm:opred-preserve-threadsubst}.  
\end{proof}
These two lemmas instantiated to initial configuration gives the second implication : if $\compose{\langle \storectx \vdash (\term,\store):\typeb\rangle}{\langle \vdash (\ectx,\gamma,\store'):(\ectxtype{\typeb}{\Unit};\namectx)\rangle}\Downarrow \pn_f$ then $(\ectx[\term\{\gamma\}],\store\cdot\store')\Downarrow\unit$.\\

For the first implication, we need a reciprocal lemma of Lemma~\ref{lm:par_recons}, however when an operational reduction is available for $\recons{\ogsconf}{\ogsconfb}$, some synchronisation steps may be required for $\compose{\ogsconf}{\ogsconfb}$ to perform the same reduction.

\begin{lemma}
  \label{lm:recons_par}
  For any composable configurations $\ogsconf,\ogsconfb$ such that $\compose{\ogsconf}{\ogsconfb}\cancel{\sosynchred}$, if $\recons{\ogsconf}{\ogsconfb} \opred \term$ then there exists two composable configurations $\ogsconf',\ogsconfb'$ such that $\term = \recons{\ogsconf'}{\ogsconfb'}$ and $\compose{\ogsconf}{\ogsconfb} \soopred \compose{\ogsconf'}{\ogsconfb'}$.
\end{lemma}

\begin{proof}
  This proof is done in multiple steps where we prove results similar to the lemma at different levels of the construction:
  \begin{enumerate}
  \item Let $\gamma,\gamma',\store$ and $\threadpool$ such that $(\threadpool;\store;\gamma')\cancel{\itrans{\mu}}$ for any player question $\mu$. If $(\threadpool[\gamma];\store;\gamma')\itrans{\oplabel}(\threadpoolb;\storeb;\gamma')$ then there exists $\threadpool'$ such that $(\threadpool;\store;\gamma')\itrans{\oplabel}(\threadpool';\storeb;\gamma')$ and $\threadpoolb=\threadpool'[\gamma]$. This is straightforward by induction on $\threadpool$ using Lemma~\ref{lm:opred-reflect-varsubst}.
  \item Let $\threadpool,\stackT,\store$ such that for each leaf $\pn$ of $\stackT, \threadpool(\pn)$ is not a value and for $\pn,\qn\in \stackT$ there exists $\ectx$ such that $\suspend \ectx \qn$ appears in $\threadpool(\pn)$ if and only if $\qn$ is a child of $\pn$ in $\stackT$.\\
    $\reconscont{\threadpool}{\stackT}$ is not a value and if $(\reconscont{\threadpool}{\stackT},\store) \opred (\term,\storeb)$ then there exists $\pn\in \stackT,\threadpoolb,\psterm,\pstermb$ such that $\threadpool = \threadpoolb\parcomp \named \pn \psterm$, $(\psterm,\store)\opred (\pstermb,\storeb)$ and $\term = \reconscont{(\threadpoolb\parcomp \named \pn \pstermb)}{\stackT}$
    \begin{itemize}
    \item If $\stackT = (\pn,[])$, then $\threadpool(\pn)=\term$ which is not a value as $\pn$ is a leaf. The result is immediate.
    \item If $\stackT = (\pn,[\stackT_1,\dots,\stackT_n])$, by induction hypotheses we have that $\reconscont{\threadpool}{\stackT_i}$ are not values, therefore if $(\reconscont{\threadpool}{\stackT},\store)\opred (\term,\storeb)$ we use Lemma~\ref{lm:opred:reflect:threadsubst} :
      \begin{itemize}
      \item Either there exists $\psterm$ such that $(\threadpool(\pn),\store)\opred(\psterm,\storeb)$ and $\psterm[\overrightarrow{\reconscont{\threadpool}{\stackT_i}}] =\term$ therefore we get the expected result with $\pn$. 
      \item Or there exists $\termb_i$ such that $(\reconscont{\threadpool}{\stackT_i},\store)\opred(\termb_i,\storeb)$ and $\term = \threadpool(\pn)[\overrightarrow{(\reconscont{\threadpool}{\stackT_j})_{j\neq i}},(\termb_i,\pn_i)]$ where $\pn_i$ is the root of $\stackT_i$. By induction hypothesis, there exists $\qn\in \stackT_i,\threadpoolb,\psterm,\pstermb$ such that $\threadpool = \threadpoolb\parcomp \named \qn \psterm$, $(\psterm,\store)\opred (\pstermb,\storeb)$ and $\termb_i = \reconscont{(\threadpoolb\parcomp \named \qn \pstermb)}{\stackT_i}$. For $j\neq i$, $\qn\notin \stackT_j$ therefore $\reconscont{\threadpool}{\stackT_j}= \reconscont{(\threadpoolb\parcomp \named \qn \pstermb)}{\stackT_j}$ and we have the expected result.
      \end{itemize}
      We now need to show that $\reconscont{\threadpool}{\stackT}$ is not a value. This is straightforward by induction on $\psterm(\pn)$ using that $n>0$ and that the $\pn_i$ such that $\suspend \ectx {\pn_i}$ appears in $\psterm(\pn)$ correspond exactly $\mathtt{snd}\ \reconscont{\threadpool}{\stackT_i}$.
    \end{itemize}
  \end{enumerate}
We can now consider the general case.
    First, we show that if $\compose{\ogsconf}{\ogsconfb}\cancel{\sosynchred}$ then for any player question $\mu$, $\threadpool_\ogsconf \cancel{\itrans{\mu}}$ and $\threadpool_\ogsconfb \cancel{\itrans{\mu}}$ and for any leaf $\pn\neq \pn_f$ of $\stackT_\ogsconfb$, $(\threadpool_\ogsconf\parcomp \threadpool_\ogsconfb)(\pn)$ is not a value.\\
    Let us assume that $\threadpool_\ogsconf \itrans{\questP{\fn}{\aval,\pn}_{\qn}}$, therefore $\fn \in \freen(\threadpool_\ogsconf(\qn))$, by composability $\threadpool_\ogsconf$ is validated by $\viewG_\ogsconfb$, therefore $\fn \in \viewG_\ogsconfb(\qn)$, hence we have that $\ogsconfb \ogstrans{\questO{\fn}{\aval,\pn}_\qn}$ and also $\ogsconf \ogstrans{\questP{\fn}{\aval,\pn}_\qn}$, thus $\compose{\ogsconf}{\ogsconfb} \sosynchred$ which contradicts our hypothesis. The case for $\threadpool_\ogsconfb \itrans{\questP{\fn}{\aval,\pn}_\qn}$ is similar.\\
    Let $\pn\neq\pn_f$ a leaf of $\stackT_\ogsconfb$ therefore it is also a leaf of $\stackT_\ogsconf$, without loss of generality we assume $\named \pn \psterm \in \threadpool_\ogsconf$, if $\psterm$ is a value then $\ogsconf \ogstrans{\ansP{\pn}{\aval}}$ and $\ogsconfb \ogstrans{\ansO{\pn}{\aval}}$ thus $\compose{\ogsconf}{\ogsconfb} \sosynchred$ which contradicts our hypothesis.
    \begin{itemize}
    \item If $\pn_f$ is not a leaf of $\stackT_\ogsconfb$, we have that for each leaf $\pn$ of $\stackT_\ogsconfb$, $(\threadpool_\ogsconf \parcomp \threadpool_\ogsconfb)(\pn)$ is not a value, we can then combine steps 1 and 2 to build $\ogsconf'$ and $\ogsconfb'$ after the reduction (only the interactive part of the configuration is affected by an operational reduction).
    \item Otherwise, we have $\stackT_\ogsconfb = (\pn_f,[])$ therefore there exists $\term$ such that $\threadpool_\ogsconfb = \named \pn_f \term$, $\threadpool_\ogsconf = \mathbf{0}$ and $\recons{\ogsconf}{\ogsconfb} = \term$, the expected result is immediate.
    \end{itemize}
\end{proof}
The next lemma means that whenever both a synchronisation and an operational reduction are possible we can start with the synchronisation.

\begin{lemma}
  \label{lm:synchro-eager}
  For any pairs of composable configurations $(\ogsconf,\ogsconfb),(\ogsconf_1,\ogsconfb_1)$ and $(\ogsconf_2,\ogsconfb_2)$ such that $\compose{\ogsconf}{\ogsconfb}\sosynchred \compose{\ogsconf_1}{\ogsconf_1}$ and $\compose{\ogsconf}{\ogsconfb}\soopred \compose{\ogsconf_2}{\ogsconf_2}$. Then there exists $(\ogsconf_3,\ogsconfb_3)$ such that $\compose{\ogsconf_1}{\ogsconf_1} \soopred \compose{\ogsconf_3}{\ogsconf_3}$ and $\compose{\ogsconf_2}{\ogsconf_2} \sosynchred \compose{\ogsconf_3}{\ogsconf_3}$
\end{lemma}
\begin{proof}
  The operational reduction is not affected by and does not change the type-context configuration and the visible configuration, therefore we only consider the interactive configuration in this proof.\\
  We start by showing that for a term $\term$ and a store $\store$, we cannot have both $(\term,\store)\opred(\termb,\storeb)$ and $\term = \ectx[\fn\ \val]$, the first implication is done by induction on $\term$ and the second one is a case analysis on $\opred$.\\
  Next, without loss of generality $\compose{\ogsconf}{\ogsconfb}\soopred \compose{\ogsconf_2}{\ogsconf_2}$ gives us that there exists $\pn$ such that $(\threadpool_\ogsconf(\pn),\store_\ogsconf)\opred (\psterm,\storeb)$ and $\compose{\ogsconf}{\ogsconfb}\sosynchred \compose{\ogsconf_1}{\ogsconf_1}$ gives us that there exists $\mu$ such that $\langle \threadpool_\ogsconf;\store_\ogsconf;\gamma_\ogsconf\rangle \itrans{\mu} \langle \threadpool;\store_\ogsconf;\gamma\rangle$ and $\langle \threadpool_\ogsconfb;\store_\ogsconfb;\gamma_\ogsconfb\rangle \itrans{\switchio{\mu}} \langle \threadpool';\store_\ogsconfb;\gamma'\rangle$. By cases on $\mu$ :
  \begin{itemize}
  \item If $\mu = \questO{\fn}{\aval,\qn_1}_{\qn_2}$ or $\mu = \ansP{\qn_1}{\aval}$, we know that $\qn_1\neq \pn$, in the first case because $\qn_1$ is fresh and in the second because $\threadpool_\ogsconf(\qn_1)$ is a value. Therefore, the two reductions do not interact with each other and can be done in any order.
  \item If $\mu = \questP{\fn}{\aval,\qn_2}_{\qn_1}$ or $\mu = \ansO{\qn_1}{\aval}$, if $\qn_1\neq \pn$ we have the same argument as in the previous case, otherwise we proceed by induction on $\threadpool_\ogsconf(\pn)$. In the base case we use the result on terms from the beginning of the proof and in the other cases either the reductions appear in different branches and can happen in either order, or we can apply the induction hypothesis.
  \end{itemize}
\end{proof}

We can then "saturate" the execution with all the synchronisations possible. However, we need to only have a finite number of possible synchronisations between two operational reduction. For this we define a notion of accessible names, these are the names on which a synchronisation may happen without requiring any operational reduction.
\begin{definition}
  We consider partially suspended evaluation contexts of the form :
  \[\psectx \Coloneqq \hole \mid \app \val \psectx \mid \parop \psterm \ectx \mid \parop \psectx \psterm \]
  Let $\ogsconfP,\ogsconfO$ two OGS configurations
  \begin{itemize}
  \item A function name $\fn$ is accessible if there exists $\psectx,\gn,\val$ such that $\psectx[\app \gn \val]\in \threadpoolP\parcomp \threadpoolO$ and $\fn \leq_{\gammaP,\gammaO}\gn$
  \gj{I think $\leq_{\gammaP,\gammaO}$ has not yet been introduced};
  \item A continuation name $\pn$ is accessible if there exists $\qn$ a descendant of $\pn$ in $\stackTO$ or $\qn=\pn$ such that $(\threadpoolP\parcomp \threadpoolO)(\qn)$ is a value.
  \end{itemize}
  We write $\acc{\ogsconfP}{\ogsconfO}$ the set of accessible names in $\ogsconfP$ and $\ogsconfO$
\end{definition}

\begin{lemma}
  \begin{itemize}
  \item If $\compose{\ogsconfP}{\ogsconfO}\sosynchred\compose{\ogsconfP'}{\ogsconfO'}$ then $\acc{\ogsconfP'}{\ogsconfO'}\subseteq \acc{\ogsconfP}{\ogsconfO}$
  \gj{What about fresh names introduced by the synchronization step ?}.
  \item If $\compose{\ogsconfP}{\ogsconfO}\sosynchred$, then the subject of the reduction is in $\acc{\ogsconfP}{\ogsconfO}$
  \gj{a formal definition of the subject of the reduction would be useful}.
  \end{itemize}
\end{lemma}

\begin{proof}
  The proof is done by case analysis on the synchronisation and in the case of a question by induction on $\threadpool_{\ogsconfP}(\pn)$ if $\mu = \questP{\fn}{\aval,\qn}_\pn$ or on $\threadpool_{\ogsconfO}(\pn)$ if $\switchio{\mu} = \questP{\fn}{\aval,\qn}_\pn$.
\end{proof}
We can now define a measure on the configurations
\begin{definition}
  Let $\ogsconfP,\ogsconfO$ two configurations, $S$ a set of names such that $S\supseteq \acc{\ogsconfP}{\ogsconfO}$.\\
  Let $n=|S|$, and $L$ the $n$-uplet of names of $S$ sorted with $>_{\gammaP,\gammaO}$ and $\stackTO$.\\
  The measure $\accmeas{S}(\ogsconfP,\ogsconfO)$ is the $n$-uplet where the $i$-th element of $\accmeas{S}(\ogsconfP,\ogsconfO)$ is number of free occurences of $L_i$ in $\threadpoolP$ and $\threadpoolO$. We equip the measure with the lexicographical order.
\end{definition}

\gj{An example of the measure on a given pair of configurations
would be quite useful.}

\begin{lemma}
If $\compose{\ogsconfP}{\ogsconfO}\sosynchred\compose{\ogsconfP'}{\ogsconfO'}$ then for any $S\supseteq\acc{\ogsconfP}{\ogsconfO}$,  $\accmeas{S}(\ogsconfP,\ogsconfO) > \accmeas{S}(\ogsconfP,\ogsconfO)$
\end{lemma}

\begin{proof}
  The proof is done by case analysis on the synchronisation and in the case of a question by induction on $\threadpool_{\ogsconfP}(\pn)$ if $\mu = \questP{\fn}{\aval,\qn}_\pn$ or on $\threadpool_{\ogsconfO}(\pn)$ if $\switchio{\mu} = \questP{\fn}{\aval,\qn}_\pn$. 
\end{proof}

\begin{lemma}
  \label{lm:saturate_sync}
  Let $\ogsconf,\ogsconfb$ two composable configurations, there exists $\ogsconf',\ogsconfb'$ such that $\compose{\ogsconf}{\ogsconfb} {\sosynchred}^* \compose{\ogsconf'}{\ogsconfb'}$ and $\compose{\ogsconf'}{\ogsconfb'}\cancel{\sosynchred}$
\end{lemma}
\begin{proof}
  Let $S=\acc{\ogsconf}{\ogsconfb}$, then we use the well-foundedness of the lexicographical order on $n$-uplets and that any synchronisation decreases $\accmeas{S}$.
  Hence, we have that between two operational reduction, only a finite number of synchronisations are possible.
\end{proof}

We can now show Theorem~\ref{thm:obs-adequacy}

\begin{proof}
  \begin{itemize}
  \item Assume there exists $\term_1,\dots,\term_n$ such that $\recons{\ogsconfP}{\ogsconfO} \opred \term_1\opred\dots\opred\term_n \opred \unit$, we show by induction on $n$ that there exists ${\ogsconfP}_1,\dots,{\ogsconfP}_n,{\ogsconfP}_f$ and ${\ogsconfO}_1,\dots,{\ogsconfO}_n,{\ogsconfO}_f$ such that $\compose{\ogsconfP}{\ogsconfO}{\sosynchred}^*\soopred \compose{{\ogsconfP}_1}{{\ogsconfO}_1}{\sosynchred}^*\soopred \dots {\sosynchred}^*\soopred \compose{{\ogsconfP}_n}{{\ogsconfO}_n}{\sosynchred}^*\compose{{\ogsconfP}_f}{{\ogsconfO}_f}$ and ${\ogsconfO}_f\ogstrans{\ansP{\pn_f}{\unit}}$
    \begin{itemize}
    \item If $n=0$, by lemma~\ref{lm:saturate_sync} there exists $\ogsconfP',\ogsconfO'$ such that $\compose{\ogsconfP}{\ogsconfO} {\sosynchred}^* \compose{\ogsconfP'}{\ogsconfO'}$ and
      $\compose{\ogsconfP'}{\ogsconfO'}\cancel{\sosynchred}$, by lemma~\ref{lm:sync_eq} $\recons{\ogsconfP'}{\ogsconfO'}=\recons{\ogsconfP}{\ogsconfO} = \unit$, using lemma~\ref{lm:par_recons} $\compose{\ogsconfP'}{\ogsconfO'}\cancel{\soopred}$, the only possibility for this is $\threadpool_{\ogsconfO'}= \named {\pn_f} ()$, $\threadpool_{\ogsconfP'}=\mathbf{0}$ and $\stackT_{\ogsconfO'} = (\pn_f,[])$ meaning that $\ogsconfO'\ogstrans{\ansP{\pn_f}{\unit}}$
    \item For $n+1$, by lemma~\ref{lm:saturate_sync} there exists $\ogsconfP',\ogsconfO'$ such that $\compose{\ogsconfP}{\ogsconfO} {\sosynchred}^* \compose{\ogsconfP'}{\ogsconfO'}$ and $\compose{\ogsconfP'}{\ogsconfO'}\cancel{\sosynchred}$, by lemma~\ref{lm:sync_eq} $\recons{\ogsconfP'}{\ogsconfO'}=\recons{\ogsconfP}{\ogsconfO}$. Therefore, $\recons{\ogsconfP'}{\ogsconfO'} \opred \term_1$, by lemma~\ref{lm:recons_par} there exists ${\ogsconfP}_1,{\ogsconfO}_1$ such that $\term_1= \recons{{\ogsconfP}_1}{{\ogsconfO}_1}$ and $\compose{\ogsconfP'}{\ogsconfO'}\soopred \compose{{\ogsconfP}_1}{{\ogsconfO}_1}$, and we conclude with the induction hypothesis on ${\ogsconfP}_1,{\ogsconfO}_1$ and $\term_2,\dots,\term_{n+1}$
    \end{itemize}
  \item Assume there exists ${\ogsconfP}_1,\dots,{\ogsconfP}_n,{\ogsconfP}_f,{\ogsconfO}_1,\dots,{\ogsconfO}_n,{\ogsconfO}_f$ such that $\compose{\ogsconfP}{\ogsconfO} {\sosynchred}^*\soopred \compose{{\ogsconfP}_1}{{\ogsconfO}_1}{\sosynchred}^*\soopred\dots {\sosynchred}^*\soopred \compose{{\ogsconfP}_n}{{\ogsconfO}_n}{\sosynchred}^*\compose{{\ogsconfP}_f}{{\ogsconfO}_f}$ and ${\ogsconfO}_f \ogstrans{\ansP{\pn_f}{\unit}}$, by a straightforward induction on $n$ using lemma~\ref{lm:sync_eq} and lemma~\ref{lm:par_recons} that $\recons{\ogsconfP}{\ogsconfO} \opred \recons{{\ogsconfP}_1}{{\ogsconfO}_1}\opred\dots \opred \recons{{\ogsconfP}_n}{{\ogsconfO}_n}=\recons{{\ogsconfP}_f}{{\ogsconfO}_f}$. And using the composability of ${\ogsconfP}_f$ and ${\ogsconfO}_f$, we can show that $\recons{{\ogsconfP}_f}{{\ogsconfO}_f} = \named {\pn_f} \unit$.
  \end{itemize}
\end{proof}

From it, we deduce the proof of Corollary~\ref{cor:ciu_par}.

\begin{proof}
Let $\pectx,\term,\gamma,\store,\store',\storectx,\typeb,\namectx$ and $\ogsconfP$ such that $\langle \storectx \vdash (\term,\store):\typeb\rangle \ogstrans{\movePI{\overrightarrow{\aval_i},\pn}} \ogsconfP$ and $\ogsconfO$ such that $\langle \vdash (\pectx,\gamma,\store'):(\ectxtype{\typeb}{\Unit};\namectx)\rangle \ogstrans{\moveOI{\overrightarrow{\aval_i},\pn}} \ogsconfO$. We have that $\ogsconfP$ and $\ogsconfO$ are composable and $\recons{\ogsconfP}{\ogsconfO} = (\ectx[\term\{\gamma\}],\store\cdot\store')$. Therefore, we can show $\recons{\ogsconfP}{\ogsconfO} \Downarrow\unit$ if and only if $\compose{\ogsconfP}{\ogsconfO}\Downarrow \pn_f$.
\end{proof}

}{}
\iftoggle{fullappendix}{
\section{Completeness}
\label{sec:app_completeness}

\ds{see if the lemma below has to be added} 

\begin{lemma}
  \label{lm:mk_trace}
  For all $\term,\store,\typeb$, given a complete trace $\tr \in \ogsCTr{\langle \namectx \vdash (\term,\store):\typeb\rangle}$,  we have $\overline{\tr\cdot \pn_f}\in \ogsCTr{\langle \vdash (\ectx_{\overline{\tr\cdot \pn_f}},\delta_{\overline{\tr\cdot \pn_f}},\store_{\overline{\tr\cdot \pn_f}}):(\typeb\to\Unit;\namectx)\rangle}$
\end{lemma}
 \begin{proof}
   The proof is split into three parts,  for each component of the configuration:
   \begin{itemize}
   \item For the interactive configuration, we build interactive configurations for each step of the trace and show that they can make the next step.
   \item For the typing configuration, we show by induction on traces that if $\ctuple{\namectxO\mid \namectxP}\tTrans{\tr}$ then $\ctuple{\namectxP\mid \namectxO} \tTrans{\overline{\tr}}$.
   \item For the visible configuration we need to show that $\overline{\tr\cdot \pn_f}$ is visible. To prove this we build a sequence of visible configurations that will validate each step of the reduction $\ogsCTr{\langle \namectx \vdash (\term,\store):\typeb\rangle} \tTrans{\tr}$.
   \end{itemize}
 \end{proof}

\subsection{Preliminary notions}
%\label{sec:app_completeness_prel}

The goal of this section is to prove theorem~\ref{thm:completeness}

In order to reason on traces made by configurations we need to be more precise on the threads where the moves are made. To do so, we introduce the notion of labeling:

\begin{definition}[Labelling]
  A labeling $\lab$ is a partial function from questions to continuation names.\\
  We write a labelled move, $\move_\lab = \move_{\lab(\move)}$ if $\move$ is a question and $\move_\lab = \move$ if $\move$ is an answer.\\
  We write a labelled trace, $(\move_0,\cdots,\move_n)_\lab = (\move_{0,\lab}\cdots \move_{n,\lab})$ \\
  Two labelings $\lab_1,\lab_2$ are Proponent-equivalent, written $\lab_1\eqPl \lab_2$ if $\dom{\lab_1}=\dom{\lab_2}$ and for all player questions $\pmove\in \dom{\lab_1}$, we have $\lab_1(\pmove)=\lab_2(\pmove)$.\\
  Similarly, two labelings $\lab_1,\lab_2$ are Opponent-equivalent, written $\lab_1\eqOp \lab_2$ if $\dom{\lab_1}=\dom{\lab_2}$ and for all opponent questions $\pmove\in \dom{\lab_1}$, we have $\lab_1(\pmove)=\lab_2(\pmove)$
\end{definition}
With label lings, the exact definition of visible traces is the following :

\begin{definition}[Visible traces]
  Let $\vconf$ a visible configuration and $\tr$ a trace, we say that $\tr$ is visible from $\vconf$ if $\vconf \vTrans{\tr}$\\
  Let $\vconf$ a visible configuration, $\tr$ a trace and $\lab$ a labeling, we say that $\tr_\lab$ is visible from $\vconf$ if $\vconf \vTrans{\tr_\lab}$\\
  Let $\tr$ a trace starting with an initial move $\tr_0$, if $\tr_{\geq 1}$ is visible from the visible configuration after $\tr_0$, we say that $\tr$ is visible.
\end{definition}
% \begin{lemma}
%   \label{lm:CTr_lttr}
%   If $\tr \in \ogsCTr{\ogsconf}$ then for all $\tr'\lttr\tr$ such that $\tr'$ is visible from $\vconf_\ogsconf$, $\tr'\in \ogsCTr{\ogsconf}$
% \end{lemma}
We give a more precise sketch of the proof of lemma~\ref{lm:CTr_lttr}
\begin{proof}
  To prove this lemma we use a series of smaller lemmata that explicit the label lings used.\\
  We define the proposition $P(\tr,\lab) := \forall \questP{\fn}{\aval,\pn}\in \tr, \lab(\questP{\fn}{\aval,\pn}) = \qn \Rightarrow$ the answer on $\qn$ can only happen after the answer on $\pn$ in $\tr$.\\
  \begin{enumerate}
  \item If $\tr_{\lab_1}$ is visible from $\vconf$ and $P(\tr,\lab_2)$, then $\tr_{\lab_3}$ is visible from $\vconf$ where $\lab_3 := \pmove \mapsto \lab_2(\pmove); \omove\mapsto \lab_1(\omove)$
  \item If $\tr'\lttr \tr$ and $\tr_\lab$ is visible from $\vconf$ then $P(\tr',\lab)$
  \item Let $\iconf$ an interactive configuration, if $\tr_\lab \in \LtsTr{\Intera}{\iconf}$ and $\lab'\eqPl \lab$ then $\tr_{\lab'}\in \LtsTr{\Intera}{\iconf}$
  \item If $\tr_\lab\in \LtsTr{\Intera}{\iconf}$ and $\tr'\lttr \tr$ then $\tr'_\lab\in\LtsTr{\Intera}{\iconf}$
  \item If $\tr \in \LtsTr{\Ty}{\tconf}$ then for any labeling $\lab$, $\tr_\lab\in \LtsTr{\Ty}{\tconf}$
  \item If $\tr\in \LtsTr{\Ty}{\tconf}$ and $\tr'\lttr \tr$ then $\tr'\in\LtsTr{\Ty}{\tconf}$
  % \item Let $\vconf$ a visible configuration,$\tr,\tr'$ two traces such that $\tr'\lttr \tr$, and $\lab$ a labelling. If $\tr_\lab$ is visible from $\vconf$ and $\tr'$ is visible from $\vconf$, then there exists $\lab'$ such that $\lab'\eqPl \lab$ and $\tr'_{\lab'}$ is visible from $\vconf$.
  \end{enumerate}
  \begin{enumerate}
  \item This proof is done by induction on $\tr$. We just need to notice that in the visible LTS, the labels of player move only affect the continuation tree, so as long as the labeling respects proposition $P$, we can change the labels of player moves
  \item This proof is done by induction on $\tr'\lttr \tr$ and by noticing that we already have $P(\tr,\lab)$ and that $\lttr$ will only exchange player and opponent moves to put the opponent move before the player move, not the other way around, and therefore $P$ is invariant.
  \item This proof is done by a straightforward induction on $\tr$
  \item This proof is by induction on $\tr'\lttr \tr$ and then (for the non trivial cases), if the two actions occur on the same thread $\pn$, by induction on $\iconf(\pn)$
  \item This is a straightforward induction on $\tr$ as the typing configurations do not depend on the labeling.
  \item This is a straightforward induction on $\tr'\lttr \tr$.
  \end{enumerate}
  Let $\tr\in \ogsTr{\ogsconf}$, $\tr'\lttr\tr$ such that $\tr'$ is visible from $\vconf_\ogsconf$. There exists a labeling $\lab$ such that $\tr_{\lab}\in \ogsTr{\ogsconf}$, hence $\tr_{\lab}$ is visible from $\vconf_\ogsconf$. Using 2, we have that $P(\tr',\lab)$. $\tr'$ is visible from $\vconf_\ogsconf$ therefore there exists $\lab'$ such that $\tr'_{\lab'}$ is visible from $\vconf_\ogsconf$, therefore, using 1, we have that there exists $\lab''$ such that $\tr'_{\lab''}$ is visible from $\vconf_\ogsconf$ and $\lab \eqPl \lab''$. Furthermore, using 3 and 4, we have that $\tr'_{\lab''} \in \LtsTr{\Intera}{\iconf_\ogsconf}$.
  Last, using 5 and 6, we have that $\tr'_{\lab''} \in \LtsTr{\Ty}{\tconf_\ogsconf}$.
  Hence, combining the different results we get that $\tr'_{\lab''}\in \ogsTr{\ogsconf}$, thus $\tr'\in \ogsTr{\ogsconf}$.\\
  It is easy to see that the size of the continuation tree does not change when reordering the different moves, therefore if $\tr\in \ogsCTr{\ogsconf}$ and $\tr'\lttr \tr$ where $\tr'$ is visible from $\vconf_\ogsconf$ then $\tr'\in \ogsCTr{\ogsconf}$
\end{proof}

As we are only considering visible traces, we need to ensure that all the traces generated by some term is a visible trace from the point of view of the context.

\begin{lemma}
  \label{lm:dual_vis}
  For all $\term,\store,\typeb$ and complete trace $\tr \in \ogsCTr{\langle \namectx \vdash (\term,\store):\typeb\rangle}$. $\overline{\tr\cdot\pn_f}$ is a complete visible trace.
\end{lemma}
\begin{proof}
  The proof is done by building a sequence of visible configurations such that at each step of the trace we have $\freen(\threadpool(\pn)) \subseteq \viewG(\pn)$ for any $\pn\in\dom{\viewG}$, $\freen(\gamma(\fn))\subseteq \viewF(\fn)$ for any $\fn\in \dom{\viewF}$ and the trees coincide apart from the root $\pn_f$. This is first done for a trace $\tr \in \ogsTr{\langle \namectx \vdash (\term,\store):\typeb\rangle}$, by induction on $\tr$. And then we show for complete traces that the final answer $\ansP{\pn_f}{()}$ is available.
\end{proof}

A technical lemma to manipulate the permutations of traces :

\begin{lemma}
  \label{lm:lttr_dual}
  Let $\tr,\tr'$ be two traces, if $\tr \lttr \tr'$ then $\overline{\tr'} \lttr \overline{\tr}$
\end{lemma}

\subsection{Construction}

We now have all the components to build the context corresponding to a complete visible trace and show lemma~\ref{lm:ex_context}.
In this subsection, we fix a complete trace $\tr$ ending with the player answer $\moveP{\pn_f}{()}$ with a labeling $\lab$.

We introduce a structure that characterizes the order $\lttr$. This will allow us to define precisely which moves need to occur before others.

\begin{definition}
  A \textbf{dependency structure} is a triple $\depst =(\depstsup,\depstjust,\depstdep)$ where $\depstsup$ is a finite set, $\depstjust$ and $\depstdep$ are partial orders on $\depstsup$ such that $\forall \move_1 \depstdep \move_2, \forall \move_2 \depstjust \move_3, \move_1\depstdep \move_3$
\end{definition}

For any trace $\tr$ we define a dependency structure $\depsttr$ :
\begin{itemize}
\item $\depstsuptr \defeq \{\move\in\tr\}$
% \item $\depstjusttr \defeq (\justif{}{}\cup \{(\pmove,\ansmove) | \exists \tr_1,\tr_2,\tr_3, \text{ such that }\tr = \tr_1\cdot \pmove \cdot \tr_2\cdot \ansmove \cdot \tr_3\text{ and }\close{\ansmove}{\pmove}{\tr_1}\})^*$ where $-^*$ is the reflexive and transitive closure
\item $\depstjusttr \defeq \justifstar{}{}$
\item $\depstdeptr \defeq \{ (\omove,\move)| \exists \pmove,\tr_1,\tr_2,\tr_3, \text{ such that }\tr = \tr_1\cdot\omove\cdot\tr_2\cdot\pmove\cdot\tr_3 \text{ and }\pmove\depstjusttr \move\}$
\end{itemize}

\begin{definition}
  For a dependency structure $\depsttr$ we say that a trace $\tr'$ is a trace of $\depsttr$ if for $\move,\move'\in \tr'$:
  \begin{itemize}
  \item if $\move \depstjusttr \move'$ then $\move$ appears before $\move'$ in $\tr'$
  \item if $\move \depstdeptr \move'$ then $\move$ appears before $\move'$ in $\tr'$
  \end{itemize}
  We write $\depstTr{\depsttr}$ the set of traces of $\depsttr$.\\
  A complete trace of $\depsttr$ is a trace $\tr'$ such that $\depstsuptr = \{ \move\in \tr'\}$, we write $\depstCTr{\depsttr}$ the set of complete traces of $\depsttr$.\\
\end{definition}

\begin{lemma}
  \label{lm:depsttr_lttr}
  $\depstCTr{\depsttr}=\{\tr'| \tr'\lttr \tr\}$
\end{lemma}

Hence, all the contexts have to verify is that the actions that occur and the order in which they happen respects the dependency structure.\\
We write $\Parn{i\in I}{\term_i}$ with $I$ finite, the n-ary parallel composition, if $I=\emptyset$ we define $\Parn{i\in I}{\term_i}\defeq ()$.\\
We define a non-deterministic choice between a finite set of terms $\Select{\{\term_i|1\leq i \leq n\}}$ :
\begin{align*}
  \Select{\{\term_i|1\leq i\leq n\}} &\defeq \newref {\var} 0 \letin \_ {\Parn{1\leq i\leq n}{\writeop{\var}{i}}} \Run_n(\var,\term_1,\dots,\term_n) \\
  \Run_0(\var,\term_1,\dots,\term_n) & \defeq \unit\\
  \Run_{i+1}(\var,\term_1,\dots,\term_n) &\defeq \ifte{\readop{\var}=i+1}{\term_{i+1}}{\termb_i}
\end{align*}
%Where $\loc\notin \bigcup \{\freen(\term_i)|1\leq i\leq n\}$. 
The order of the terms does not impact the choice or the possible reductions, therefore we write $\Select{\{\term_i|i\in I\}}$ for an arbitrary order on $I$.
%   % \begin{lemma}
%   %   Let $(\psterm_i)_{i\in I}, j\in I$ such that $\forall i\in I\setminus\{j\}, \psterm_i=\unit$ and $\storectx,namectx,\typeb$ such that $\storectx,\namectx \vdash \psterm_j : \typeb$. For all $\storectx',\gamma,\ectx,\store$ such that :
%   %   \begin{itemize}
%   %   \item $\storectx'\disjoint\storectx$;
%   %   \item $\storectx\cdot \storectx';\emptyset\vdash \gamma:\namectx$;
%   %   \item $\storectx \cdot \storectx';\emptyset \vdash \ectx:\ectxtype{\typeb}{\Unit}$;
%   %   \item $\vdash \store:\storectx\cdot \storectx'$;
%   %   \end{itemize}    
%   % \end{lemma}
  
We define a translation $\llbracket\cdot\rrbracket_{\depsttr,\lab}$ of moves $\move\in \depstsuptr$ and a value assignment $\gamma$ of player names by mutual induction :

\begin{itemize}
\item For each opponent move $\omove$ we set a reference $\loc_\omove$ to $0$, this means that it did not occur yet. The initial store is $\store_\tr= \{[\loc_\omove \mapsto 0] | \omove\in \depstsuptr\}$
\item Before each player move we will need to check that the opponent moves that are supposed to occur beforehand have indeed occurred. For this we define a term $\mathtt{check\_lock}_\pmove$ that tests their locks and diverge if one of them did not occur :
  \[\mathtt{check\_lock}_\pmove \defeq \Parn{\loc \in \{\loc_\omove| \omove \depstdeptr \pmove\}}{\ifte{!\loc}{()}{\Omega}}\]
\item When multiple opponent questions have the same subject, we will need to match on the abstract values received to know which behaviour we should have, for this we match on the structure of the abstract value and chose non-deterministically if the abstract values cannot be distinguished by their structure (they can each be renamed in the other), we write $\aval \equiv\avalb$ when two abstract values are undistinguishable, for example :
  \[
    \begin{array}[h]{ll}
      \mathtt{match}\ \var \ \mathtt{with}&\\
      |\, \aval_1 \to& \Select{\{\llbracket \moveO{\fn}{\aval,\pn} \rrbracket_{\depsttr}\ \aval_1 |\aval\equiv \aval_1\}}\\
      |\, \_ \to & \Omega
    \end{array}\]
\end{itemize}

\begin{align*}
  \llbracket \moveO{\fn}{\aval,\pn} \rrbracket_{\depsttr,\lab} \defeq& \lambda \var.
                                                                  \begin{array}[t]{l}
                                                                   \mathtt{match}\ \var\ \mathtt{with}\\
                                                                   |\,\aval \to \ifte{\cas{\loc_{\moveO{\fn}{\aval,\pn}}}{0}{1}}{\letin \_ {\Parn{\{\pmove| \lab(\pmove)=\pn\}}{\llbracket \pmove\rrbracket_{\depsttr,\lab}}}{\llbracket \moveP{\pn}{\val}\rrbracket_{\depsttr}}}{\Omega} \\
                                                                   |\, \_\to \Omega\\
                                                                 \end{array}
  \\
    \llbracket \moveO{\pn}{\val} \rrbracket_{\depsttr,\lab} \defeq& \lambda \var.\ifte{x=V}{\ifte{\cas{\loc_{\moveO{\pn}{\aval}}}{0}{1}}{\unit}{\Omega}}{\Omega}\\
  \llbracket \moveP{\fn}{\aval,\pn}\rrbracket_{\depsttr,\lab}\defeq & \mathtt{check\_lock}_{\moveP{\fn}{\aval,\pn}}; \llbracket \moveO{\pn}{\val}\rrbracket_{\depsttr} \ (\fn\ \aval\{\gamma\})\\
  \llbracket \moveP{\pn}{\val}\rrbracket_{\depsttr,\lab}\defeq& \mathtt{check\_lock}_{\moveP{\pn}{\val}};\val\\
  \gamma(\fn) \defeq& \begin{array}[t]{l}
    \lambda \var.\mathtt{match}\ \var \ \mathtt{with}\\
    |\, \aval_1 \to  \Select{\{\llbracket \moveO{\fn}{\aval,\pn} \rrbracket_{\depsttr}\ \aval_1 |\aval\equiv \aval_1\}}\\
    \vdots \\
    |\, \aval_n \to \Select{\{\llbracket \moveO{\fn}{\aval,\pn} \rrbracket_{\depsttr}\ \aval_n |\aval\equiv \aval_n\}}\\
    |\, \_ \to \Omega
  \end{array}
\end{align*}

In the definition of $\gamma(\fn)$ the $(\aval_i)$ are chosen pairwise distinct such that $\{\aval_i|1\leq i\leq n\} = \{\aval | \exists \pn, \moveO{\fn}{\aval,\pn} \in \depstsuptr\}/_\equiv$. The order in which they appear does not matter as any $x$ can match with at most one $\aval_i$.\\
In these definitions $\val$ is the unique abstract value such that $\moveO{\pn}{\val}\in \depstsuptr$ or $\moveP{\pn}{\val}\in \depstsuptr$.

\begin{itemize}
\item $\ectx_{\tr,\lab} \defeq (\lambda \var. \llbracket \ansP{\pn_f}{\unit}\rrbracket_{\depsttr}) \ \hole$
\item $\delta_{\tr,\lab} \defeq \gamma_{\big| \{ \fn | \exists \aval,\pn,\imove\in \depstsuptr, \justif{\imove}{\moveO{\fn}{\aval,\pn}}\in \depstsuptr\}}$ where $\imove$ is an initial move
\end{itemize}

We show that this context can only generate complete traces in $\depstCTr{\depsttr}$, this lemma is under the assumption that it is typable. We will show it's typability later as it requires more assumptions on $\tr$.

\begin{lemma}
  \label{lm:tr_dep}
  $\ogsCTr{\langle \vdash (\ectx_{\tr,\lab},\delta_{\tr,\lab},\store_{\tr,\lab}):(\typeb\to\Unit;\namectx)\rangle}\subseteq \depstCTr{\depsttr}$
\end{lemma}
\begin{proof}
We write $[\loc \mapsto x]\term \in \ogsconf$ if there exists $\loc'$ such that $\term[\loc'/\loc]$ appears in the thread pool of $\iconf$, $\loc'$ does not appear outside of $\term[\loc'/\loc]$ in $\ogsconf$ and $\loc'$ is set to $x$ in the store of $\ogsconf$.\\
We write $\aval\in \gamma(\fn)$ if $\aval$ is one of the abstract values used to match in the definition of $\gamma(\fn)$
For any function name $fn$ and abstract value $\aval$ we define :
\begin{align*}
  \mathbb{P}(\fn,\aval) :=& \{ \gamma(\fn)\ \aval \} \cup \{\Omega\}\\
  \cup& \{ \newref \var 0 \letin \_ {\Parn{1\leq i\leq n}{\writeop{\var}{i}}} \Run_n (\var,\term_1\ \aval_j,\dots,\term_n\ \aval_j) | \{\term_i\}=Eq(\fn,\aval_j) \}\\
  \cup & \{ [\loc \mapsto 0] \letin \_ {\Parn{1\leq i\leq n}{\writeop{\loc}{i}}} \Run_n (\loc,\term_1\ \aval_j,\dots,\term_n\ \aval_j) | \{\term_i\}=Eq(\fn,\aval_j) \}\\
  \cup & \{ [\loc \mapsto k] \letin \_ {\Parn{1\leq i\leq n}{\termb_i}} \Run_n (\loc,\term_1\ \aval_j,\dots,\term_n\ \aval_j) | \{\term_i\}=Eq(\fn,\aval_j), \\
                          & \hspace{2em} I\subseteq \llbracket 1;n\rrbracket, (\forall i\in I, \termb_i = ()) \wedge (\forall i\notin I,\termb_i = \writeop{\loc}{i}) \wedge k\in I \}\\
  \cup& \{ [\loc \mapsto k] \Run_m (\loc,\term_1\ \aval_j,\dots,\term_n\ \aval_j) | \{\term_i\}=Eq(\fn,\aval_j), m\geq k\}\\
  \cup & \{ \llbracket \omove \rrbracket_{\depsttr} \ \aval_j | \omove \in Eq(\fn,\aval)\} \cup \{ \Omega \} \\
  \cup& \{\ifte{\cas{\loc_{\moveO{\fn}{\aval',\pn}}}{0}{1}}{\letin \_ {\Parn{\{\pmove| \lab(\pmove)=\pn\}}{\llbracket \pmove\rrbracket_{\depsttr}}}{\llbracket \moveP{\pn}{\val}\rrbracket_{\depsttr}}}{\Omega} | \moveO{\fn}{\aval',\pn}\in Eq(\fn,\aval)\} \\
  Eq(\fn,\aval) := &\{\moveO{\fn}{\aval',\pn}\in\depstsuptr |\aval'\equiv \aval\}
\end{align*}
  We show by induction on the size of $\tr'$ such that $\langle \vdash (\ectx_\tr,\delta_\tr,\store_\tr):(\typeb\to\Unit;\namectx)\rangle \ogsTrans{\tr'} \ogsconf$, that either $\tr'\in \depstTr{\depsttr}$ or for all $\tr''\in \ogsCTr{\langle \vdash (\ectx_\tr,\delta_\tr,\store_\tr):(\typeb\to\Unit;\namectx)\rangle}$, $\tr'$ is not a prefix of $\tr''$, and that for each $\fn\in \dom{\gamma}$:
  \begin{itemize}
  \item there exists $O_1,O_2$ such that $O_1\uplus O_2 = \{ \moveO{\fn}{\aval,\pn} \in |\tr'| \}$
  \item there exists $O_1'\subseteq \{\omove\in \depstsuptr\}$ such that $|O_1'|=|O_1|$ and $\forall \omove \in O_1',\loc_\omove \mapsto 1$
  \item there exists $\loc_1,\dots, \loc_{|O_2|}$ pairwise distinct, such that, writing $\fn(\aval_1,\pn_1),\dots,\fn(\aval_{|O_2|},\pn_{|O_2|})$ the elements of $O_2$, for $1\leq i\leq |O_2|$ there exists $\term \in \mathbb{P}(\fn,\aval_i)$ such that $\term[\loc_i/\loc] \in \ogsconf$
  \end{itemize}
\end{proof}

We now show that the context is of the right type
\begin{lemma}
  \label{lm:typ_ctx}
  Let $\term,\store,\typeb,\storectx$ and $\tr_\lab \in \ogsCTr{\langle \namectx \vdash (\term,\store):\typeb\rangle}$ a labelled complete trace. Let $\storectx' = \{\loc_\pmove : \reftype \Int | \pmove \in \tr \}$, then :
  \begin{itemize}
  \item $\typeJudge {\storectx'} \emptyset {\ectx_{\overline{\tr\cdot \pn_f},\overline{\lab}}} {\ectxtype{\typeb}{\Unit}}$
  \item $\storeJudge {\store_{\overline{\tr\cdot \pn_f}}} {\storectx'}$
  \item $\typeJudgeg[\delta_{\overline{\tr\cdot \pn_f},\overline{\lab}}] {\storectx'} \typectx$
  \end{itemize}
\end{lemma}

\begin{proof}
  $\tr_\lab \in \ogsCTr{\langle \namectx \vdash (\term,\store):\typeb\rangle}$ therefore there exists $\namectxO,\namectxP,\viewF$ such that $\langle \namectx \vdash (\term,\store):\typeb\rangle \ogsTrans{\tr_\lab} (\_,\ctuple{\namectxO\mid \namectxP},\ctuple{\viewF,\_,\varepsilon})$.\\
  We write $\namectxb = \namectxO\cup \namectxP \cup (\pn_f : \neg \Unit)$.\\
  We start by showing that if $\lab(\questO{\fn}{\aval',\qn})=\pn$ and $\questP{\gn}{\aval,\pn}\in \tr$ then $\fn \in \viewF(\gn)\cup \supp{\aval}$
  and for $\questO{\fn}{\aval,\pn}\in \tr$, $\typeJudgePat{}{\namectxb_{|\supp(\aval)}}{\aval}{\lnot{(\namectxb(\fn))}}$\\
  This is done by induction on the reduction $\langle \namectx \vdash (\term,\store):\typeb\rangle \ogsTrans{\tr_\lab} (\_,\ctuple{\namectxO\mid \namectxP},\ctuple{\viewF,\_,\varepsilon})$.\\
  Then we show by induction on the construction that :
  \begin{itemize}
  \item $\typeJudge {\storectx'} {\namectxb_{|\viewF(\fn)}} {\gamma (\fn)} {\namectxb(\fn)}$
  \item $\typeJudge {\storectx'} {\namectxb_{|\viewF(\fn)}} {\llbracket \questO{\fn}{\aval,\pn} \rrbracket_{\depsttr[\overline{\tr\cdot\pn_f}],\lab}} {\namectxb(\fn)}$
  \item $\typeJudge {\storectx'} {\namectxb_{|\viewG(\pn)}} {\llbracket \questP{\fn}{\aval,\qn} \rrbracket_{\depsttr[\overline{\tr\cdot\pn_f}],\lab}} {\Unit}$ for $\pn = \lab(\questO{\fn}{\aval,\qn}$
  \item $\typeJudge {\storectx'} {\emptyset} {\llbracket \ansO{\pn}{\val} \rrbracket_{\depsttr[\overline{\tr\cdot\pn_f}],\lab}} {\lnot{\namectxb(\pn)}\to \Unit}$
  \item $\typeJudge {\storectx'} {\emptyset} {\llbracket \ansP{\pn}{\val} \rrbracket_{\depsttr[\overline{\tr\cdot\pn_f}],\lab}} {\lnot{\namectxb(\pn)}}$
  \end{itemize}
  From this we deduce that $\typeJudge {\storectx'} \emptyset {\ectx_{\overline{\tr\cdot \pn_f},\overline{\lab}}} {\ectxtype{\typeb}{\Unit}}$ and  $\typeJudgeg[\delta_{\overline{\tr\cdot \pn_f},\overline{\lab}}] {\storectx'} \typectx$.
  And we have $\storeJudge {\store_{\overline{\tr\cdot \pn_f}}} {\storectx'}$ by definition.

\end{proof}
We now need that our context generates the intended trace, however this will not be true for any traces, we show it for traces generated by a term.
The more precise definition of lemma~\ref{lm:mk_trace} is :

\begin{lemma}
  For all $\term,\store,\typeb$ and a labelled complete trace $\tr_\lab \in \ogsCTr{\langle \namectx \vdash (\term,\store):\typeb\rangle}$,  $\overline{\tr\cdot \pn_f}\in \ogsCTr{\langle \vdash (\ectx_{\overline{\tr\cdot \pn_f},\overline{\lab}},\delta_{\overline{\tr\cdot \pn_f},\overline{\lab}},\store_{\overline{\tr\cdot \pn_f}}):(\typeb\to\Unit;\namectx)\rangle}$
\end{lemma}

We split the proof of this lemma in three parts, one for each part of the configuration after the initial move.
For the interactive configuration we build an interactive configuration for each step of the execution.\\
Let $\lab$ a labeling of $\tr$ such that $\tr_\lab$ is visible. For each non-empty prefix $\tr_1$ of $\tr$, we define the interactive configuration $\iconf_{\tr_1} = (\threadpool_{\tr_1},\store_{\tr_1},\delta_{\tr_1})$ as follow :
\begin{itemize}
\item For $\omove\in \tr_1$, let $\loc_\omove'$ a fresh location name.\\
  $\store_{\tr_1} = \{[\loc_\omove' \mapsto 1]\cdot [\loc_\omove \mapsto 1]|\omove \in \tr_1\} \cup \{[\loc_\omove \mapsto 0]|\omove \in \depstsuptr \setminus \tr_1\}$.
\item $\delta_{\tr_1} = \gamma_{|\{ \fn | \exists \aval,\pn,\pmove\in \tr_1 ,\justif{\pmove}{\questO{\fn}{\aval, \pn}}\}}$
\item For each $\qn\in \{ \lab(\pmove)| \pmove\in \tr_1 \wedge \forall \aval, \ansP{\lab(\pmove)}{\aval}\notin \tr_1\}$
  \[ \threadpool(\qn) =
    \begin{array}[t]{cl}
      \mathtt{par} & \Parn{\pmove\in PM_{\tr_1}(\qn)} {\llbracket \pmove\rrbracket_{\depsttr}}\\
      \mathtt{and} & \Parn{(\omove,\pn)\in OM_{\tr_1}(\qn)} {\suspend {\llbracket \omove \rrbracket_{\depsttr}} \pn }                     
    \end{array}\]
  where $PM_{\tr_1}(\qn) = \{ \pmove\in \depstsuptr \setminus\tr_1| \lab(\pmove)=\qn\}$ and $OM_{\tr_1}(\qn) =\{ (\ansO{\pn}{\aval},\pn)| \ansO{\pn}{\aval}\in \depstsuptr \setminus \tr_1, \exists \pmove\in \tr_1, \lab(\pmove)=\qn \wedge \justif{\pmove}{\ansO{\pn}{\aval}}\}$
  \[ \threadpool(\pn_f) = \suspend{\llbracket \ansP{\pn_f}{()} \rrbracket_{\depsttr}} \pn \text{ with }\pn \text{ such that } \exists (\moveOI{\overrightarrow{\aval_i},\pn})\in \depsttr \]  
\end{itemize}

\begin{lemma}
  \begin{itemize}
  \item Let $\tr_{1,\lab}\cdot \move_\lab$ a prefix of $\tr_\lab$ with $\tr_{1,\lab}$ non-empty. Then $\iconf_{\tr_1}\iTrans{\move_\lab} \iconf_{\tr_1\cdot\move}$
  \item Let $\imove$ the initial move of $\tr$, then there exists $\tconf,\vconf$ such that $\langle \vdash (\ectx_{\tr},\delta_{\tr},\store_{\tr}):(\typeb\to\Unit;\namectx)\rangle \ogsTrans{\imove} (\iconf_{\imove},\tconf,\vconf)$
  \end{itemize}
\end{lemma}
\begin{proof}
  \begin{itemize}
  \item Let $\tr_1\cdot\move$ a prefix of $\tr$ with $\tr_1$ non-empty. We reason by cases on $\move$ :
    \begin{itemize}
    \item $\move = \questP{\fn}{\aval,\pn}$. Let $\qn=\lab(\move)$, then $\move \in PM_{\tr_1}(\qn)$\\
      For each $\omove$ such that $\omove\depstdeptr \move$, we have $\omove\in \tr_1$ therefore $[\loc_{\omove}\mapsto 1]\in \store_{\tr_1}$, hence $\mathtt{check\_lock}_\move \Opred ()$.
      Therefore there exists a unique $\aval'$ such that $(\llbracket \move \rrbracket_{\depsttr},\store_{\tr_1},\delta_{\tr_1}) \iTrans{\move} (\llbracket \suspend{ \llbracket \ansO{\pn}{\aval'}\rrbracket_{\depsttr}} \pn, \store_{\tr_1}, \delta_{\tr_1}\cdot \gamma_{|\mathtt{supp}(\aval)})$.\\
      We have $PM_{\tr_1\cdot \move}(\qn) = PM_{\tr_1}(\qn)\setminus\{\move\}$, $OM_{\tr_1\cdot \move}(\qn) = OM_{\tr_1}(\qn)\cup \{ (\ansO{\pn}{\aval'},\pn)\}$, $\store_{\tr_1\cdot\move}=\store_{\tr_1}$, $\delta_{\tr_1\cdot\move} = \delta_{\tr_1}\cdot \gamma_{|\mathtt{supp}(\aval)}$. Hence $\iconf_{\tr_1}\iTrans{\move}\iconf_{\tr_1\cdot\move}$
    \item $\move = \questO{\fn}{\aval, \pn}$. We have $\fn\in \dom{\delta_{\tr_1}}$ therefore $(\threadpool_{\tr_1},\store_{\tr_1},\delta_{\tr_1})\itrans{\move} (\threadpool_{\tr_1}|\named {(\lab(\move))} \gamma(\fn)\ \aval,\store_{\tr_1},\delta_{\tr_1})$. $\move\in \depstsuptr$ therefore there exists $\aval'\equiv \aval$ such that $\aval'$ is used to match in the definition of $\gamma(\fn)$.\\
      Let $\loc_\move'$ a fresh location, therefore we have $(\named {(\lab(\move))} \gamma(\fn)\ \aval,\store_{\tr_1},\delta_{\tr_1}) \iTrans{} (\named {(\lab(\move))} \llbracket \move \rrbracket_{\depsttr}\ \aval',\store_{\tr_1}\cdot[\loc_\move'\mapsto 1], \delta_{\tr_1})$\\
      $\move \notin \tr_1$ therefore $[\loc_{\move} \mapsto 0]\in \store_{\tr_1}$. Hence $(\named {(\lab(\move))} \llbracket \move \rrbracket_{\depsttr}\ \aval',\store_{\tr_1}\cdot[\loc_\move'\mapsto 1], \delta_{\tr_1}) \iTrans{} (\named {(\lab(\move))} \Parn{\pmove \in \{\pmove| \justif \move \pmove\}}{\llbracket \pmove\rrbracket_{\depsttr}},\store_{\tr_1}\cdot[\loc_\move'\mapsto 1]\cdot [\loc_\move \mapsto 1], \delta_{\tr_1})$.\\
      We have that $\store_{\tr_1\cdot \move} = \store_{\tr_1}\cdot[\loc_\move'\mapsto 1]\cdot [\loc_\move \mapsto 1]$, $\delta_{\tr_1\cdot \move} = \delta_{\tr_1}$, $PM_{\tr_1\cdot\move}(\lab(\move)) = \{\pmove| \lab(\pmove)=\pn\}$ and $OM_{\tr_1\cdot\move}(\lab(\move)) = \emptyset$, hence $\iconf_{\tr_1}\iTrans{\move} \iconf_{\tr_1\cdot\move}$
    \item $\move = \ansP{\pn}{\val}$, this case is similar to the first one.
    \item $\move = \ansO{\pn}{\val}$, this case is similar to the second one, the only differences are that the action is performed in an already existing thread and that we do not need to chose between different possible actions.
    \end{itemize}
  \item This is easy to verify using the rule for the initial move.
  \end{itemize}
\end{proof}

Now for the typing configuration
\begin{lemma}
  Let $\ctuple{\namectxO\mid \namectxP}$ and $\tr$ a trace such that $\ctuple{\namectxO\mid \namectxP}\tTrans{\tr}$. Then $\ctuple{\namectxP\mid \namectxO} \tTrans{\overline{\tr}}$
\end{lemma}
\begin{proof}
  This is by an immediate induction on $\ctuple{\namectxO\mid \namectxP}\tTrans{\tr}$.
\end{proof}

For the visible history configuration we prove a version of lemma~\ref{lm:dual_vis} that preserves label lings.
\begin{lemma}
  \label{lm:dual_vis_bis}
  For all $\term,\store,\typeb$ and labelled complete trace $\tr_\lab \in \ogsCTr{\langle \namectx \vdash (\term,\store):\typeb\rangle}$. $\overline{\tr_\lab\cdot\pn_f}$ is a complete visible trace.
\end{lemma}
\begin{proof}
  The proof is similar to the one of lemma~\ref{lm:dual_vis}
\end{proof}

We can now prove lemma~\ref{lm:mk_trace}
\begin{proof}
  Let $\term,\store,\typeb, \tr_\lab \in \ogsCTr{\langle \namectx \vdash (\term,\store):\typeb\rangle}$.
  Combining the previous lemmas we have that after the initial move, each part of the configuration can make the trace $\overline{\tr_\lab\cdot\pn_f}$ which is a complete trace, therefore $\overline{\tr\cdot \pn_f}\in \ogsCTr{\langle \vdash (\ectx_{\overline{\tr\cdot \pn_f},\overline{\lab}},\delta_{\overline{\tr\cdot \pn_f},\overline{\lab}},\store_{\overline{\tr\cdot \pn_f}}):(\typeb\to\Unit;\namectx)\rangle}$  
\end{proof}

We can now prove lemma~\ref{lm:ex_context}
\begin{proof}
  We instantiate the lemma with the context we built previously, we have that $\overline{\tr\cdot \pn_f}\in \ogsCTr{\langle \vdash (\ectx_{\overline{\tr\cdot \pn_f},\overline{\lab}},\delta_{\overline{\tr\cdot \pn_f},\overline{\lab}},\store_{\overline{\tr\cdot \pn_f}}):(\typeb\to\Unit;\namectx)\rangle}$ and by lemma~\ref{lm:CTr_lttr} we have that $\{\tr'\lttr \overline{\tr\cdot \pn_f}| \tr'\text{ is visible}\}\subseteq \ogsCTr{\langle \vdash (\ectx_{\overline{\tr\cdot \pn_f},\overline{\lab}},\delta_{\overline{\tr\cdot \pn_f},\overline{\lab}},\store_{\overline{\tr\cdot \pn_f}}):(\typeb\to\Unit;\namectx)\rangle}$.\\
  Moreover, by lemma~\ref{lm:tr_dep}, for any $\tr$ we have
  $\ogsCTr{\langle \vdash (\ectx_{\tr,\lab},\delta_{\tr,\lab},\store_{\tr,\lab}):(\typeb\to\Unit;\namectx)\rangle}\subseteq \depstCTr{\depsttr}$,
  and by lemma~\ref{lm:depsttr_lttr} we have $\depstCTr{\depsttr}= \{\tr'\lttr \tr | \tr' \text{ is visible}\}$.\\
  Hence $\{\tr'\lttr \overline{\tr\cdot \pn_f}| \tr'\text{ is visible}\}= \ogsCTr{\langle \vdash (\ectx_{\overline{\tr\cdot \pn_f},\overline{\lab}},\delta_{\overline{\tr\cdot \pn_f},\overline{\lab}},\store_{\overline{\tr\cdot \pn_f}}):(\typeb\to\Unit;\namectx)\rangle}$.
\end{proof}

We can finally complete the proof of Theorem~\ref{thm:completeness}
\begin{proof}
  Let $\storectx,\namectx,\typeb,\term_1,\term_2$ such that $\storectx,\namectx \vdash \term_1\ciule \term_2 : \typeb$.\\
  Let $\store$ such that $\vdash \store:\storectx$, and $\tr \in \ogsCTr{\langle \namectx \vdash (\term_1,\store):\typeb\rangle}$.\\
  By lemma~\ref{lm:ex_context}, there exists a context $\ectx$, a value assignment $\gamma$ and a store $\store'\disjoint \store$ such that $\ogsCTr{\langle \vdash (\ectx,\gamma,\store'):(\typeb\to \Unit;\namectx)\rangle} = \{ \tr' | \tr'\lttr \overline{\tr\cdot\pn_f}\ \text{ with }\tr' \text{ visible }\}$, hence by lemma~\ref{cor:par_tr} we have that $\compose{\langle \namectx \vdash (\term_1,\store):\typeb\rangle}{\langle \vdash (\ectx,\gamma,\store'):(\ectxtype{\typeb}{\Unit};\namectx)\rangle}\Downarrow \pn_f$.\\
  Using $\storectx,\namectx \vdash \term_1\ciule \term_2 : \typeb$ we have that $\compose{\langle \namectx \vdash (\term_2,\store):\typeb\rangle}{\langle \vdash (\ectx,\gamma,\store'):(\ectxtype{\typeb}{\Unit};\namectx)\rangle}\Downarrow \pn_f$.\\
  Using lemma~\ref{cor:par_tr} again, we have that there exists $\tr'$ such that $\tr'\in \ogsCTr{\langle \namectx \vdash (\term_2,\store):\typeb\rangle}$ and $\overline{\tr'\cdot \pn_f} \in \ogsCTr{\langle \vdash (\ectx,\gamma,\store'):(\typeb\to \Unit;\namectx)\rangle} = \{ \tr' | \tr'\lttr \overline{\tr\cdot\pn_f}\ \text{ with }\tr' \text{ visible }\}$, therefore $\overline{\tr'\cdot \pn_f}\lttr \overline{\tr \cdot \pn_f}$ now by lemma~\ref{lm:lttr_dual} we have that $\tr \lttr \tr'$.\\
  Furthermore, $\tr$ is visible therefore by lemma~\ref{lm:CTr_lttr} we have that $\tr\in \ogsCTr{\langle \namectx \vdash (\term_2,\store):\typeb\rangle}$.
\end{proof}
}{}

\end{document}